\documentclass[preprint]{aa_arxiv}
\pdfoutput=1

\usepackage{amsmath}
\usepackage{amsfonts}
\usepackage{dsfont}
\usepackage{amsxtra}
\usepackage{hyperref}
\usepackage{amssymb}
\usepackage{upgreek}

\usepackage{multirow}
\usepackage{url}

\usepackage{graphicx,epsfig}

\usepackage{xspace} 

\usepackage{color}
\usepackage{xcolor}

\usepackage[mathlines, switch]{lineno}
\usepackage{comment}

\usepackage{subcaption}
\usepackage{graphicx}
\usepackage{siunitx}

\newcommand{\be}{\begin{equation}}
\newcommand{\ee}{\end{equation}}
\newcommand{\bea}{\begin{eqnarray}}
\newcommand{\eea}{\end{eqnarray}}
\newcommand{\beaa}{\begin{eqnarray*}}
\newcommand{\eeaa}{\end{eqnarray*}}
\newcommand{\ba}{\begin{array}}
\newcommand{\ea}{\end{array}}
\newcommand{\bi}{\begin{itemize}}
\newcommand{\ei}{\end{itemize}}
\newcommand{\ben}{\begin{enumerate}}
\newcommand{\een}{\end{enumerate}}

\newcommand{\td}{\tilde}

\newcommand{\lb}{\label}
\newcommand{\g}{\ensuremath{\gamma}\xspace}
\newcommand{\G}{\Gamma}

\newcommand{\al}{\alpha}

\newcommand{\p}{\partial}

\newcommand{\ld}{\lambda}

\newcommand{\Om}{\Omega}

\newcommand{\sm}{\sigma}

\newcommand{\La}{{\mathcal{L}}}

\newcommand{\Fermi}{\textsl{Fermi}\xspace}

\newcommand{\DAMPE}{\textsl{DAMPE}\xspace}
\newcommand{\Healpix}{HEALPix\xspace}

\newcommand{\onepic}{0.49}
\newcommand{\twopic}{0.49}

\newcommand{\threepic}{0.33}

\newcommand{\red}{\textcolor{red}}

\definecolor{darkgreen}{rgb}{0.0, 0.7, 0.0}

\newcommand{\cmt}[1]{}

\newcommand{\low}{\text{low}}

\newcommand{\east}{\text{east}}
\newcommand{\west}{\text{west}}
\newcommand{\de}{\text{d}}
\newcommand{\IC}{\text{IC}}

\newcommand{\el}{\text{e}}
\newcommand{\pr}{\text{p}}
\newcommand{\cut}{\text{cut}}

\newcommand{\Hy}{\text{H}}
\newcommand{\tot}{\text{tot}}

\begin{document}

   \title{Hard and bright gamma-ray emission at the base of the \Fermi bubbles}

   \author{L. Herold \thanks{\email{laura.herold@fau.de}}
          \inst{1}
          \and
          D. Malyshev \thanks{on leave of absence from ITEP, B. Cheremushkinskaya st. 25, Moscow, Russia 117218, 
          \email{dmitry.malyshev@fau.de}}
          \inst{1}
          }

   \institute{
             Erlangen Centre for Astroparticle Physics, Erwin-Rommel-Str. 1, Erlangen, Germany
             }


\abstract
{The \Fermi bubbles (FBs) are large gamma-ray emitting lobes extending up to $55^\circ$ in latitude above and below the Galactic center (GC).
Although the FBs were discovered 8 years ago, their origin and the nature of the gamma-ray emission are still unresolved.
Understanding the properties of the FBs near the Galactic plane may provide a clue to their origin.
Previous analyses of the gamma-ray emission at the base of the FBs, what remains after subtraction of Galactic foregrounds,
have shown an increased intensity compared to the FBs at high latitudes,
a hard power-law spectrum
without evidence of a cutoff up to approximately 1 TeV,
and a displacement of the emission to negative longitudes relative to the GC.
}
{We analyze 9 years of  \Fermi Large Area Telescope data in order to study in more detail than the previous analyses
the gamma-ray emission at the base of the FBs,
especially at energies above 10 GeV.
}
{ We use a template analysis method to model the observed gamma-ray data
and calculate the residual emission after subtraction of the expected foreground and background emission components.
Since there are large uncertainties in the determination of the Galactic gamma-ray emission towards the GC,
we use several methods to derive Galactic gamma-ray diffuse emission as well as the contribution from point sources
in order to estimate the uncertainties in the emission at the base of the FBs.
}
{We confirm that the gamma-ray emission at the base of the FBs is well described by a simple power law up to 1 TeV energies.
The 95\% confidence lower limit on the cutoff energy is about 500 GeV.
It has larger intensity than the FBs emission at high latitudes and is shifted to the west (negative longitudes) from the GC.
If the emission at the base of the FBs is indeed connected to the high-latitude FBs, 
then the shift of the emission to negative longitudes disfavors models where the FBs are created by
the supermassive black hole at the GC.
We find that the gamma-ray spectrum can be explained either by gamma rays produced in hadronic interactions or by 
leptonic inverse Compton scattering.
In the hadronic scenario, the emission at the base of the FBs can be explained either by several hundred supernova remnants (SNRs) near the Galactic center
or by about 10 SNRs at a distance of $\sim$ 1 kpc.
In the leptonic scenario, the necessary number of SNRs that can produce the required density of CR electrons
is a factor of a few larger than in the hadronic scenario.
}
{}


\maketitle

\tableofcontents

\section{Introduction}
\lb{sec:intro}

The \Fermi bubbles (FBs) are one of the most spectacular and unexpected discoveries 
in the \Fermi Large Area Telescope (LAT) data \citep{2010ApJ...724.1044S}.
The FBs extend up to $55^\circ$ in latitude above and below the Galactic center (GC),
they have a well-defined edge and a relatively uniform intensity across the surface, apart from a ``cocoon'' in the south eastern part of the bubbles
\citep{2012ApJ...753...61S, 2014ApJ...793...64A}.
The intensity spectrum is $\sim E^{-2}$ at GeV energies with a cutoff or a softening around 100 GeV at latitudes $|b| > 10^\circ$ \citep{2014ApJ...793...64A}.
The origin of the FBs is attributed either to an emission from the supermassive black hole (SMBH) at
the center of our Galaxy
or to a period of starburst activity which resulted in a combined wind
from supernova (SN) explosions of massive stars
\citep{2010ApJ...724.1044S}.
The gamma-ray signal up to 100 GeV can be produced either by interactions of hadronic cosmic rays (CR) with gas (hadronic model)
or by inverse Compton (IC) scattering of high energy electrons and positrons and the interstellar radiation (leptonic model).

Although the FBs were detected about 8 years ago, their origin is still unresolved.
Important insights into their origin can be obtained from the study of the morphology and the spectrum of the FBs near the GC.
The spectrum of gamma rays can provide information on the composition of the 
CR that produce the gamma-ray signal (hadronic vs leptonic CR),
as well as the age of the CR (through a cooling cutoff in the leptonic scenario or a break due to escape of high energy CR)
and the spectrum of the CR at the source.
The morphology of the emission can point to the source of the bubbles: either the SMBH Sgr A* or a recent star-forming region.
Previous analyses of the FBs at low latitudes indicated higher intensity of emission near the Galactic plane (GP) and a displacement
to negative longitudes \citep{2016ApJS..223...26A, 2017ApJ...840...43A, 2017JCAP...08..022S}.
The spectrum of the FBs for $|b| < 10^\circ$ is consistent with a power-law $\propto E^{-2}$ 
without a cutoff up to 1 TeV \citep{2017ApJ...840...43A}.

The study of the FBs in the GP is complicated due to bright Galactic diffuse emission components.
The $\pi^0$ component of the gamma-ray emission is well traced by the distribution of gas,
but it has large uncertainties towards the GC due to a lack of kinematic information from the motion of the gas in the 
Galaxy, which is used to reconstruct the gas distribution
(the velocity of the gas in the direction of the GC is perpendicular to the line of sight)
as well as the uncertainties in the CO emission along the line of sight to the GC (which is used as a tracer of molecular hydrogen)
and large dispersion of velocities of some molecular clouds near the GC. 
There are also large uncertainties in the distribution of the CR sources and the propagation model near the GC,
which make it rather difficult to predict a priori 
the distribution of the propagated CR in the Galaxy.
For the IC component of the gamma-ray emission, 
there is a significant uncertainty in the interstellar radiation field (ISRF) density near the GC 
\citep[][]{2017MNRAS.470.2539P, 2017ApJ...846...67P,  2019APh...107....1N} in addition to 
the uncertainties in the CR distribution.
There should also be undetected point-like and extended sources, which nevertheless contribute to the total flux.
Distribution of CR in the Galaxy can be computed with CR propagation tools, such as GALPROP \citep{2007ARNPS..57..285S}.
The maps of gamma-ray emission from interactions of CR with gas in different Galactocentric rings and from interaction
of CR electrons and positrons with the ISRF can be used as templates for the corresponding components
of gamma-ray emission.
The agreement of the gamma-ray data with models based on templates derived with the CR propagation tools
is rather poor in the GP
\citep[e.g.,][]{2017ApJ...846...67P}.

In this paper, we analyze the gamma-ray emission at the base of the FBs and 
estimate the uncertainties
related to modeling of the Galactic foreground and background components.
We focus on morphology and spectrum of the FBs at energies from 10 GeV to 1 TeV,
where the intensity of the gamma-ray emission from the FBs 
relative to the other Galactic components
is higher than at low energies due to softer spectra of the Galactic components.
The study of the FBs at high energies will be also important for future searches with neutrino and Cherenkov telescopes.

We use several methods to determine the Galactic foreground/background emission.
Unfortunately the notion of the foreground and background emission may not be well defined if we search for
an extended gamma-ray emitting source based on the gamma-ray data itself
(rather than using the multi-wavelength data, e.g., to trace the distribution of gas).
Here and in the following by foreground/background emission we will mean a steady state 
(or average) diffuse emission of gamma-rays.
The steady state distribution of CR is obtained by averaging in time over many sources, it results in the local CR proton spectrum
of $\sim E^{-2.7}$ for energies from a few GeV to about a PeV and $\sim E^{-3.3}$ for the local CR electrons from GeV to TeV energies.
If the spectrum of the CR at the source is $\sim E^{-2.0 - 2.2}$, then the softening of the spectrum of CR protons by $\sim E^{-0.3} - E^{-0.6}$ 
is due to energy-dependent escape from the Galaxy,
while the softening for the CR electrons by $\sim E^{-1}$ is due to cooling.
A distinguishing property of a source of CR is that the spectrum of CR at or near the source is much harder than the average (propagated)
spectrum of CR.
Thus, one can look for the presence of a population of freshly accelerated CR by searching for areas of gamma-ray emission
with spectrum harder than average. In particular, one can use an ``on-off'' analysis to subtract the steady state component of the gamma-ray emission.
One of the caveats of this analysis is that even the steady state gamma-ray emission in the ``on'' region can be more intense than the emission in the ``off'' region,
but in this case the difference of the fluxes would have a soft spectrum characteristic of the propagated CR,
unless the difference in flux is much smaller than the flux in both ``on'' and  ``off''  regions: in this case the difference can have a harder spectrum than the two terms.
We use the ``on-off'' technique as a preliminary check in a search for a population of freshly accelerated CR:
the hard spectrum of the difference is a necessary condition for the presence of a population of CR with a spectrum harder than the steady state distribution of CR.
It is also a sufficient condition for the presence of a population of CR with a hard spectrum, if the difference
has intensity comparable to the overall intensity in the ``on'' and  ``off''  regions.

Consequently, on consideration that there is a tentative asymmetry in diffuse gamma-ray emission near the GC at high energies with a spectrum
harder to the west of the GC relative to the emission to the east of the GC,
as a first step,
we estimate the amount of the asymmetry by taking the difference in the gamma-ray data to the west and to the east of the GC 
after masking bright point sources.
Second, since the spectrum of the FBs at high latitudes as well as the inferred spectrum at low latitudes
is harder than the spectra of the other components of diffuse emission,
the contribution of the FBs at energies $\lesssim 1$ GeV is relatively small. 
We use the data below 1 GeV as a template of the Galactic foreground emission, 
which we fit to the data at energies above 1 GeV.
The residual emission is used to estimate the contribution from the components with spectra harder than the typical 
spectra of the steady state diffuse gamma-ray components.
Finally, we determine a model for the Galactic gamma-ray diffuse emission using templates for the emission 
components based on GALPROP calculation%
\footnote{\url{http://galprop.stanford.edu}} 
\citep{Moskalenko:1997gh, Strong:1998fr, Strong:2004de, Ptuskin:2005ax, 2007ARNPS..57..285S, Porter:2008ve,Vladimirov:2010aq}. 
In this model, we allow many free parameters, such as rescaling of the $\pi^0$ and bremsstrahlung emission in Galactocentric rings and refitting of bright point sources near the GC.
With the many free parameters, this model absorbs as much of the gamma-ray emission at the base of the FBs as it can. As a result, the residual emission at the base of the FBs in this model is smaller than the residuals in the other models considered in the paper.

The similarity of the energy spectrum below 100 GeV of the hard and bright emission at the base of the FBs 
and the FBs at high latitudes as well as the spatial coincidence of the emission with a continuation of the FBs from high latitudes 
make the physical correspondence of the two objects very plausible.
Nevertheless, their alignment along the line of sight can be accidental and the distance to the two objects can be different, i.e., the FBs may be
above and below the GC while the hard component may be at a closer distance, e.g., 1 kpc, from us so that the two objects would be physically unrelated.
Thus, we will refer to the hard and bright component as ``emission at the base of the FBs'' to keep both possibilities open: 
that the emission is the base of the FBs and that the emission is at a different location along the line of sight towards the base of the FBs.
We discuss possible origins of the hard component of emission in Section \ref{sec:Interpretation}.
Section \ref{sect:concl} contains conclusions.

\section{Data selection}
\lb{sec:data}

\begin{figure*}[h]
\includegraphics[width=\threepic\textwidth]{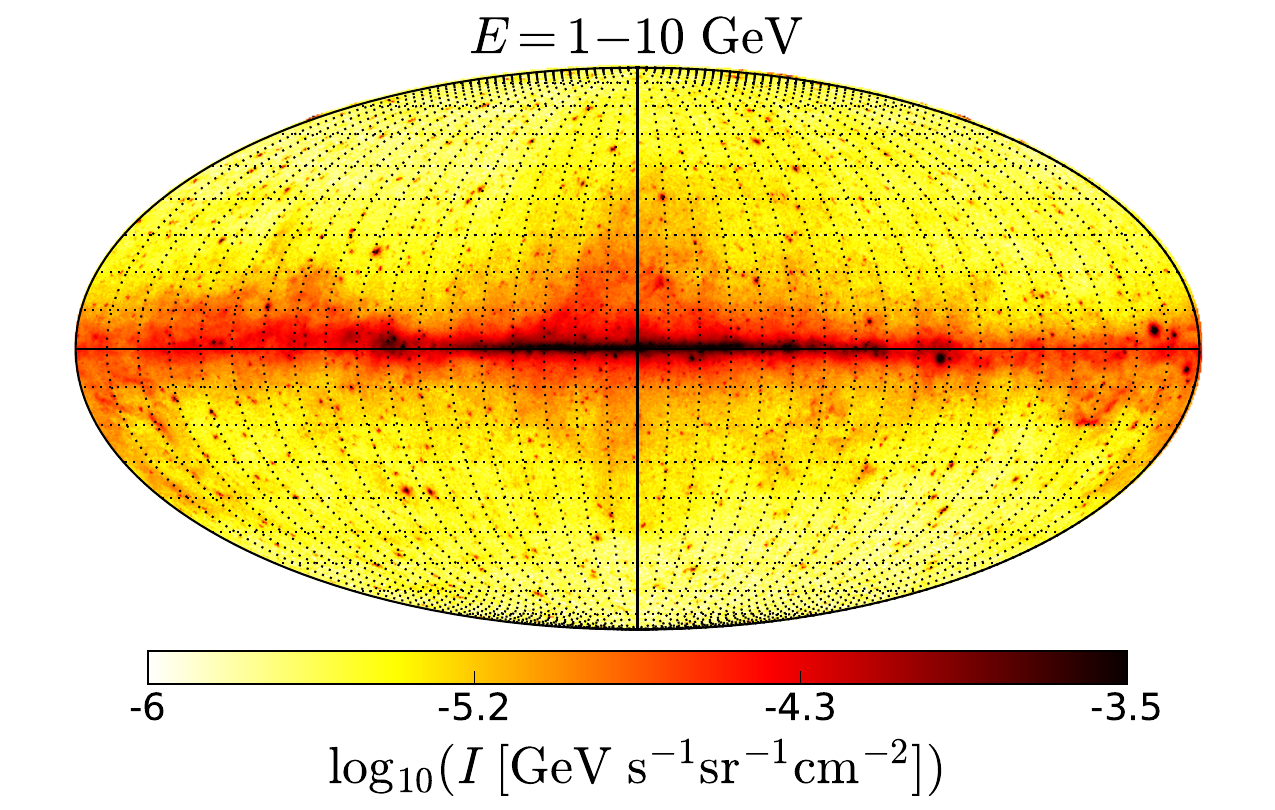}
\includegraphics[width=\threepic\textwidth]{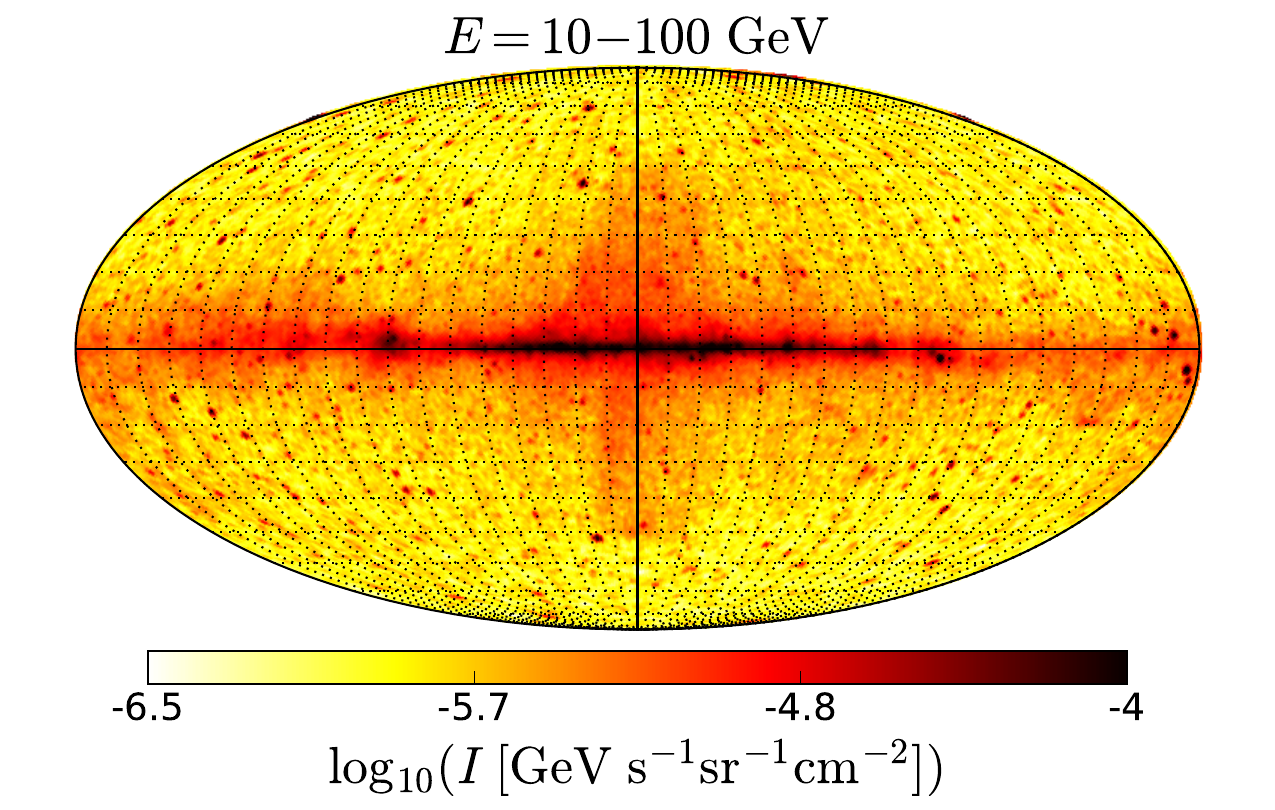}
\includegraphics[width=\threepic\textwidth]{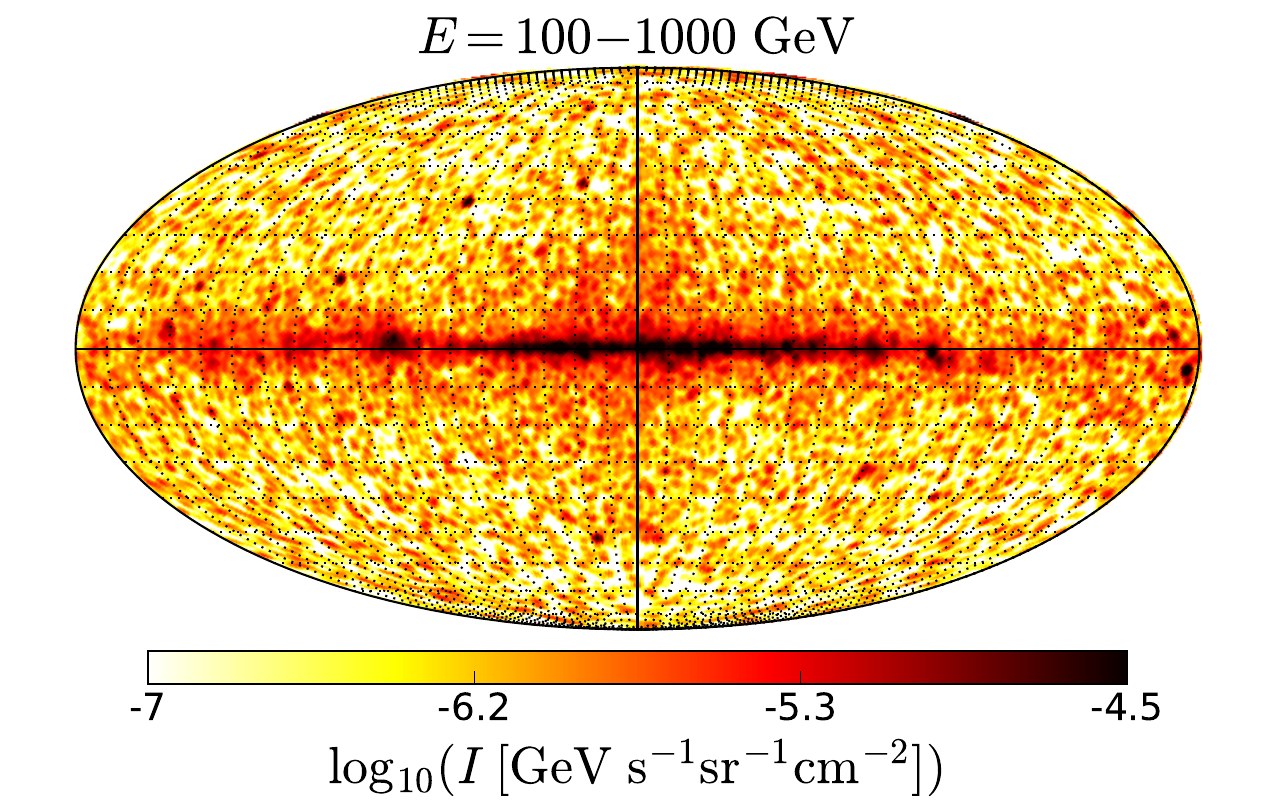}
\caption{
Energy flux of the \Fermi-LAT 9 years of Source class data integrated in energy ranges 1 -- 10 GeV, 10 -- 100 GeV, and 100 GeV -- 1 TeV.
The maps are represented in logarithmic scale. 
The map in the energy range 10 -- 100 GeV (100 -- 1000 GeV) 
is smoothed with a Gaussian kernel with radius $\sigma = 0{\fdg}5$ ($\sigma = 0{\fdg}7$).
The plots are created with the \Healpix package \citep{2005ApJ...622..759G} in Mollweide projection with 
nside = 128 ($\approx 0{\fdg}46$ pixel size). 
}
\label{fig:Maps_data}
\end{figure*}

We use 9 years of the \Fermi-LAT Pass 8 Source class events
between August 4, 2008  and August 3, 2017 ({\Fermi} Mission Elapsed Time 239557418\,s -- 523411376\,s)
with energies between 316 MeV $ = 10^{2.5}$ MeV
and 1 TeV separated in 21 logarithmic energy bins (6 bins per decade).
The selection of the events is performed with the standard quality cuts (DATA\_QUAL$>$0)\&\&(LAT\_CONFIG==1).
In order to avoid contamination from gamma rays produced in interactions of CR in the Earth atmosphere, 
we select events with zenith angles $\theta < 100^{\circ}$,
which is sufficient for energies above 316 MeV.
We calculate the exposure and point-spread function (PSF) using the {\Fermi}-LAT Science Tools package version 
10-01-01 available from the {\Fermi} Science Support Center\footnote{\url{http://fermi.gsfc.nasa.gov/ssc/data/analysis/}} 
with the P8R2\_SOURCE\_V6 instrument response functions.
Figure \ref{fig:Maps_data} shows the energy flux of the gamma-ray emission
integrated in three energy ranges%
\footnote{The integral of the energy flux intensity between $E_0$ and $E_1$ is defined as
$I = \int_{E_0}^{E_1} E \frac{dN}{dE} dE.$}.
For all maps shown in this paper, we use Galactic coordinates centered on the GC in Mollweide projection.
For spatial binning we use the \Healpix\footnote{\url{http://sourceforge.net/projects/healpix/}} \citep{2005ApJ...622..759G} scheme with a pixelization of order 7  
($\approx 0{\fdg}46$ pixel size).

\section{Modeling of the \Fermi bubbles at low latitudes}
\label{sec:Modeling}

\subsection{East-west  asymmetry in the data}
\label{sec:data_diff}

In order to investigate the asymmetry of the emission at the base of the FBs relative to the GC, 
we compare the \Fermi-LAT data east and west of the GC. 
We mask the 200 point sources (PS) in the Third \Fermi-LAT source catalog \citep[3FGL,][]{2015ApJS..218...23A}
with the largest flux above 1 GeV.
Each PS is masked with a circle of radius $\frac{\delta}{\sqrt{2}} + 0{\fdg}5$, where $\delta = 0{\fdg}46$ is the characteristic size of the pixels
so that ${\delta}/{\sqrt{2}}$ is half of the pixel diagonal (if it were a square), while $0{\fdg}5$ is comparable to 
the 68\% containment radius of the point-spread function above $\approx 2$ GeV.
Thus the total radius is about $0{\fdg}82$.
In order to avoid possible bias by masking more pixels on one side of the GC in the Galactic plane,
we symmetrize the PS mask relative to the GC, i.e., we set $m_{-b, -\ell} = 0$ if $m_{b, \ell} = 0$ and vice versa.
This PS mask is also used in the following sections (apart from Section \ref{sec:galprop_model}).
The data are averaged over regions to the east ($0^\circ < \ell < 10^\circ$) and to the west ($-10^\circ < \ell  <  0^\circ$) 
of the GC at different latitudes. 
The regions have a latitude width of $4^\circ$. 
The fraction of masked pixels is about 50\% within $|b| < 2^\circ$, about 10\% for $2^\circ < |b| < 6^\circ$, and less than  5\% 
for $|b| > 6^\circ$.

\begin{figure}[h]
\hspace{-2mm}
 \includegraphics[width=\onepic\textwidth]{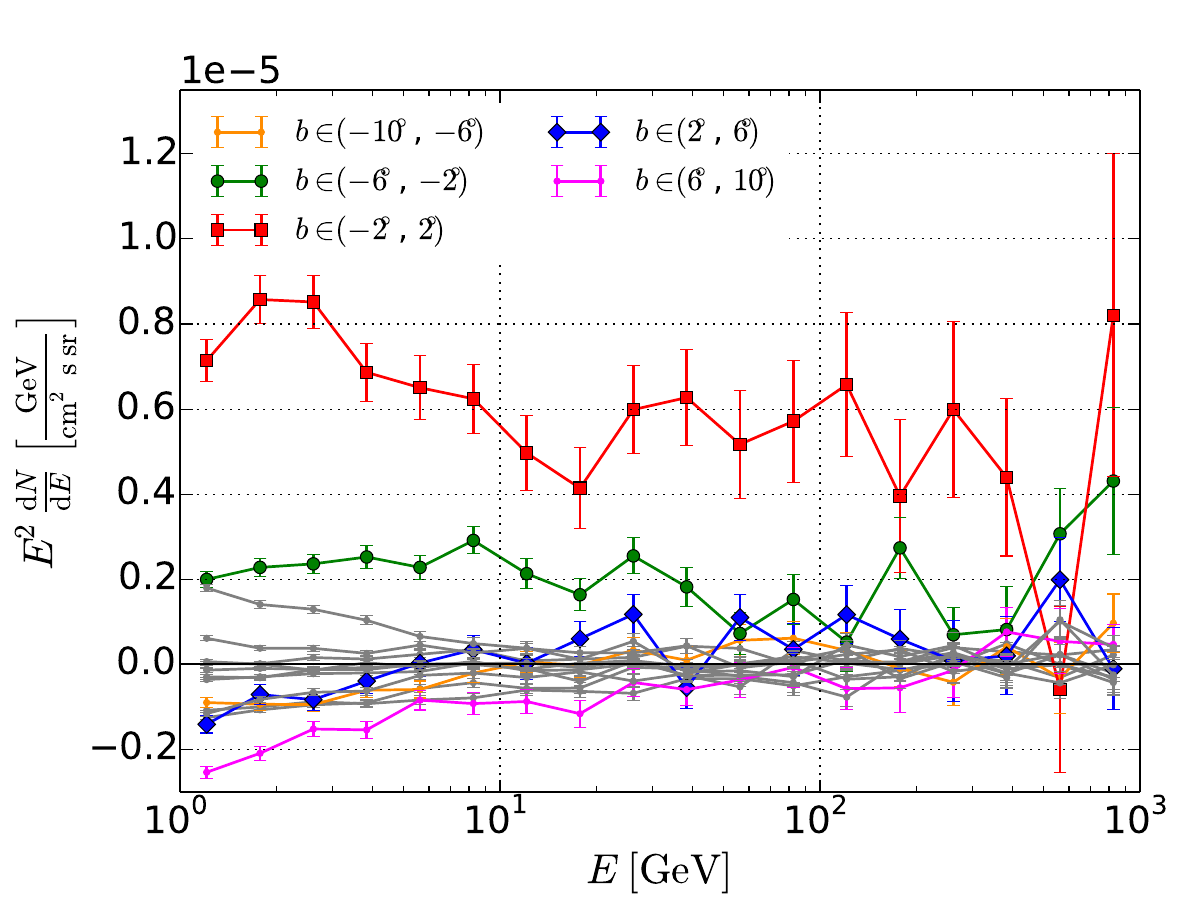}
 \caption{Difference west minus east in the \Fermi-LAT intensity relative to the GC after masking bright PSs.
 The PS mask is symmetrized relative to the GC.
 The west (east) region is defined between $-10^\circ < \ell <0^\circ$ ($0^\circ < \ell <10^\circ$).
 Grey lines show spectra for latitudes $|b| > 10^\circ$. 
 }
 \label{fig:data_diff}
\end{figure}

The difference of the averaged intensity of emission west minus east of the GC is shown in Fig. \ref{fig:data_diff}. 
The error bars represent the statistical errors.
The emission for latitudes $b \in (-6^\circ, -2^\circ)$ and $b \in (-2^\circ, 2^\circ)$ shows excess emission to the west of the GC, 
which remains significant at high energies. 
The Fermi-LAT exposure within $|b| < 10^\circ$ and $|\ell| < 10^\circ$ is rather uniform; 
the maximal fractional variation of the exposure in this region for $E > 1$ GeV is less than 4\%.

\subsection{Low-energy data as a background model}
\label{sec:le_data_model}

\begin{figure*}[t]
\includegraphics[width=\threepic\textwidth]{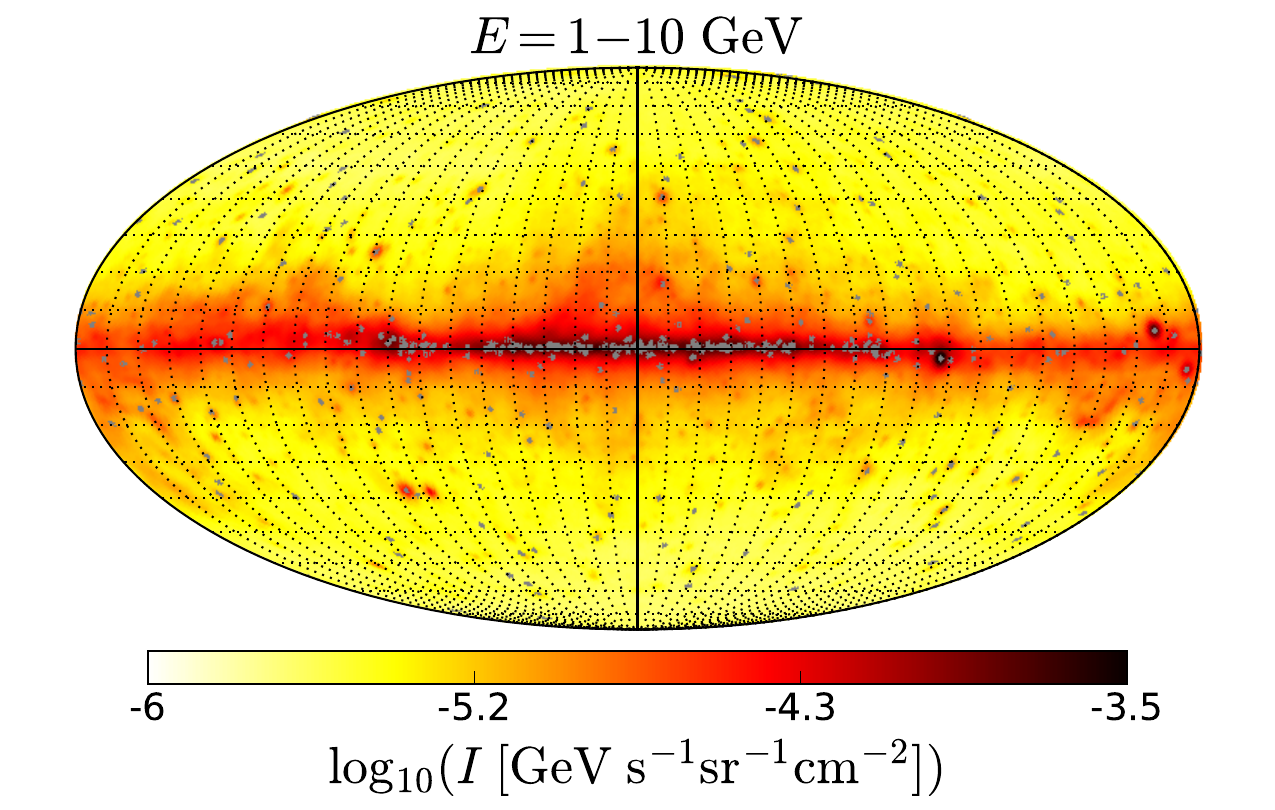}
\includegraphics[width=\threepic\textwidth]{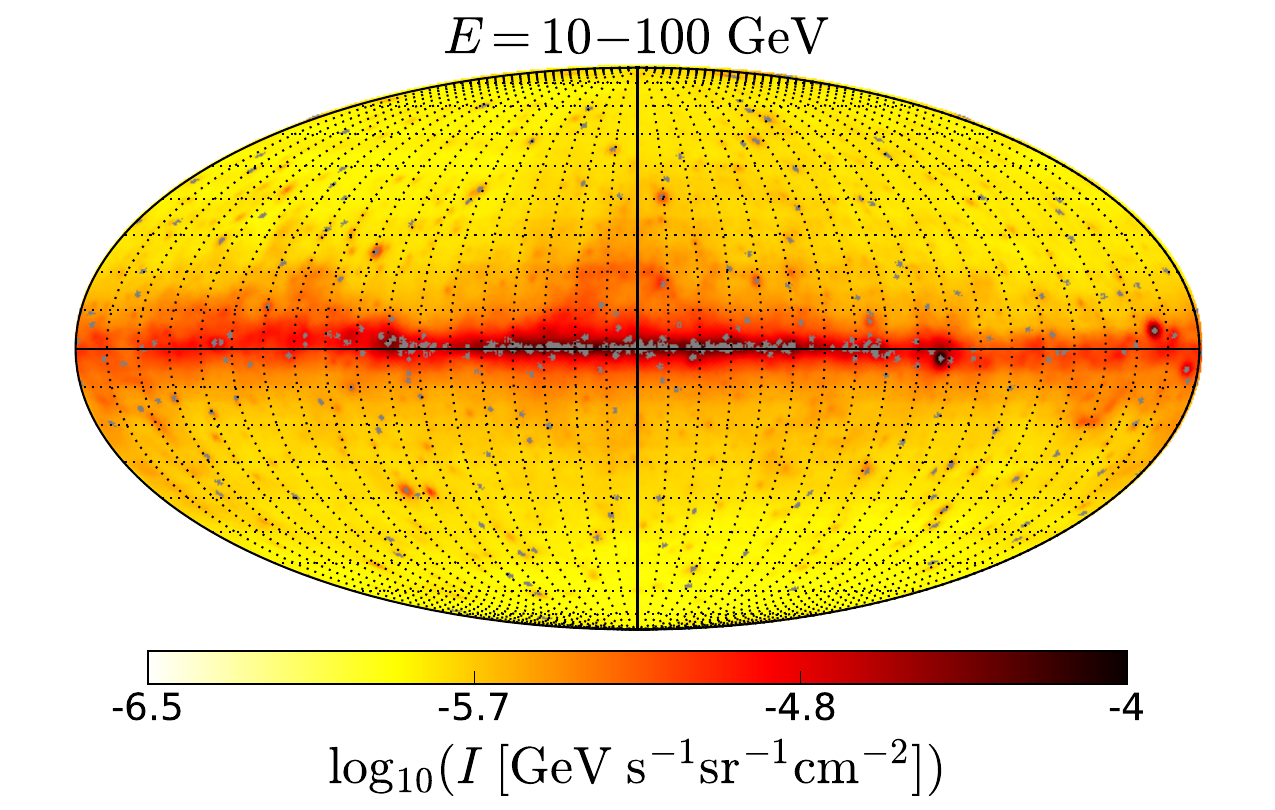}
\includegraphics[width=\threepic\textwidth]{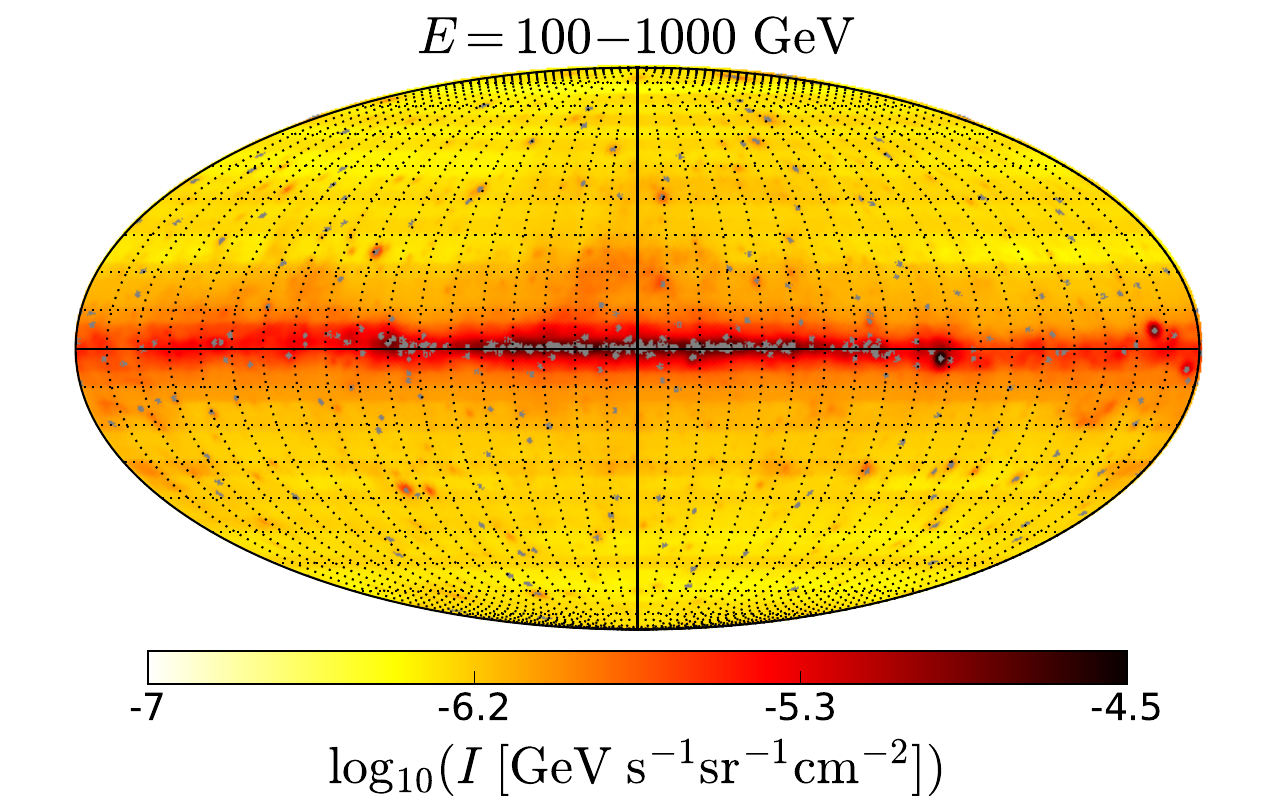}
\caption{Energy flux of the diffuse emission model based on the low-energy data (Section \ref{sec:le_data_model})
integrated over three energy ranges. }
\label{fig:Maps_lowE_model}
\end{figure*}

\begin{figure*}[t]
\includegraphics[width=\threepic\textwidth]{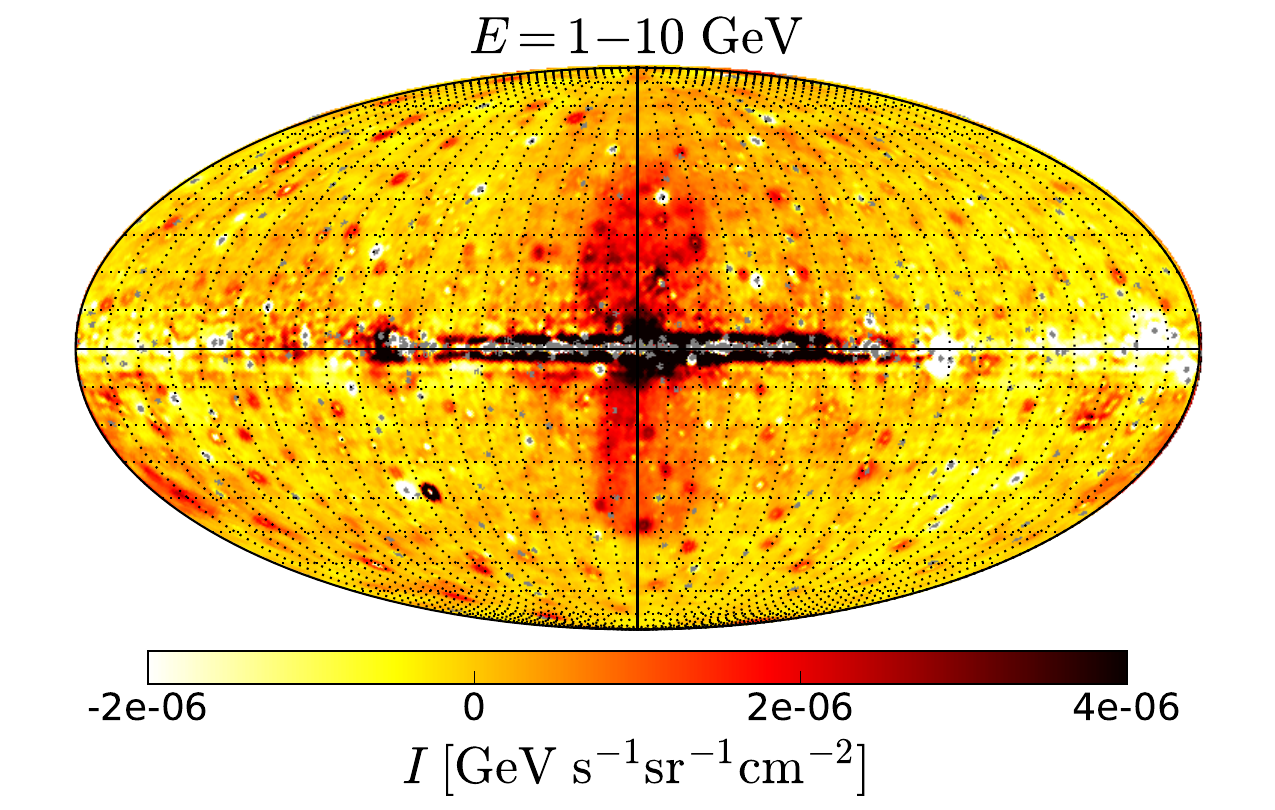}
\includegraphics[width=\threepic\textwidth]{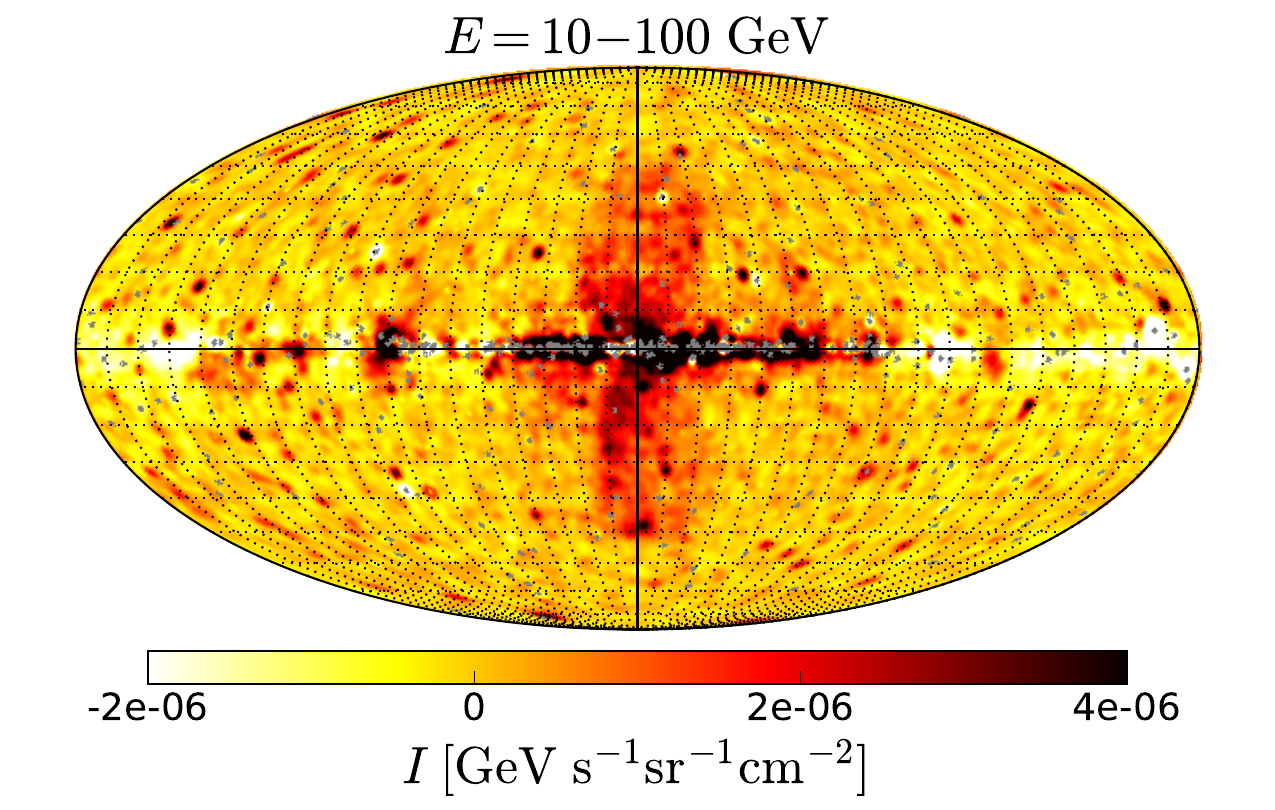}
\includegraphics[width=\threepic\textwidth]{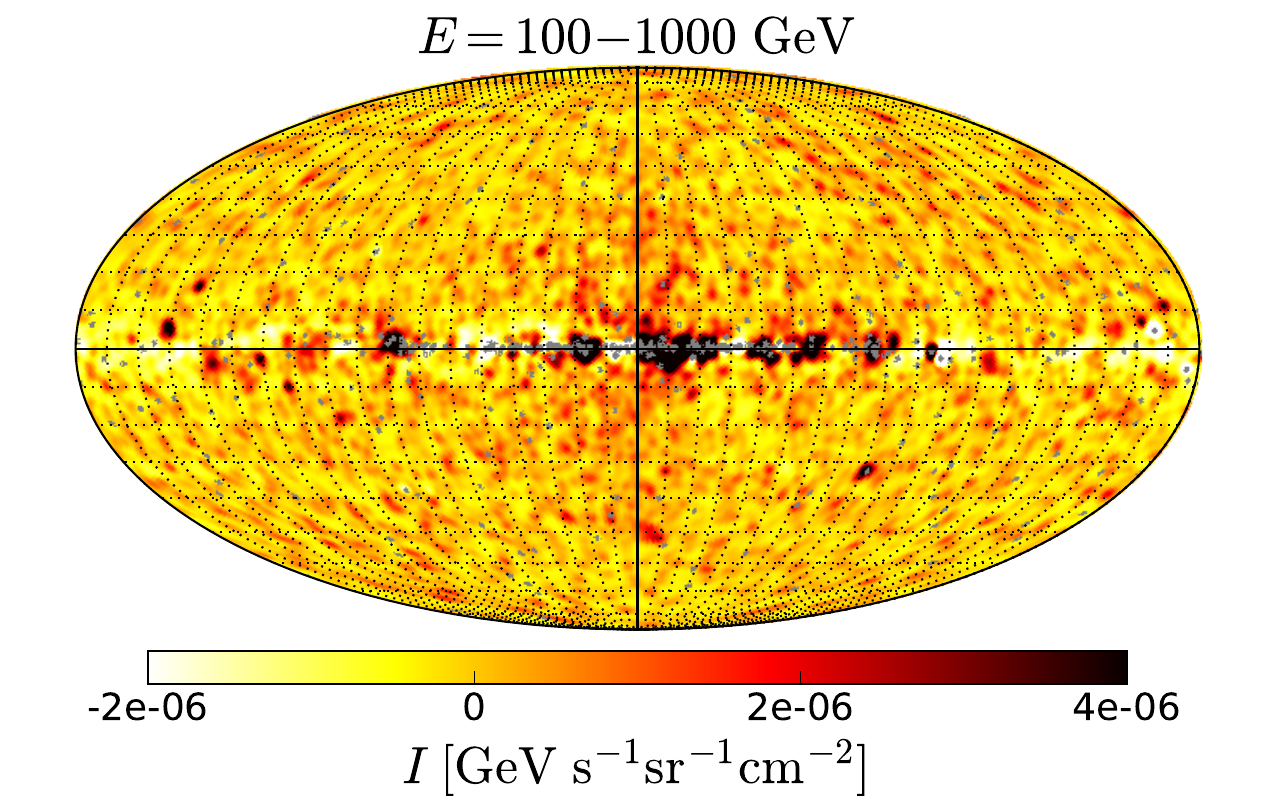}
\caption{Energy flux of the residuals of the low-energy model derived in Section \ref{sec:le_data_model}
integrated over three energy ranges. }
\label{fig:Maps_lowE}
\end{figure*}

In the previous section we have shown that the west-minus-east difference in the \Fermi-LAT data in the Galactic plane relative to the GC has a hard 
spectrum $\sim E^{-2}$ up to 1 TeV.
One of the simplest ways to separate the soft component of emission from the hard one is to use the low-energy data as
a model of the soft component and subtract it from the data at higher energies.
Gamma rays produced in interactions of the steady state Galactic population of CR with gas and ISRF,
i.e., the $\pi^0$, bremsstrahlung, and IC components,
dominate the gamma-ray emission in the Galactic plane in the energy range $E \lesssim \SI{1}{GeV}$. 
Consequently, the low-energy \Fermi-LAT data is a good tracer for the soft components of the
diffuse gamma-ray emission in the Galactic plane and can be used to create a spatial template for the Galactic foreground.
The 68\% containment for the \Fermi-LAT photons between $\SI{316}{MeV}$ and $\SI{1}{GeV}$ (averaged with an $E^{-2}$ spectrum) is about 
$1{\fdg}5$,
which is much larger than the sub-degree angular resolution above $\SI{1}{GeV}$.
In order to compensate for the difference in the angular resolution, 
we smooth the data in each high-energy bin above 1 GeV
with a Gaussian kernel of $\sigma = \ang{1}$ (which corresponds to $68\%$ containment angle of
$1{\fdg}5$ in 2 dimensions).%
\footnote{In general, the width of the Gaussian kernel $\sigma$ can be obtained by subtracting in quadrature
the equivalent widths at low and at high energies $\sigma^2 = \sigma_{\rm low}^2 - \sigma_{\rm high}^2$.
Since $\sigma_{\rm low}^2 \gg \sigma_{\rm high}^2$, we neglect $\sigma_{\rm high}^2$ and smooth the 
data at high energies with $\sigma_{\rm low} = 1^\circ$.}

We separate the whole sky in latitude stripes of $\ang{4}$ width where the \Healpix pixels are assigned to stripes that contain the centers of the pixels.
We parametrize the model in each stripe and each energy bin independently, i.e., 
in the latitude stripe $b$ for energy bin $E$ and \Healpix pixel with index $i$ the model consists of two terms:
\be
\lb{eq:le_model}
M_{b}(E, i) = k_{b}(E) \cdot \tilde N^\low_{b}(E, i) + c_b(E) \cdot \tau(E, i),
\ee
where $k_{b}(E)$ and $c_b(E)$ are coefficients to be determined from the fit to the data.
The term $c_b(E) \cdot \tau(E, i)$ is proportional to the exposure $\tau(E, i)$: it accounts for the isotropic extragalactic background and partially compensates for 
the latitude-dependent IC emission.
$k_{b}(E) \cdot \tilde N^\low_b(E, i)$ is proportional to the low-energy photon counts summed over 
$n_\low = 3$ energy bins between $\SI{316}{MeV}$ and $\SI{1}{GeV}$ 
rescaled by the ratio of exposures at low and at high energies:
\be
\tilde N^\low_b(E, i) = \frac{1}{n_\low} \left(\sum_{\alpha = 1}^{n_\low} \frac{N(E_\al, i)}{\tau(E_\al, i)} \right)\tau(E, i),
\ee
where $N(E_\al, i)$ is the number of photons in energy bin $E_\al$ and in pixel $i$.

We determine the parameters $c_{b}(E)$ and $k_{b}(E)$ by fitting the model to the \Fermi-LAT data in energy bins 
$E > \SI{1}{GeV}$.
We maximize the log likelihood of the gamma distribution 
with respect to parameters $c_{b}(E)$ and $k_{b}(E)$
using the Python iminuit optimizer%
\footnote{
The gamma distribution $p(x; \td{k} + 1) = \frac{x^{\td{k}}}{\G (\td{k} + 1)}e^{-x}$ is the probability density function
for the parameter $x$ given the observed counts $\td{k}$.
The counts $\td{k}$ are integer for unsmoothed maps and non-integer for the smoothed maps.
We discuss the optimization of the model parameters using smoothed maps in 
Appendix \ref{app:gamma}.
}.
In order to avoid an overcompensation 
for the flux from the FBs, the region $-20^\circ < \ell < 20^\circ$
is excluded from the fit, i.e., the fit is performed for 
the total remaining length of the stripe $20^\circ < \ell < 340^\circ$.
In the following, the region $-20^\circ < \ell < 20^\circ$ will be refered to as the FBs region of interest (ROI).
Bright PSs are masked with the mask described in Section \ref{sec:data_diff}.

After we fit the model in each latitude stripe, we interpolate it inside the bubbles ROI.
The intensity of the model energy flux integrated in three energy ranges
is shown in Figure \ref{fig:Maps_lowE_model},
while the residuals after subtracting the model from the data are presented in Figure \ref{fig:Maps_lowE}.
The FBs are clearly visible in the first two energy ranges, $E = 1 - \SI{10}{GeV}$ and $E = 10 - \SI{100}{GeV}$.
For $E = \SI{100}{GeV} - \SI{1}{TeV}$ the statistics are low, but one can still see an excess near the GP.
There are other regions of large residuals in the Galactic plane as well as at high latitudes.
These are diffuse and point-like sources with spectra harder than the average spectra of Galactic and
extragalactic sources.
In particular, there is a residual in diffuse emission along the Galactic plane within $|\ell | \lesssim 60^\circ$ at energies below 100 GeV.
This residual can be due to IC emission, which has a harder spectrum than the average spectrum of the $\pi^0$ component,
or to a harder spectrum of cosmic rays in the inner Galaxy \citep{2015PhRvD..91h3012G, 2016ApJS..223...26A, 2016PhRvD..93l3007Y},
or to a population of hard PSs in the Galactic plane in the inner parts of the Galaxy.
At energies above 100 GeV the residuals are resolved into localized sources, some of which are extended.
Apart from the Galactic foreground and background components, there are gamma rays emitted by the interactions of CRs with 
the Sun and its radiation field \citep{2011ApJ...734..116A} and the Moon \citep{2016PhRvD..93h2001A}.
The intensities of the emission from the Sun and the Moon are about an order of magnitude less than 
the intensity of the emission from the FBs. 
As a result, these components are not visible on  the residual maps at energies above 10 GeV.

\subsection{Rectangles model of the bubbles}
\label{sec:box_model}

\begin{figure*}[h]
\centering
 \includegraphics[width=\twopic\textwidth]{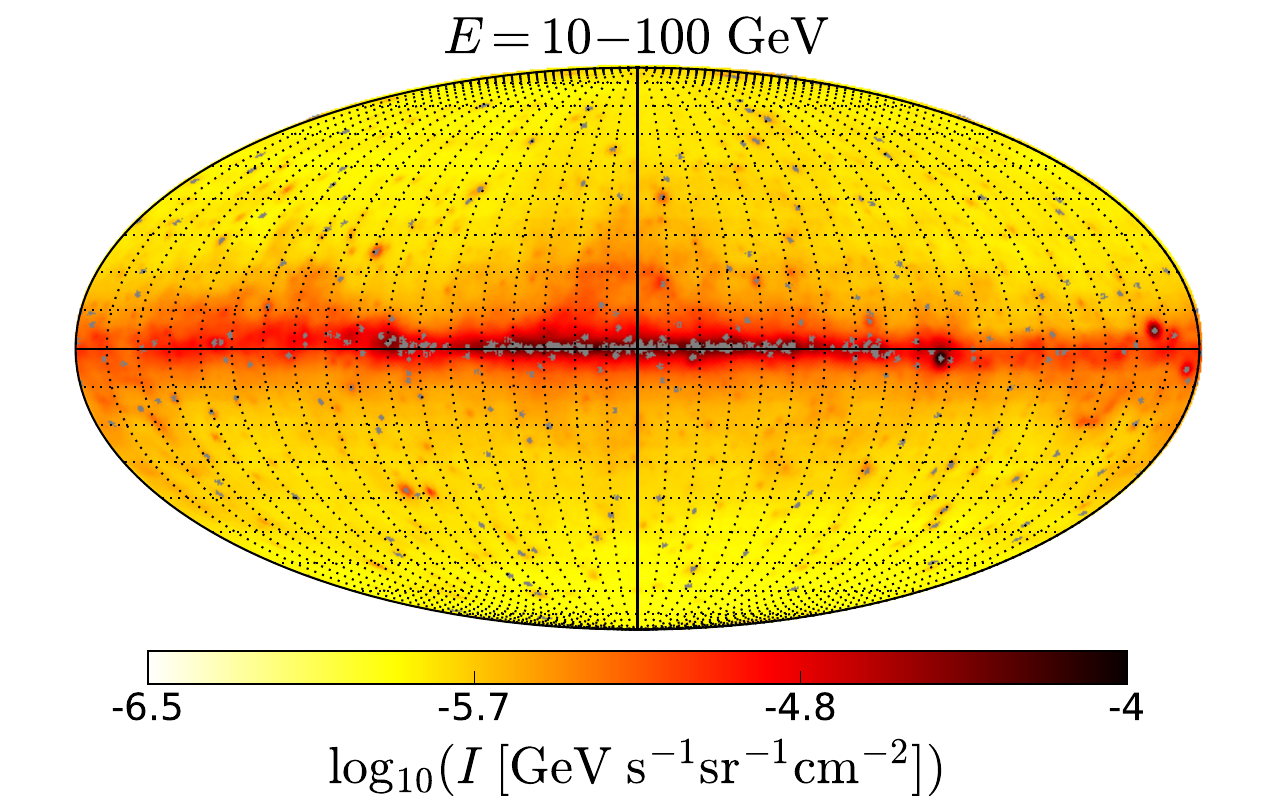}
\includegraphics[width=\twopic\textwidth]{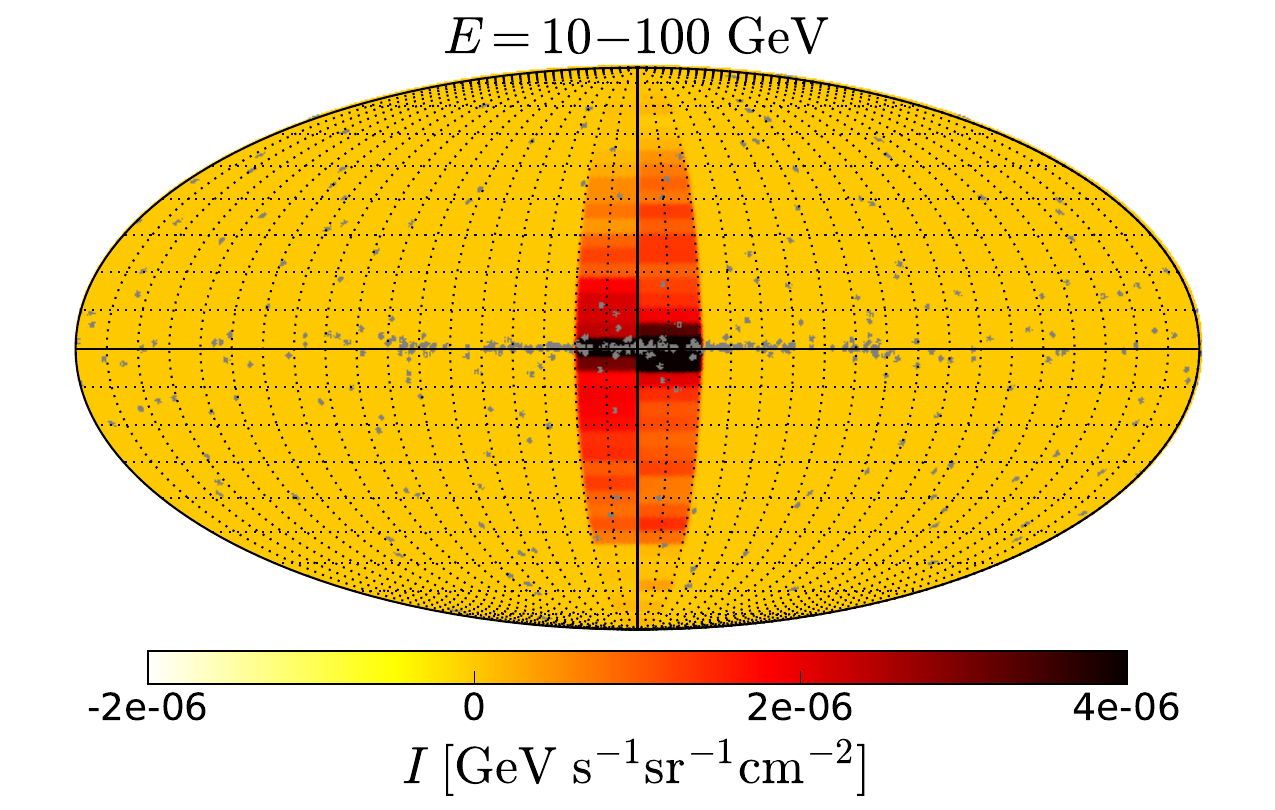}\\
\includegraphics[width=\twopic\textwidth]{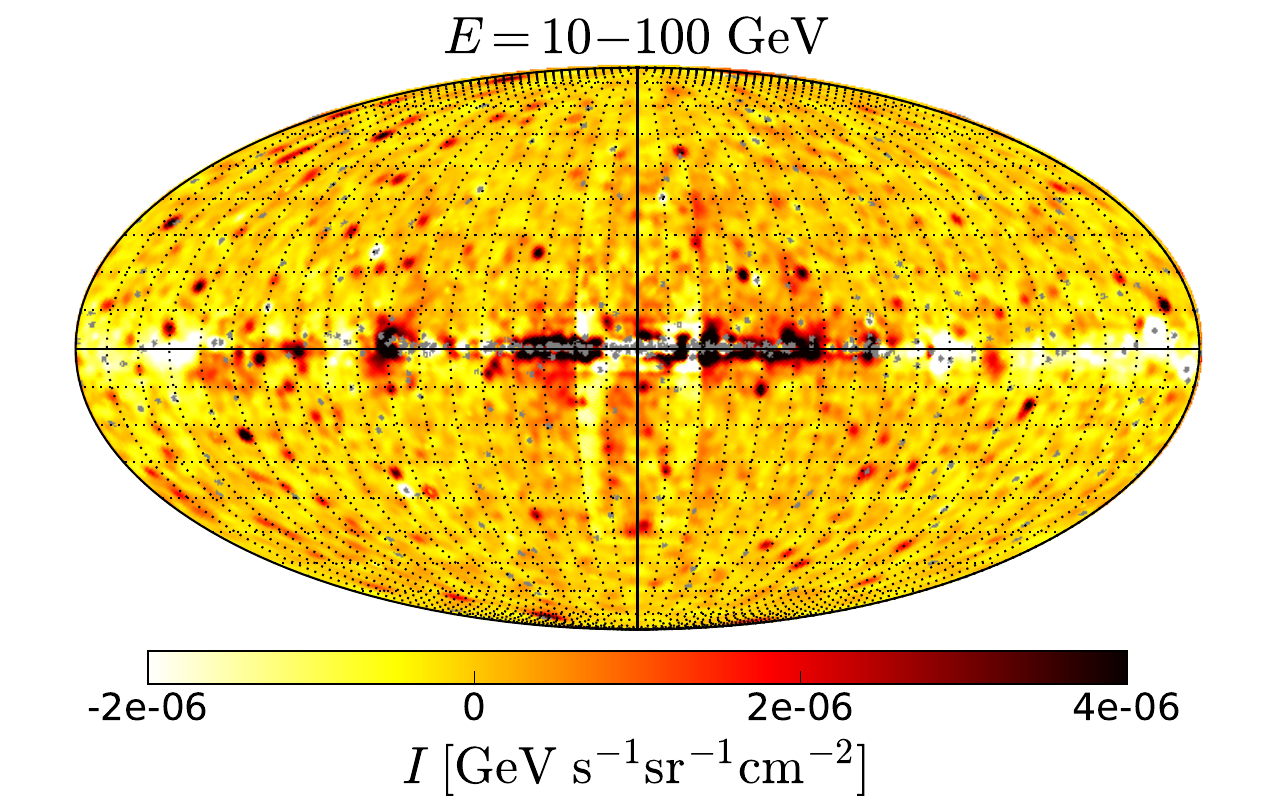}
 \includegraphics[width=\twopic\textwidth]{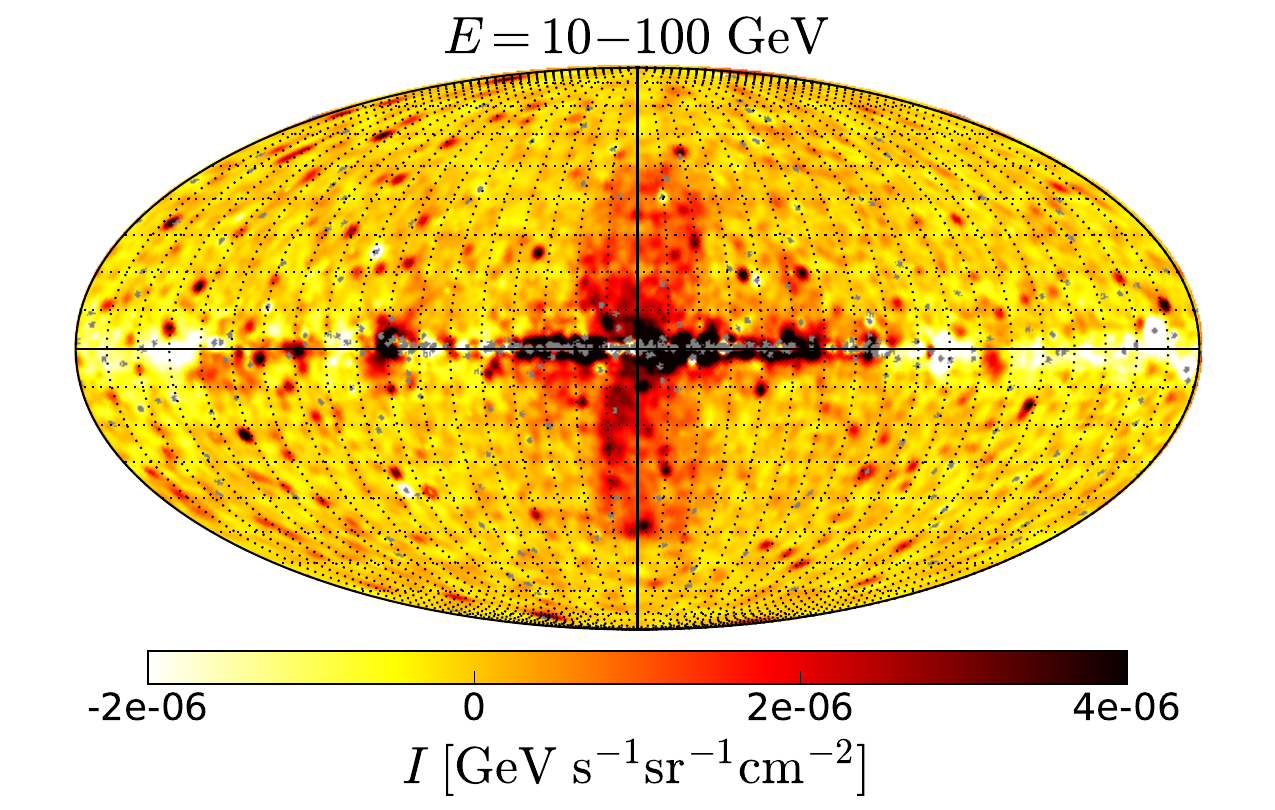}
 \caption{Energy flux of the diffuse model excluding the rectangles model of the FBs (top left),
 the rectangles model of the FBs (top right), 
the residual after subtracting the total model (bottom left),
and residuals plus the rectangles model of the FBs (bottom right)
 derived in Section \ref{sec:box_model}
 integrated between 10 and 100 GeV.}
 \label{fig:Maps_Rectangles}
\end{figure*}

In the previous subsection, we exclude the ROI of the FBs.
In this subsection, we perform the fit over the whole sky and model the emission from the FBs using rectangles.
In order to explore the east-west asymmetry of the FBs, 
we introduce two rectangular templates in each latitude stripe $b$ and energy bin $E$: 
one east, $\ell \in (0^\circ, 20^\circ)$, and one west, $\ell \in (-20^\circ, 0^\circ)$, of the GC.
The width of the rectangles is $4^\circ$, i.e., the same as the width of the latitude stripes.
We use the same foreground model as in Section \ref{sec:le_data_model} plus the isotropic template.
Overall, the model has four terms in each energy bin:
\be
\begin{split}
M_{b}(E, i) &= k_{b}(E) \cdot \tilde N^\low_{b}(E, i) + c_b(E) \cdot \tau(E, i)\\
& + R^\east_b(E) S^\east_b(i)+ R^\west_b(E) S^\west_b(i).
\end{split}
\ee
where the scaling parameters $k_{b}(E)$ and $c_{b}(E)$
and the FBs model parameters $R^\east_b(E)$ and $R^\west_b(E)$ are determined independently 
in each $4^\circ$ latitude stripe and in each energy bin $E > 1$ GeV
by fitting the model to the smoothed \Fermi-LAT data
(see Section  \ref{sec:le_data_model} for details on the smoothing of the data).
$S_b^\east(i)$  and $S_b^\west(i)$ are step function templates, which are equal to 1 in each latitude stripe $b$
for $0^\circ < \ell < 20^\circ$ and $0^\circ > \ell > - 20^\circ$ respectively and 0 otherwise.
The energy fluxes of the model and the residuals integrated in the range $E = 10 - \SI{100}{GeV}$ are shown in Fig. \ref{fig:Maps_Rectangles}.

\subsection{GALPROP model of the foreground and PS refitting}
\label{sec:galprop_model}

\begin{figure*}[h]
\centering
\includegraphics[width=\twopic\textwidth]{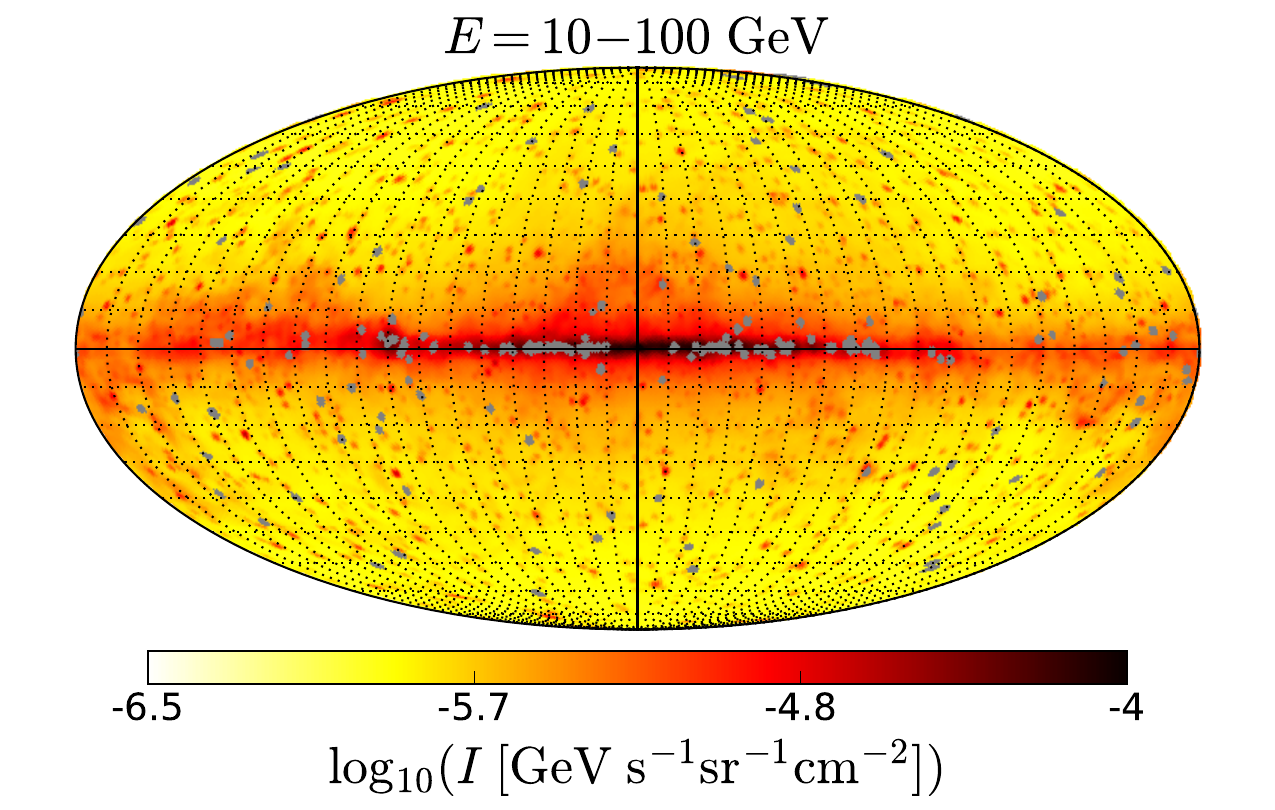}
 \includegraphics[width=\twopic\textwidth]{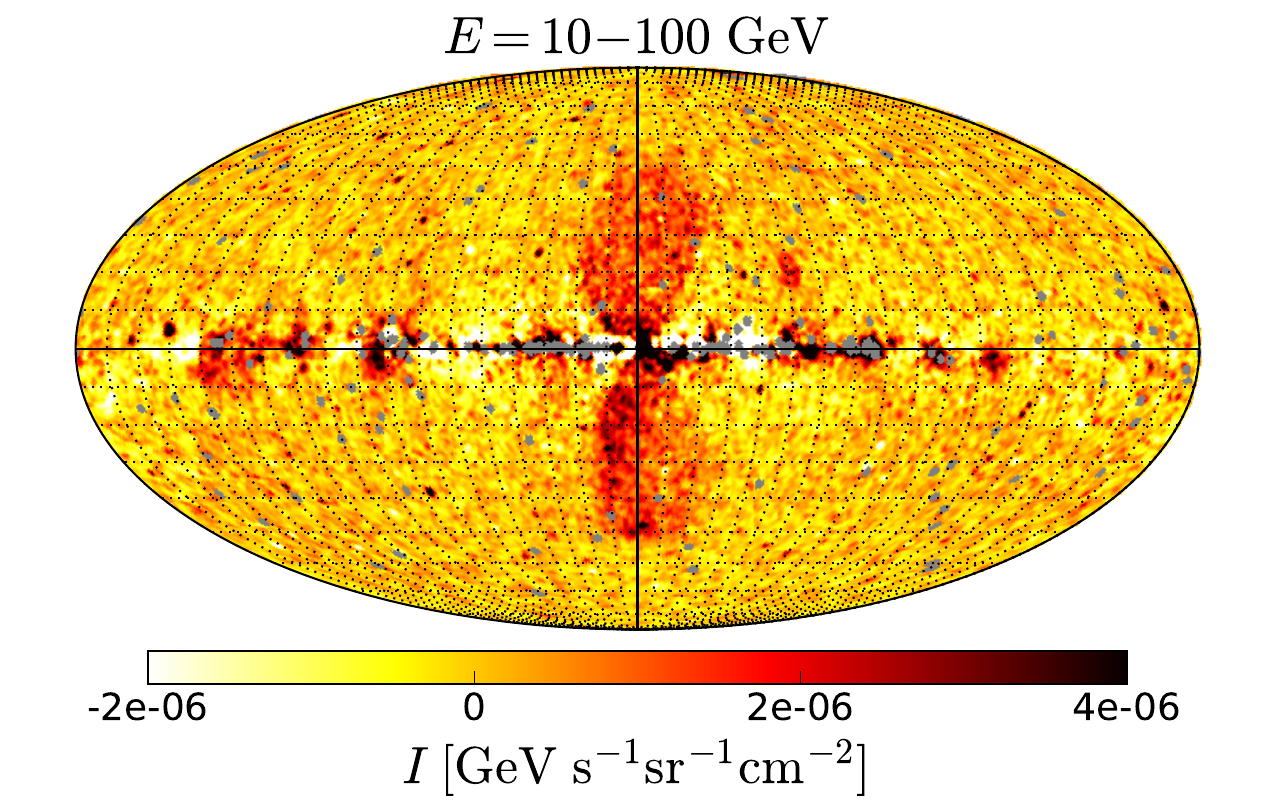}
 \caption{Energy flux of the background model based on rescaled GALPROP templates (left)
 and of the residuals plus the FBs and the gNFW component (right) derived in Section \ref{sec:galprop_model}
 integrated between 10 and 100 GeV.}
 \label{fig:Maps_GALPROP}
\end{figure*}

In this section we fit the gamma-ray data using a model derived with the GALPROP
Galactic cosmic-ray propagation code v54.1.
The model for the diffuse emission components is the same as the Sample model in \cite{2017ApJ...840...43A}.
In particular, we use the GALPROP code to calculate templates for the Galactic diffuse emission components.
We determine 5 templates in each energy bin for gamma rays produced in 
interactions of hadronic CR with gas and bremsstrahlung emission corresponding to 5 Galactocentric rings: 
0 -- 1.5\,kpc, 1.5 -- 3.5\,kpc, and 3.5 -- 8\,kpc; 8 -- 10\,kpc, and 10 -- 50\,kpc.
We assume propagation halo height of 10 kpc and radius of 20 kpc and spin temperature 150 K.
We use 3 inverse Compton templates corresponding to the three ISRF components: cosmic microwave background (CMB), 
infrared (IR) emission of dust, and starlight (SL).
In addition to GALPROP templates, we use a flat template for the \Fermi bubbles at latitudes $|b| > 10^\circ$~\citep{2014ApJ...793...64A}. 
We model the Loop~I feature using a geometric template \citep[e.g., Figure 2 of][]{2014ApJ...793...64A}
based on a polarization survey at 1.4 GHz~\citep{Wolleben:2007pq}.
Templates for the gamma-ray emission from the Sun  and the Moon
\citep{2008A&A...480..847O, 2011ApJ...734..116A, 2013arXiv1307.0197J, 2016PhRvD..93h2001A}
are obtained with the \Fermi Science Tools%
\footnote{\url{http://fermi.gsfc.nasa.gov/ssc/data/analysis/scitools/solar_template.html}}.
We also include, as a GC excess template, 
DM annihilation with a generalized Navarro-Frenk-White (gNFW) profile with index $\gamma = 1.25$ 
\citep{Goodenough:2009gk,Abazajian:2014fta, Calore:2014xka}
and scaling radius $r_{\rm s} = 20\;{\rm kpc}$, as well as the isotropic template to model the non-resolved 
extragalactic sources.

For the PS, we add sources from the 3FGL catalog~\citep{2015ApJS..218...23A} to form a template
in each energy bin.
We create independent templates for the Large Magellanic cloud and the Cygnus region.
The remaining extended sources are also added in a separate template.
The spatial templates for the extended sources are provided by the 3FGL catalog.
The PS mask in this section is different from the mask that we have used in the previous sections.
We mask the 200 sources with largest flux above 1 GeV outside a $10^\circ$ circle from the GC.
Each source is masked with a circle of $\frac{\delta}{\sqrt{2}} + 1^\circ$ radius ($\delta = 0{\fdg}46$ is the characteristic size of the pixels),
which is larger than the radius of  $\frac{\delta}{\sqrt{2}} + 0{\fdg}5$ used in the previous sections.
We do not mask PS within $10^\circ$ from the GC.
Instead, we include independent templates for 40 sources with largest flux between 10 and 100 GeV
taken from the column ``Flux10000\_100000'' in the 3FGL catalog.
Each source is fit in each energy bin independently, i.e., we do not use the spectral information from the 3FGL catalog.
In Figure \ref{fig:Maps_GALPROP} we show the residual emission plus the \Fermi bubbles model and the gNFW model summed over energies between 10 and 100 GeV.
Due to the model having more free parameters in the Galactic plane relative to the models considered in the previous sections,
there are fewer positive residuals in the Galactic plane.
There are, however, negative residuals in several locations, which indicates errors in the model. 
In particular, there is a rather significant negative residual to the east of the GC.
Nonetheless, the FBs are clearly visible as well as the bright emission at the base of the FBs which also displays a shift to negative
longitudes relative to the GC.

\section{Morphology and spectrum of the gamma-ray emission at the base of the FBs}

\subsection{Latitude and longitude profiles}
\label{sec:Latitude_profiles}

\begin{figure*}[h]
\centering
\includegraphics[width=\twopic\textwidth]{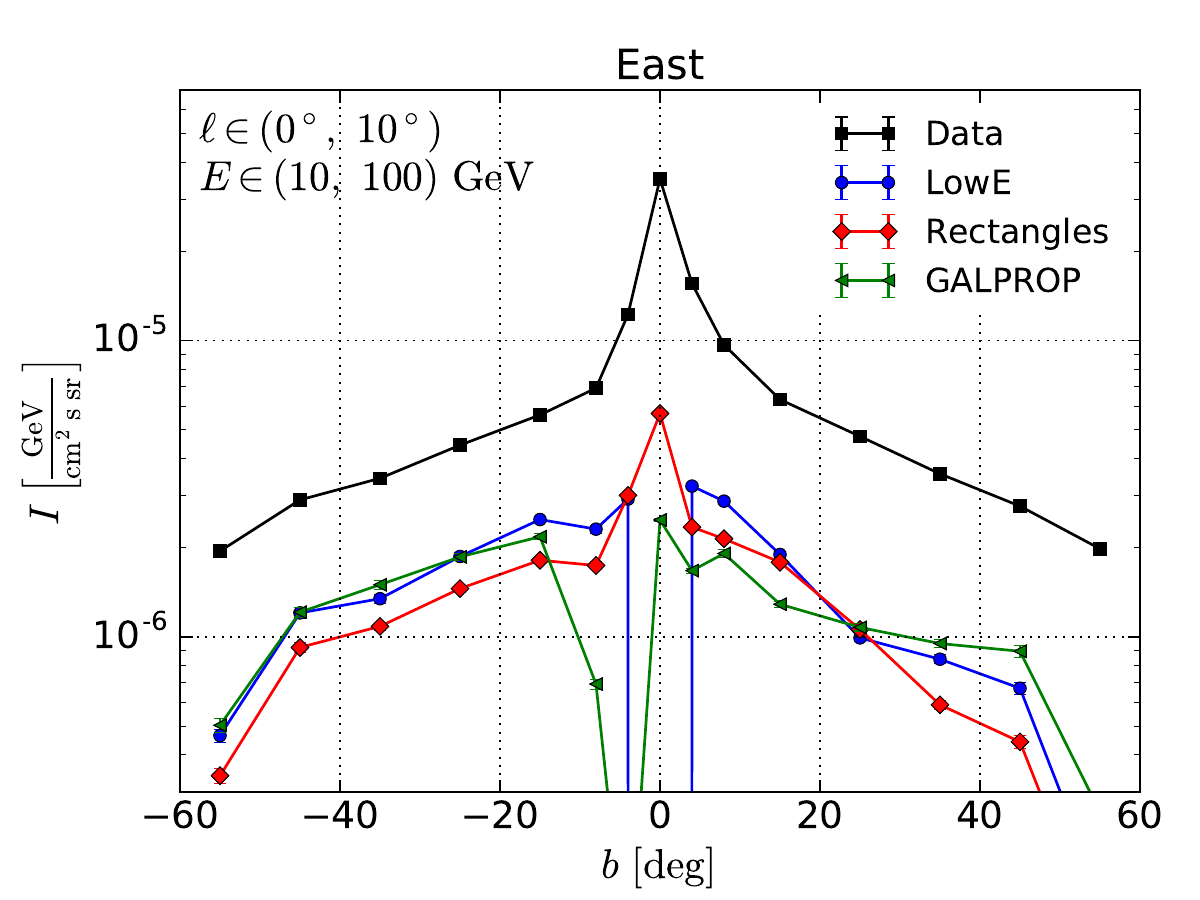}
\includegraphics[width=\twopic\textwidth]{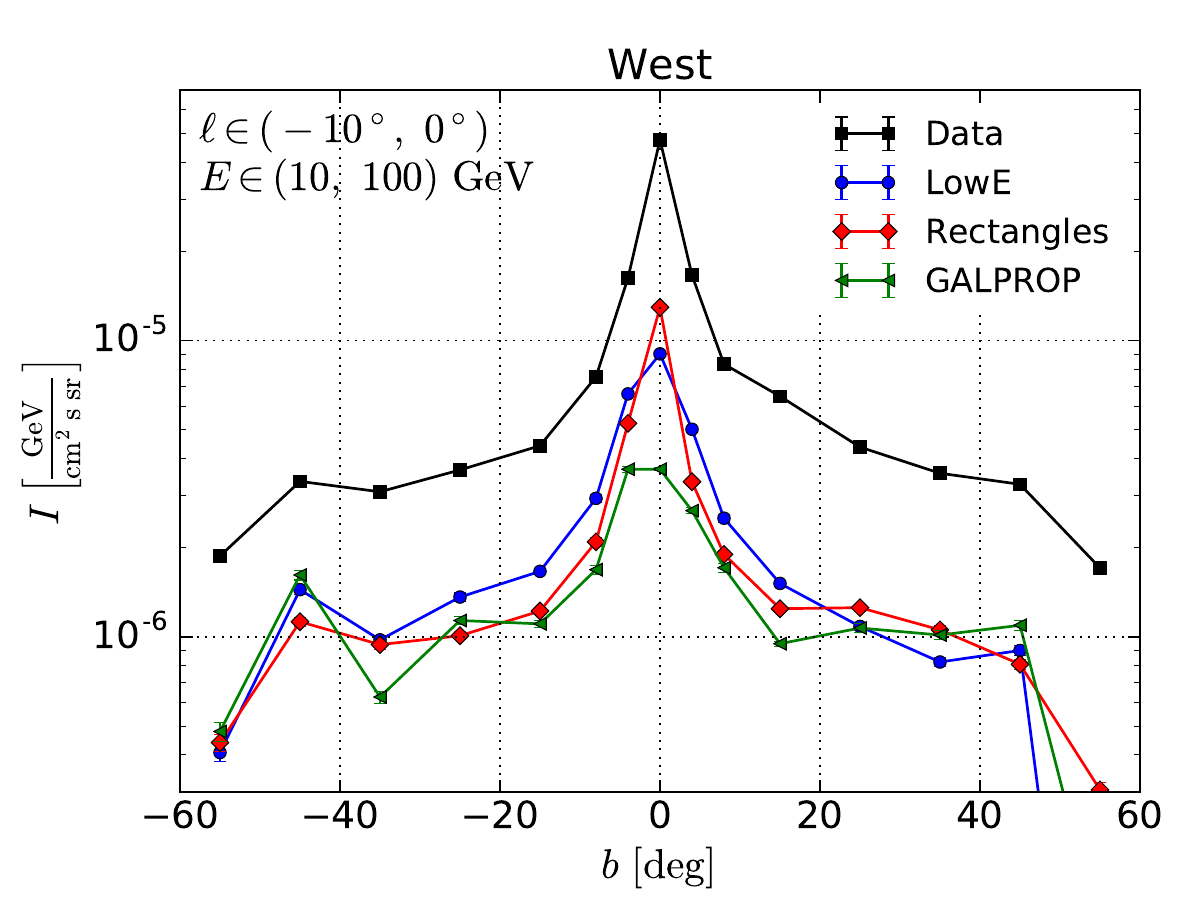}
  	\caption{Latitude profiles of the energy flux between 10 and 100 GeV for the total data excluding the PS mask and for 
	the FBs models for different foreground diffuse emission models to the east and to the west of the GC.}
  	\label{fig:Profiles}
\end{figure*}

In order to quantitatively estimate the difference in the intensity of the emission of the FBs at high and at low latitudes, as well as the asymmetry
of the gamma-ray emission at the base of the FBs,
we plot the latitude profiles to the east and to the west of the GC.
In Figure \ref{fig:Profiles} we show the profiles of the energy flux between 10 -- 100 GeV as a function 
of the Galactic latitude for different models of Galactic foreground emission.
For the low-energy model, the profiles show the residual gamma-ray emission, 
for the rectangles model -- the intensity of the rectangles templates model.
In case of the GALPROP model, we show the sum of the residual emission, the FBs template, 
and the GC excess template, as described in Section \ref{sec:Modeling}. 
For comparison we show the latitude profile of the total gamma-ray data (excluding pixels in the PS mask).
For positive longitudes,
some models overpredict the gamma-ray data near the GP, which leads to negative residuals.
For negative longitudes, there is an increase of flux at the base of the FBs 
by at least a factor of 2 to 6 for all models relative to the gamma-ray emission from the FBs at high latitudes. 
The results for the GALPROP model differ from the low-energy and rectangles model in the Galactic plane by a factor of 2 to 3. 
This is due to additional freedom in the GALPROP model related to the usage of several templates in the Galactic plane.

\begin{figure*}[h]
\centering
\includegraphics[width=\twopic\textwidth]{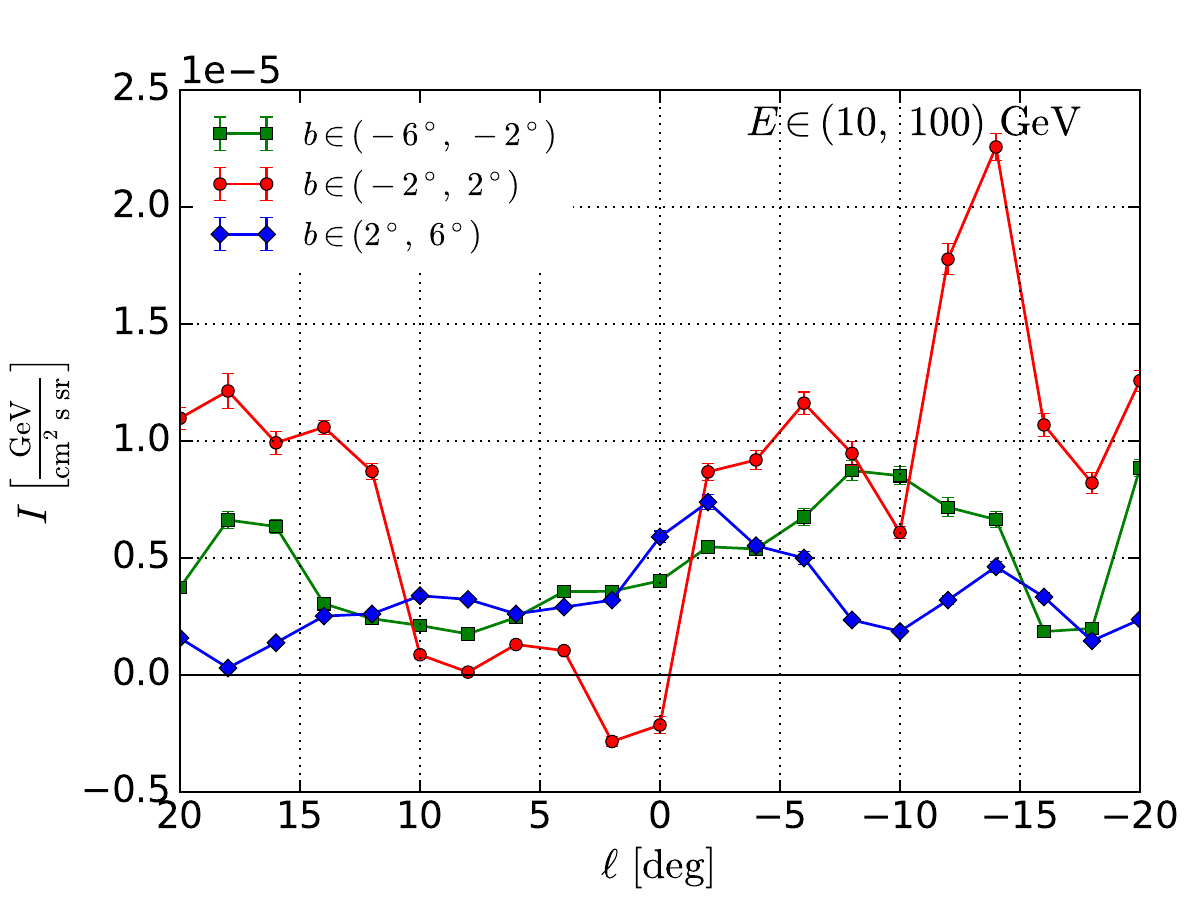}
\includegraphics[width=\twopic\textwidth]{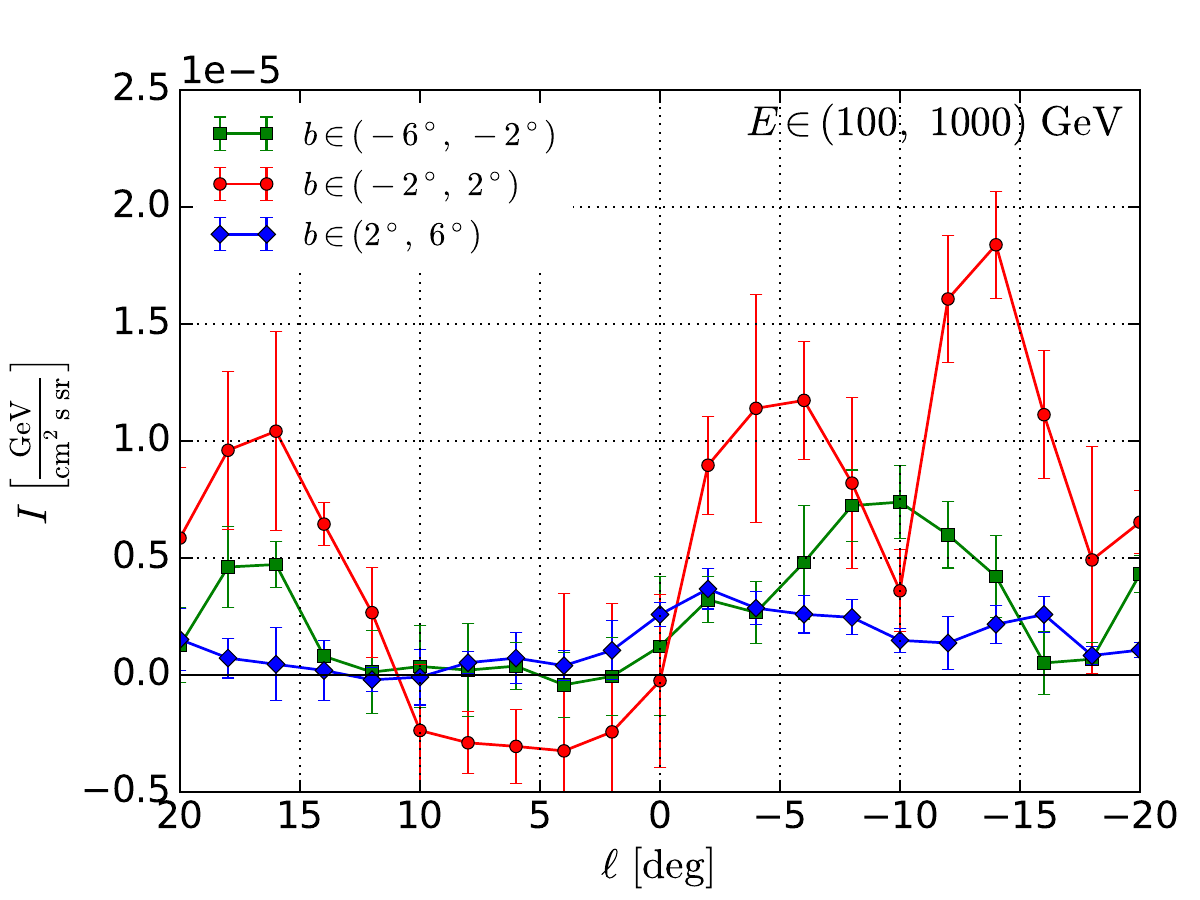}
  	\caption{Longitude profiles of the energy flux between 10 and 100 GeV and between 100 GeV and 1 TeV 
	for the residuals in the low-energy data background model (Section \ref{sec:le_data_model}).}
  	\label{fig:lon-profiles}
\end{figure*}

In Figure \ref{fig:lon-profiles} we show the longitude profiles of the residual flux 
for the background model based on low-energy data.
The excess at $\ell \approx -13^\circ$ for $|b| < 2^\circ$ is probably due to residual emission around the RX J1713 SNR \citep{2011ApJ...734...28A}.
Consequently, we do not consider it as a part of the emission at the base of the FBs.
We associate the excess between $-8^\circ \lesssim \ell \lesssim -2^\circ$ to the base of the FBs. 
In this paper, we use $\ell \in (0^\circ,\ -10^\circ)$ and $|b| < 6^\circ$ in the calculations of the spectrum of the gamma-ray
emission at the base of the FBs.

\subsection{Comparison of the spectra at different latitudes}

\begin{figure*}[h]
\centering
\includegraphics[width=\twopic\textwidth]{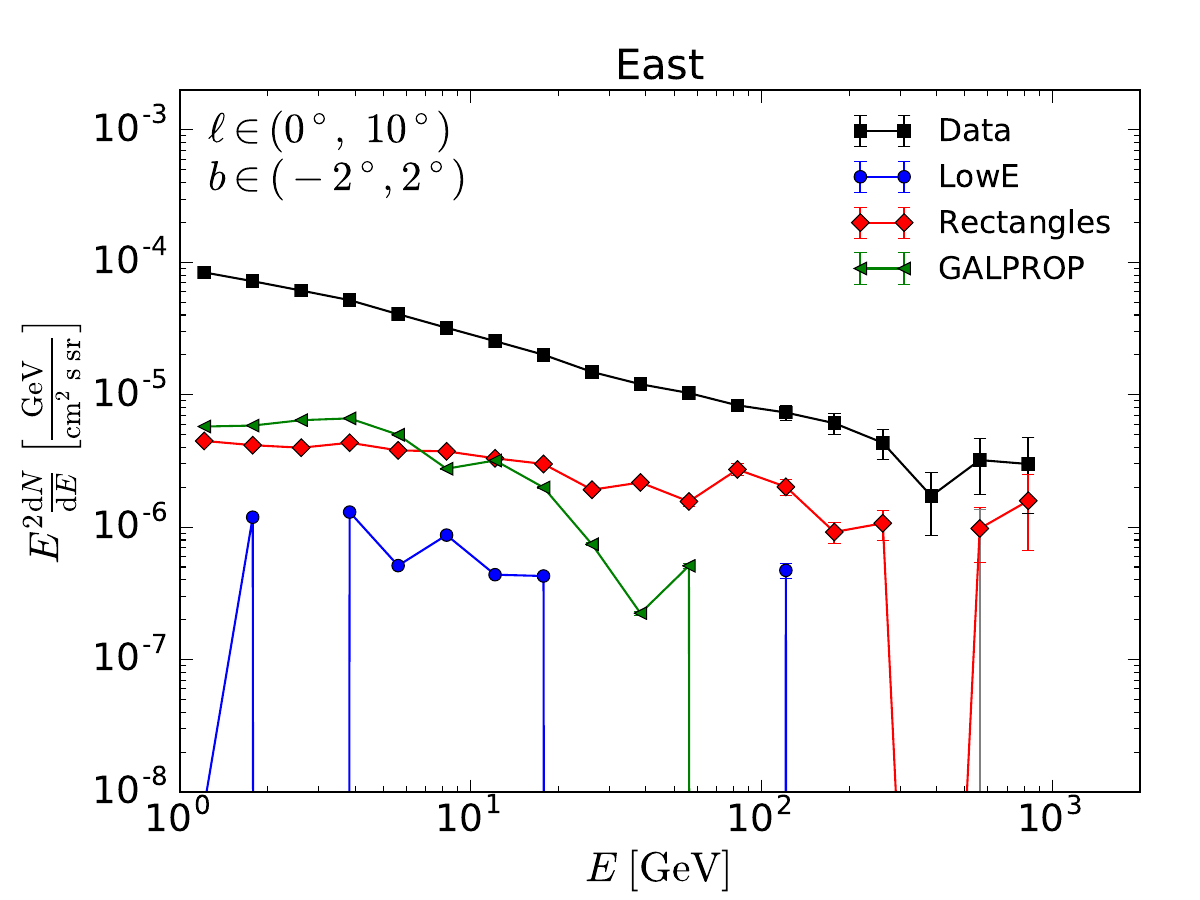}
\includegraphics[width=\twopic\textwidth]{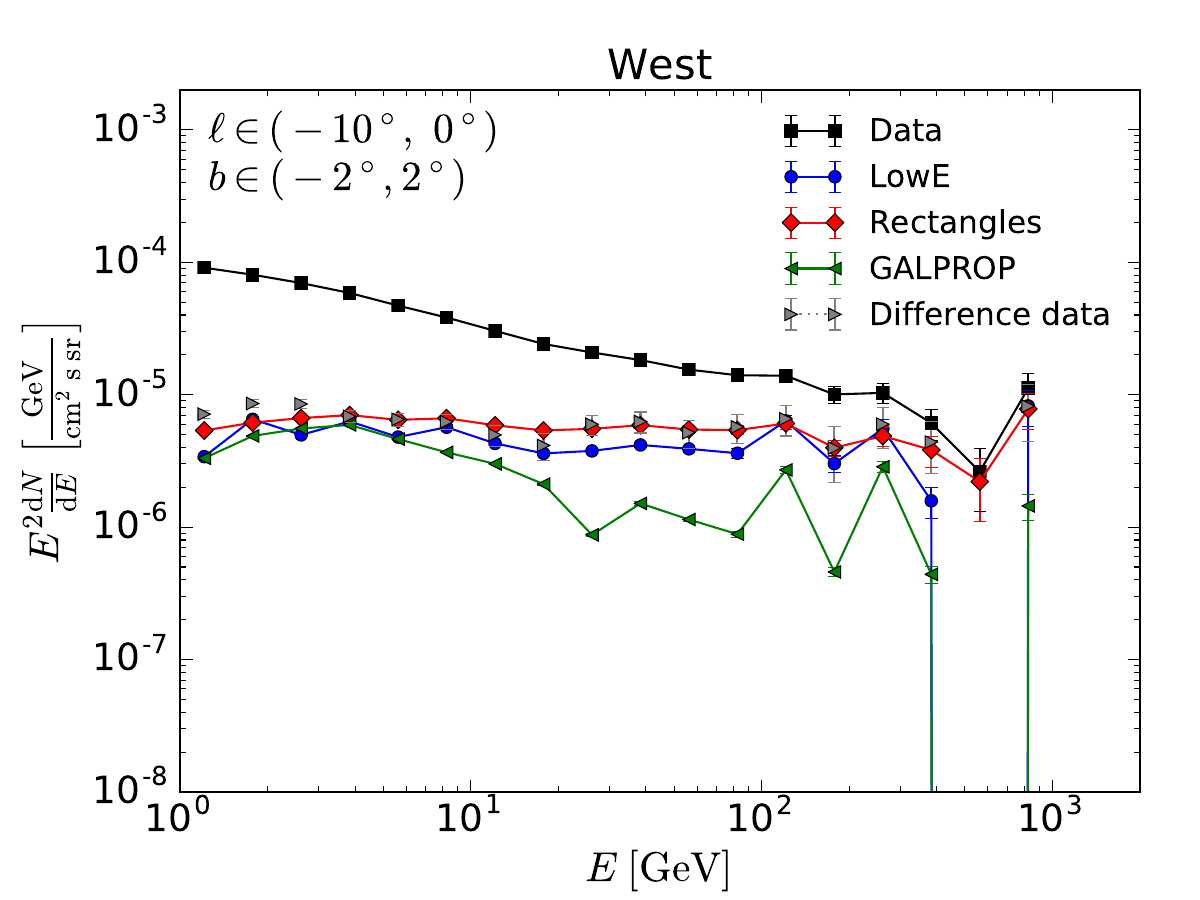}
  	\caption{SEDs of the gamma-ray data, of the residuals (``LowE'' model), 
	rectangles model of the FBs (``rectangles'' model),
	residuals plus FBs model plus gNFW DM annihilation template (``GALPROP'' model),
	and of the difference in the data west minus east of the GC (excluding pixels in the PS mask). 
	}
  	\label{fig:SED_all}
\end{figure*}

In this section we quantify the hardening of the gamma-ray spectrum at the base of the FBs. 
We first compare the spectral energy distribution (SED) of the emission at the base of the FBs for different foreground models in Figure \ref{fig:SED_all}. 
The differential flux is averaged over regions to the east, $\ell \in (\ang{0},\ \ang{10})$, and to the west, $\ell \in (\ang{-10},\ \ang{0})$, of the GC
in a thin stripe covering the Galactic plane $b \in (\ang{-2},\ \ang{2})$. 
For comparison, we show the total data (excluding pixels masked by the PS mask) and, on the plot for $\ell \in (\ang{-10},\ \ang{0})$, the difference in the data west minus east
of the GC.

For negative longitudes, all models give similar results. 
The differential flux of the GALPROP model is smaller than the differential flux of the other models above 10 GeV, 
which is consistent with the profile plots in Figure \ref{fig:Profiles}.
The difference of the data west minus east of the GC is similar to the fluxes at the base of the FBs in the low-energy model and in the rectangles model. 
The spectra at positive longitudes show large oversubtractions and softer spectra. 

\begin{figure*}[h!]
\centering
\includegraphics[width=\twopic\textwidth]{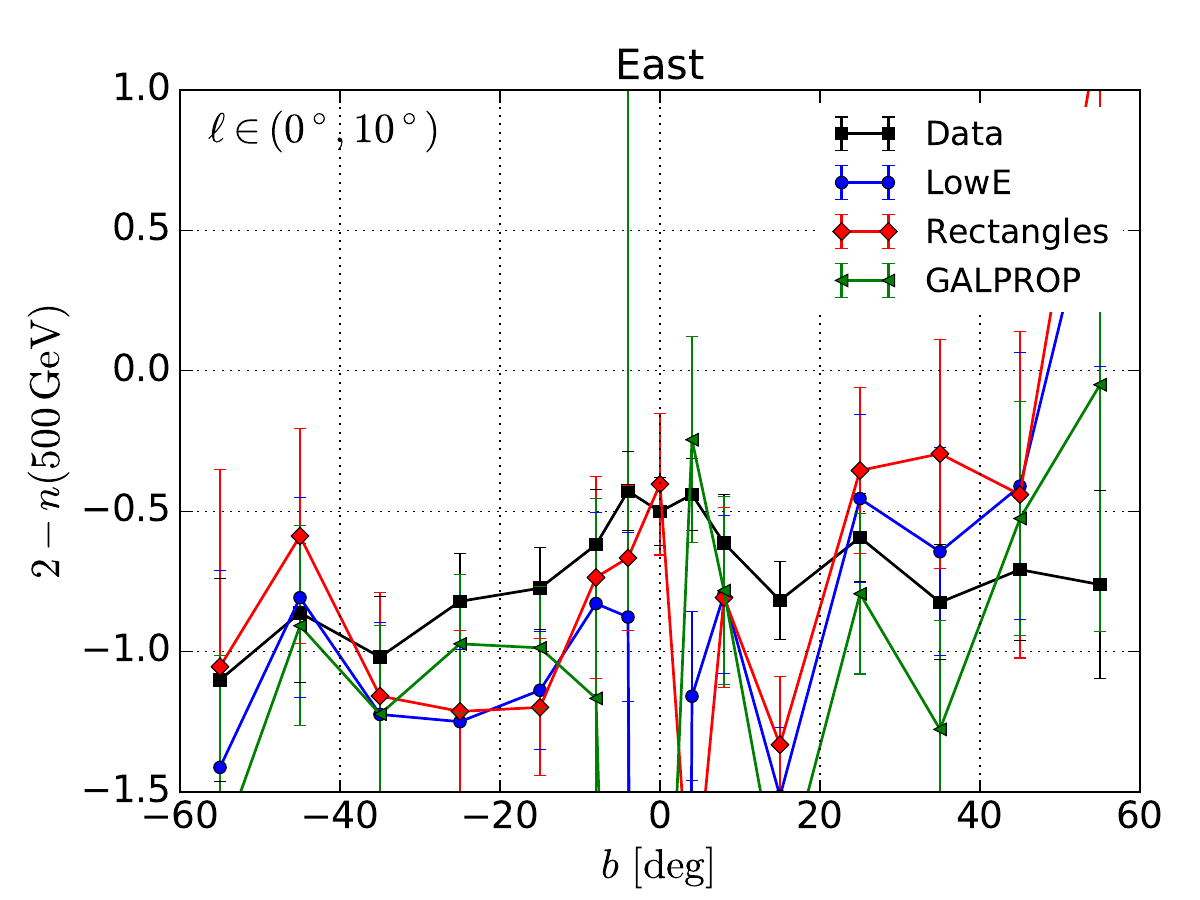}
\includegraphics[width=\twopic\textwidth]{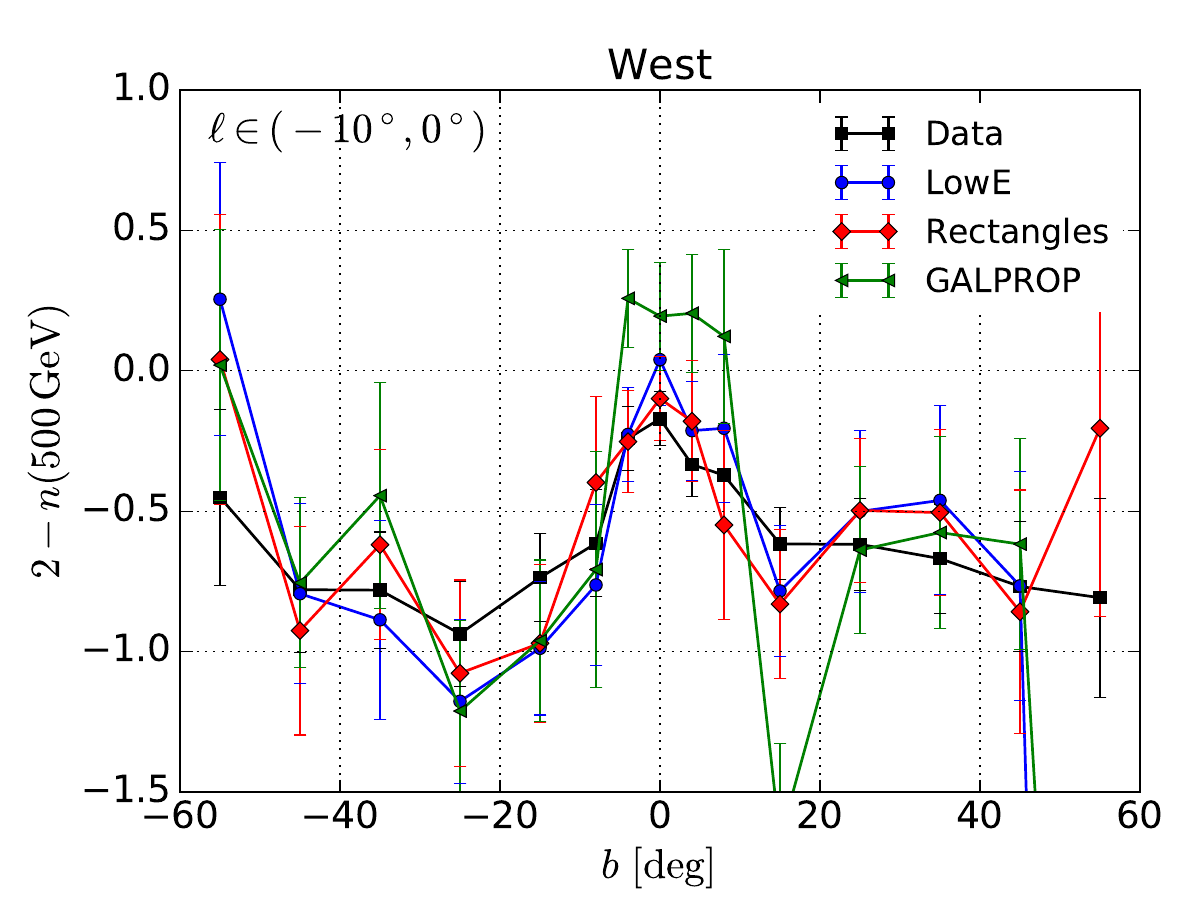}
\caption{
Index of the log-parabola defined in Equation (\ref{eq:log_par}) at  $E = \SI{500}{GeV}$ as a function of latitude for the FBs in 
the different foreground diffuse emission models.
}
\label{fig:logpar_index}
\end{figure*}

To compare the behavior of the energy spectra at high energies for different latitudes, 
we fit a log-parabola plus the fixed foreground model counts to the 
total smoothed gamma-ray counts in each latitude stripe using the likelihood based on the gamma distribution.
We use the following parametrization of the log-parabola function:
 \begin{equation}
 f(E) = N_0 \left(\frac{E}{\SI{1}{GeV}}\right)^{-\alpha - \beta \ln(E / \SI{1}{GeV})}.
 \end{equation}
The local ``index'' of the spectrum at energy $E$ is
\begin{equation}
\lb{eq:log_par}
n \equiv - \frac{\de \ln f}{\de \ln E} = \alpha + 2 \beta \ln\left(\frac{E}{\SI{1}{GeV}}\right).
\end{equation}
In Figure \ref{fig:logpar_index} we show this log-parabola index $n$ as a function of latitude at $E = \SI{500}{GeV}$. 
We plot $(2 - n)$, which corresponds to the SED index.
For positive longitudes, the index is relatively soft ($n > 2$) for most of the latitudes
except high latitudes where the gamma-ray statistics are small.
For negative longitudes (i.e., West), the index near the GC
is significantly harder ($n \approx 2$) than the index at higher latitudes.

\subsection{Parametric model of the gamma-ray spectrum at low latitudes}
\label{sec:param_model}

In this section we study the spectrum at the base of the FBs at latitudes $|b| < 6^\circ$.
As a baseline model we use the rectangles model of the FBs.
We compare a power-law model of the energy spectrum with a power-law and an exponential cutoff model
\begin{equation}
\frac{dN}{dE} = N_0 \left( \frac{E}{1\:{\rm GeV}} \right)^{-n} e^{-E / E_{\rm cut}}.
\end{equation}
We fit the spectral model added to the fixed background to the total gamma-ray data counts.
The best-fit parameters for the rectangles model are reported in Table \ref{tab:param}.
If the improvement in $2 \Delta \log \La$ is less than 1, then we report the power-law model parameters
(it turns out that the improvement in $2 \Delta \log \La$ is less than 0.1 in the cases when the cutoff is not significant).
For the model with a cutoff, we also find the 95\% statistical confidence lower limit for the cutoff energy.
In the last column we report the minimum among the 95\% confidence lower limits for the three models considered in the paper
(the low-energy data model, the rectangles model, and the GALPROP templates model).
The spectrum of the residuals including the FBs at  $\ell \in (\ang{-10},\ \ang{0})$
is consistent with a power law with an index $2.1 - 2.2$ up to $E \approx 1$ TeV.
The 95\% confidence lower limit on the cutoff energy is $\sim$ 500 GeV for latitudes $|b| < 6^\circ$ within $-10^\circ < \ell < 0^\circ$.

As one can see from the spectra in Figure \ref{fig:SED_all}, there are large (relative to statistical uncertainty) fluctuations in some of the models.
These fluctuations are due to modeling uncertainty of the Galactic diffuse emission.
As a result, a fit with a simple function may be sensitive to the starting point (there can be several local minima).
In the GALPROP model of the foreground, we also had to constrain the range of the index of the FB's spectrum
between 1.5 and 2.5 in the derivation of the 95\% confidence lower limit on the cutoff energy.

\begin{table*}
  \begin{center}
    \caption{Best-fit parameters of the gamma-ray spectrum and the significance of the cutoff for the rectangles model
    of the FBs.
    $E_{\rm cut, 95\%}$ is the statistical 95\% confidence lower limit on the energy cutoff of the FBs spectrum, 
    $E_{\rm cut, 95\%}^{\rm min}$ is the minimal 95\% statistical lower limit among the models of the FBs in 
    Sections \ref{sec:le_data_model} -- \ref{sec:galprop_model}.
    There are four cases when the introduction of a cutoff does not lead to improved fit: in all of these cases
    $2 \Delta \log \La < 0.1$ while the formal best-fit values of $E_{\rm cut}$ are larger than 1 TeV.
    }
    \label{tab:param}
    \begin{tabular}{|c|c|c|c|c|c|c|c|}
     	\hline
		 Lat & Lon  & $N_0$ & $n$ & $E_{\rm cut}$ &  $2 \Delta \log \La$ & $E_{\rm cut, 95\%}$ & $E_{\rm cut, 95\%}^{\rm min}$ \\ 
		       &        &  {\small $\SI{e-6}{\rm GeV^{-1}cm^{-2}s^{-1} sr^{-1}}$ }&  & {\small $\SI{}{GeV}$ }& &{\small  $\SI{}{GeV}$ }&{\small  $\SI{}{GeV}$ }\\ 
		\hline
  		$(\ang{2}, \ang{6})$ & $(\ang{0}, \ang{10})$ & $1.5\pm 0.3$  & $1.9\pm 0.1$ & $45\pm 17$ & 4.8 & 25& 25\\ 
		& $(\ang{-10}, \ang{0})$ & $3.1 \pm 0.4$  & $2.2 \pm 0.05$  & $-$ \cmt{5.2e3} & $< 0.1$ \cmt{0.027} & {510} & {510}  \\ 
 		\hline
  		$(\ang{-2}, \ang{2})$ & $(\ang{0}, \ang{10})$  & $6.4 \pm 1.1$  & $2.3 \pm 0.09$ & $-$ \cmt{8.3e3} &  $< 0.1$ \cmt{0.010} & {300}  & {2.4}  \\ 
		& $(\ang{-10}, \ang{0})$  & $7.9 \pm 0.7$  & $2.1 \pm 0.04$ & $-$ \cmt{8.3e6} &  $< 0.1$ \cmt{3.3e-5} & {{1000}} & 600   \\ 
 		\hline
  		$(\ang{-6}, \ang{-2})$ & $(\ang{0}, \ang{10})$  & $2.5 \pm 0.4$  & $2.1 \pm 0.07$ & $260 \pm 150$ & 3.8 & 130& {2.3} \\ 
		& $(\ang{-10}, \ang{0})$ & $4.0 \pm 0.3$  & $2.2 \pm 0.04$ &  $-$ \cmt{77e3} &  $< 0.1$ \cmt{0.020} & {560}  & {560}  \\ 
 \hline
    \end{tabular}
  \end{center}
\end{table*}

\subsection{IC model of the gamma-ray emission}
\label{sec:IC_model}

\begin{figure}[h!]
\centering
\includegraphics[width=\onepic\textwidth]{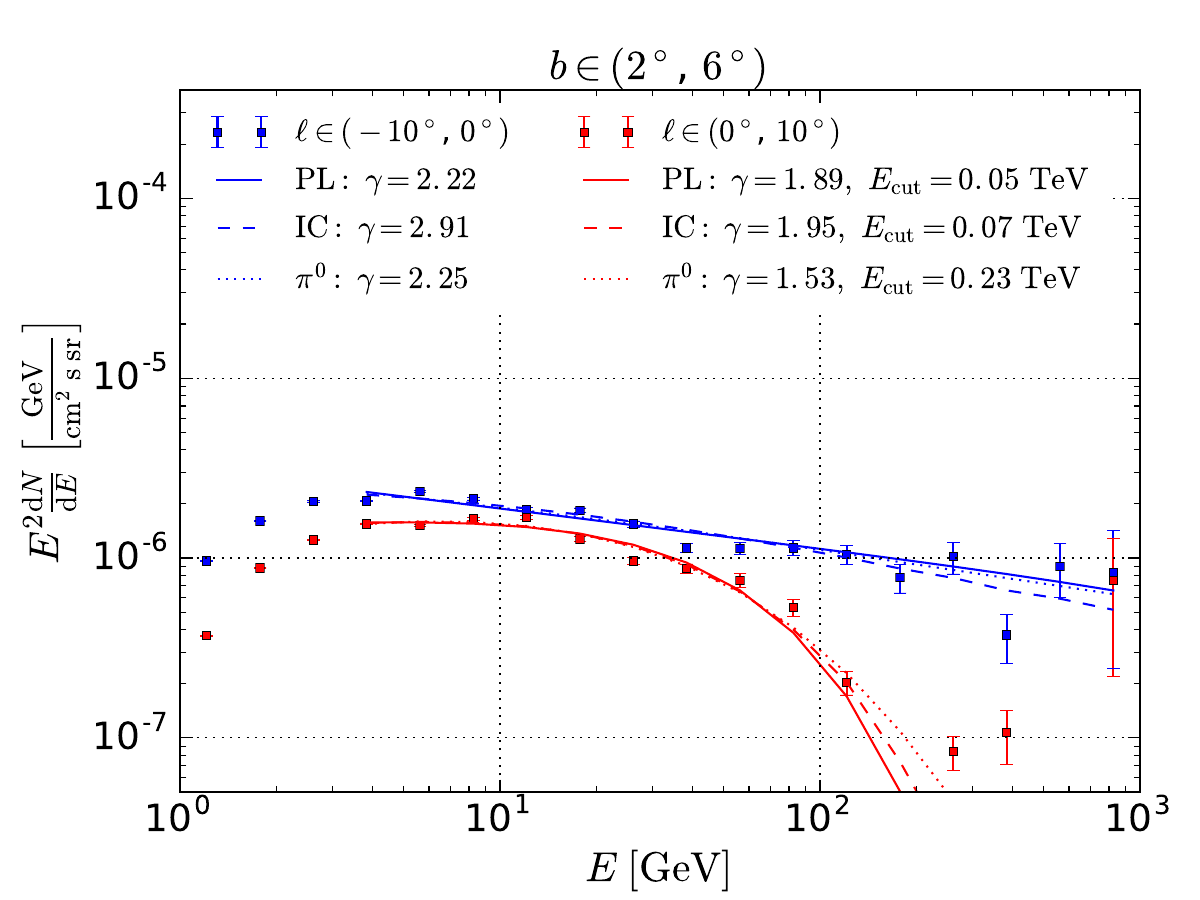} \\
\includegraphics[width=\onepic\textwidth]{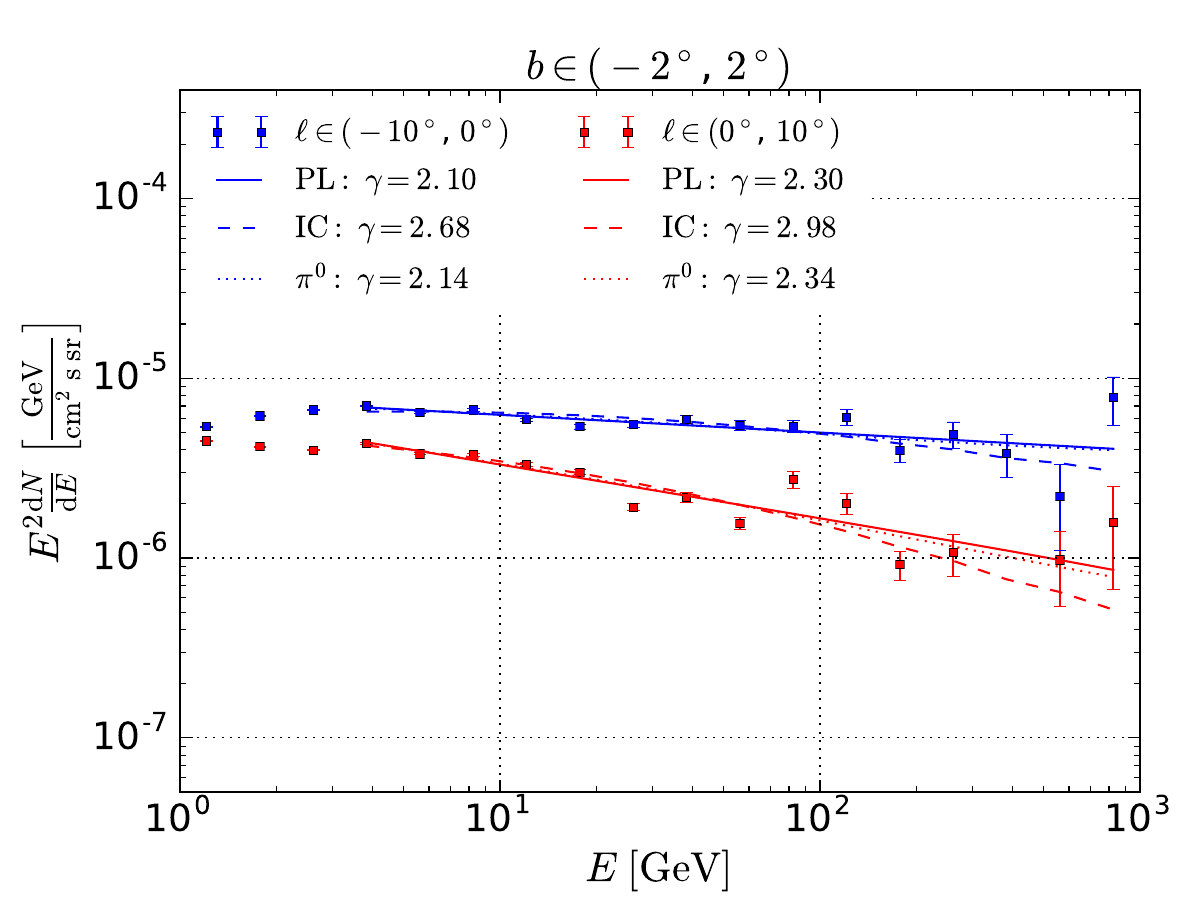} \\
\includegraphics[width=\onepic\textwidth]{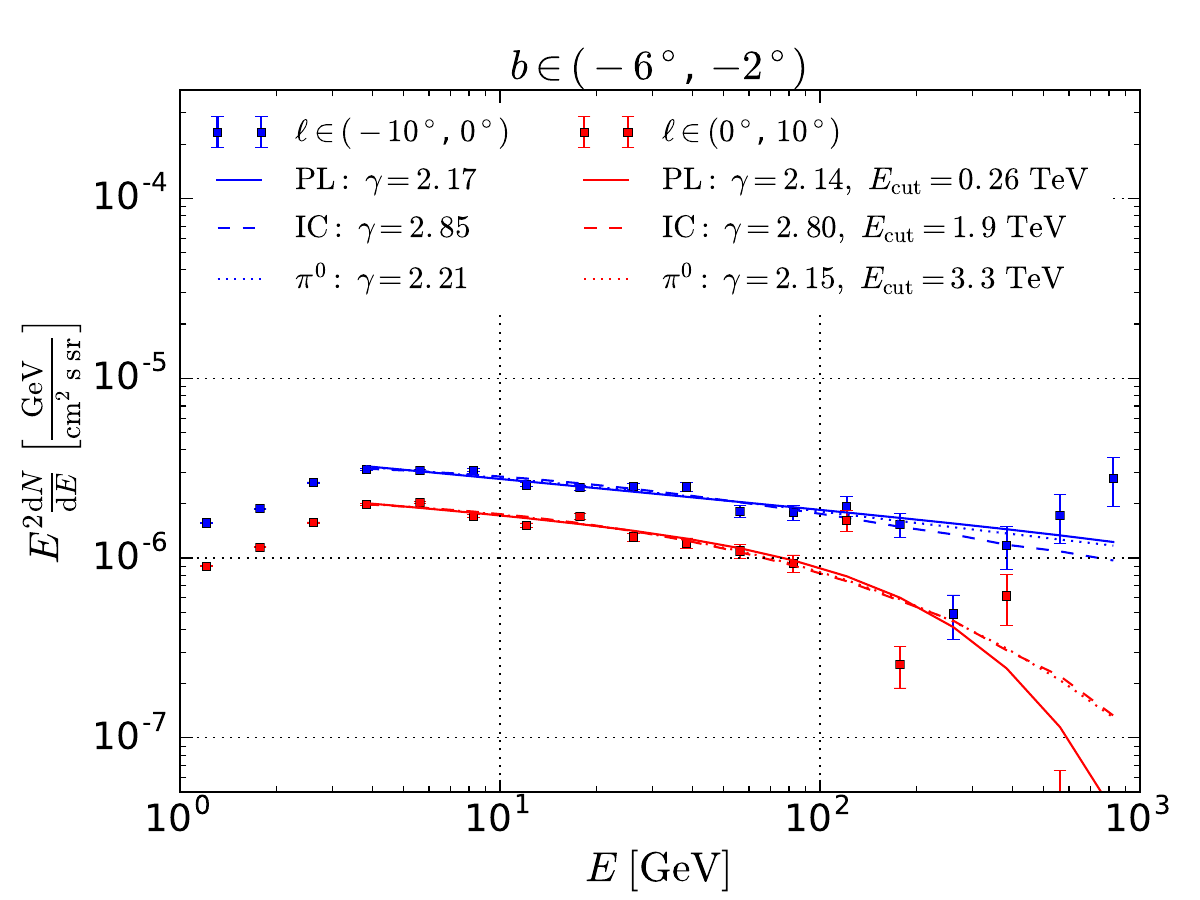}
  	\caption{SEDs of the rectangles model of the FBs in the latitude stripes $(\ang{2}, \ang{6})$, $(\ang{-2}, \ang{2})$ and $(\ang{-6}, \ang{-2})$ for negative (blue) and positive (red) longitudes. We compare the parametric model of the SED with the leptonic (Section \ref{sec:IC_model}) 
	and the hadronic (Section \ref{sec:Pion_model}) models.}
  	\label{fig:SED_with_fits}
\end{figure}

In this section we model the gamma-ray emission at the base of the FBs with an IC scattering model.
The SED of the gamma-ray intensity is parametrized as
\begin{equation}
E^2_\g \frac{dF_\g}{dE_\g} = 
\frac{E_\g c}{4\pi}\int \frac{\de n_{h\nu}}{\de E_{h\nu}} \sigma_\IC\ \frac{\de \Sigma_{\el}}{\de E_{\el}} \de E_{h\nu}\, \de E_\el,
\end{equation}
where $\frac{\de \Sigma_{\el}}{\de E_{\el}} = \int \frac{\de n_{\el}}{\de E_{\el}} dR$ is the column density 
of CR electrons integrated along the line-of-sight distance $R$,
$\frac{\de n_{h\nu}}{\de E_{h\nu}}$ is the number density of ISRF photons of energy $h\nu$,
and $\sigma_\IC(E_\gamma, E_{h\nu}, E_\el)$
is the differential IC scattering cross section in units of $E_\g\frac{d\sigma}{d E_\g}$ \citep{1970RvMP...42..237B}.
For details on the parametrization of $\sigma_\IC$ see Appendix B of \cite{2014ApJ...793...64A}.
The ISRF number density for SL and IR photons is taken from 
\cite{Porter:2008ve} (available with the distribution of GALPROP v54.1),
assuming that the emission at the base of the FBs is near the GC.
For the CMB, we use the thermal spectrum with a temperature of $\SI{2.73}{K}$.
There is a significant uncertainty in the Galactic ISRF energy density near the GC \citep[e.g.,][]{2017MNRAS.470.2539P, 2017ApJ...846...67P, 2019APh...107....1N}.
We find (Appendix \ref{app:ISRF}) that this uncertainty can lead to a factor $\sim$ 3 uncertainty in the overall
normalization of the inferred CR electrons (CRe) energy density.
Additional uncertainty on the CRe distribution and the associated IC emission comes from the distribution of the Galactic magnetic field
\citep[e.g.,][]{2018MNRAS.475.2724O, 2019arXiv190108604O}.
We model the column density of electrons as a power law with a cutoff
\begin{equation} 
\label{eq:e_spectrum}
 \frac{\de \Sigma_{\el}}{\de E_{\el}} = n_\el \left(\frac{E_\el}{\SI{1}{GeV}}\right)^{-\gamma_\el} e^{- E_\el / E_{\cut}}.
\end{equation}
In order to determine the normalization $n_\el$, the spectral index $\gamma_\el$, and the cutoff  $E_{\cut}$, 
we fit the IC model of the FBs plus the (fixed) foreground model counts to the 
total smoothed gamma-ray counts in different latitude stripes using likelihood based on the gamma distribution.
As a baseline case, we take the gamma-ray spectrum derived in the rectangles model of the FBs in Section \ref{sec:box_model}.
The best-fit spectra for the rectangles model of the bubbles are shown in Figure \ref{fig:SED_with_fits}
in latitude stripes $b \in (\ang{2}, \ang{6})$, $b \in (-\ang{2}, \ang{2})$ and $b \in (-\ang{6}, -\ang{2})$. 
If the improvement in $2 \Delta \log \La$ with and without the cutoff is less than 0.1, then we show only the parameters for the power-law model without a cutoff.
For example, for negative longitudes the cutoff is not significant.

The 95\% statistical lower limit on the cutoff energy for the rectangles model and the minimum 
among the three models presented in Sections \ref{sec:le_data_model} -- \ref{sec:galprop_model}
of the 95\% confidence lower limits 
for the cutoff energy are presented in Table \ref{tab:IC}.
For negative longitudes,
the 95\% confidence lower limit on the cutoff in the spectrum of electrons in the rectangles model is about 4 TeV,
while the minimal value of the 95\% confidence limit for the three models of the foreground emission is about 3 TeV.

\begin{table*}
  \begin{center}
    \caption{Energy cutoff values and the significance of the cutoff in the IC model of the FBs at low latitudes.
The lower bounds for $E_\cut$ at the 95\% confidence level for our baseline model and the minimum among the models 
presented in Sections \ref{sec:le_data_model} -- \ref{sec:galprop_model}
are shown in the last two columns respectively.
}
    \label{tab:IC}
    \begin{tabular}{|c|c|c|c|c|}
     	\hline
		 Lat & Lon  & $2 \Delta \log \La$ & \multicolumn{2}{c|}{Lower bound on $E_\cut$ (TeV) } \\ 
		       &        &                                  &  \multicolumn{1}{c}{Rectangles model} & All models \\ 
		\hline
  		$(\ang{2}, \ang{6})$ & $(\ang{0}, \ang{10})$ & 2.6  & {0.20} & {0.17}\\ 
		& $(\ang{-10}, \ang{0})$ & $ < 0.1$  & 4.0  & 4.0 \\ 
 		\hline
  		$(\ang{-2}, \ang{2})$ & $(\ang{0}, \ang{10})$ & $ < 0.1$ & {1.4} & 0.01 \\ 
		& $(\ang{-10}, \ang{0})$ & $ < 0.1$ & 13  & 2.9  \\ 
 		\hline
  		$(\ang{-6}, \ang{-2})$ & $(\ang{0}, \ang{10})$ & 1.1 & {0.82} & 0.03 \\ 
		& $(\ang{-10}, \ang{0})$& $ < 0.1$ & {7.0} & {7.0} \\ 
 \hline
    \end{tabular}
  \end{center}
\end{table*}

\subsection{Hadronic model of gamma-ray emission}
\label{sec:Pion_model}

In the hadronic model, the gamma rays are produced as a result of collisions of hadronic CR with the interstellar gas.
The gamma-ray intensity in the hadronic model is parametrized as
\be
E^2_\g \frac{dF_\g}{dE_\g} = \frac{E_\g}{4\pi} \int n_\Hy\ \sigma_\pr v_\pr \left(\frac{\de \Sigma}{\de T}\right)_{\!\!\pr} \de T_\pr,
\ee
where $\left(\frac{\de \Sigma}{\de T}\right)_{\!\!\pr} = \int \left(\frac{\de n}{\de T}\right)_{\!\!\pr} dR$ is the column density 
of CR protons, $v_\pr$ is the velocity of the protons,
$T_\pr = \sqrt{(q c)^2 + (m_p c^2)^2} - m_p c^2$ is the kinetic energy of the protons as a function of momentum $q$,
$\sigma_\pr (E_\gamma, T_\pr)$ is 
the differential cross section in units of $E_\g\frac{d\sigma}{d E_\g}$
for gamma rays in proton-proton collisions \citep{2006ApJ...647..692K, 2008ApJ...674..278K},
and $n_\Hy$ is the density of gas.
We will use $n_\Hy = \SI{1}{cm^{-3}}$ as a reference density,
which is comparable with the gas surface density of $\sim 30 M_\odot {\rm pc}^{-2}$ \citep{2017ApJ...834...57M}
averaged over the approximate height of the enhanced emission at the base of the bubbles, e.g., $\sim 900$ pc above and below the GC.
Since most of the gas density is concentrated within 100 -- 200 pc from the midplane \citep[e.g.,][]{2013pss5.book..985S, 2017ApJ...834...57M},
the gas density in the plane near the GC is 5 -- 10 ${\rm cm^{-3}}$, which is consistent with the enhancement of the intensity of the emission
at the base of the FBs at $\ell = 0^\circ$.
We model the proton spectrum as a power law of the momentum $\frac{d \Sigma}{d qc} = n_\pr q^{-\g_p}$ 
(note, that $ v_\pr \frac{\de \Sigma}{\de T_\pr} = c \frac{d \Sigma}{d qc}$).

We fit the hadronic model of the gamma-ray emission at the base of the FBs analogously to the leptonic model in 
Section \ref{sec:IC_model}.
The dash-dotted line in Figure \ref{fig:SED_with_fits} represents the best-fit hadronic spectrum (labeled as $\pi^0$). 
The index of the proton spectrum is relatively hard $\g_\pr \lesssim 2.3$, especially to the west of the GC.

We calculate the significance of a cutoff in the CR proton spectrum by adding an exponential cutoff factor and refitting the model to the gamma-ray data.
The improvement in the model and the 95\% confidence lower bound on the cutoff values are presented in Table \ref{tab:pi0}.
Within $\pm 6^\circ$ from the Galactic plane west of the GC, the 95\% confidence level for the lower bound on the cutoff among the models
of the foreground emission that we have considered in Sections \ref{sec:le_data_model} -- \ref{sec:galprop_model} is about 6 TeV.

\begin{table*}
  \begin{center}
    \caption{\label{tab:pi0} 
Energy cutoff values and the significance of the cutoff in the hadronic model of the FBs at low latitudes.
The lower bounds for $E_\cut$ at the 95\% confidence level for our baseline model and the minimum 
among the models presented in Sections \ref{sec:le_data_model} -- \ref{sec:galprop_model}
 are shown in the last two columns respectively. 
}
    \begin{tabular}{|c|c|c|c|c|}
     	\hline
		 lat & lon  & $2 \Delta \log \La$ & \multicolumn{2}{c|}{Lower bound on $E_\cut$ (TeV) } \\
		      &        &                                  &       \multicolumn{1}{c}{Rectangles model} & All models \\ 
		\hline
  		$(\ang{2}, \ang{6})$ & $(\ang{0}, \ang{10})$ & 4.4 & {0.88} & {0.48} \\ 
		& $(\ang{-10}, \ang{0})$ &  $ < 0.1$ & {7.4} & {7.4}\\ 
 		\hline
  		$(\ang{-2}, \ang{2})$ & $(\ang{0}, \ang{10})$ & $ < 0.1$ & {3.8} & 0.023 \\ 
		& $(\ang{-10}, \ang{0})$ & $ < 0.1$ & {29} & 6.3 \\ 
 		\hline
  		$(\ang{-6}, \ang{-2})$ & $(\ang{0}, \ang{10})$ & 2.7 & 1.6 & 0.05 \\ 
		& $(\ang{-10}, \ang{0})$ & $ < 0.1$ & {12} & {12}\\ 
 \hline
    \end{tabular}
  \end{center}
\end{table*}

\subsection{Summary of the spectral analysis}

\begin{figure*}[h]
\centering
 \includegraphics[width=\twopic\textwidth]{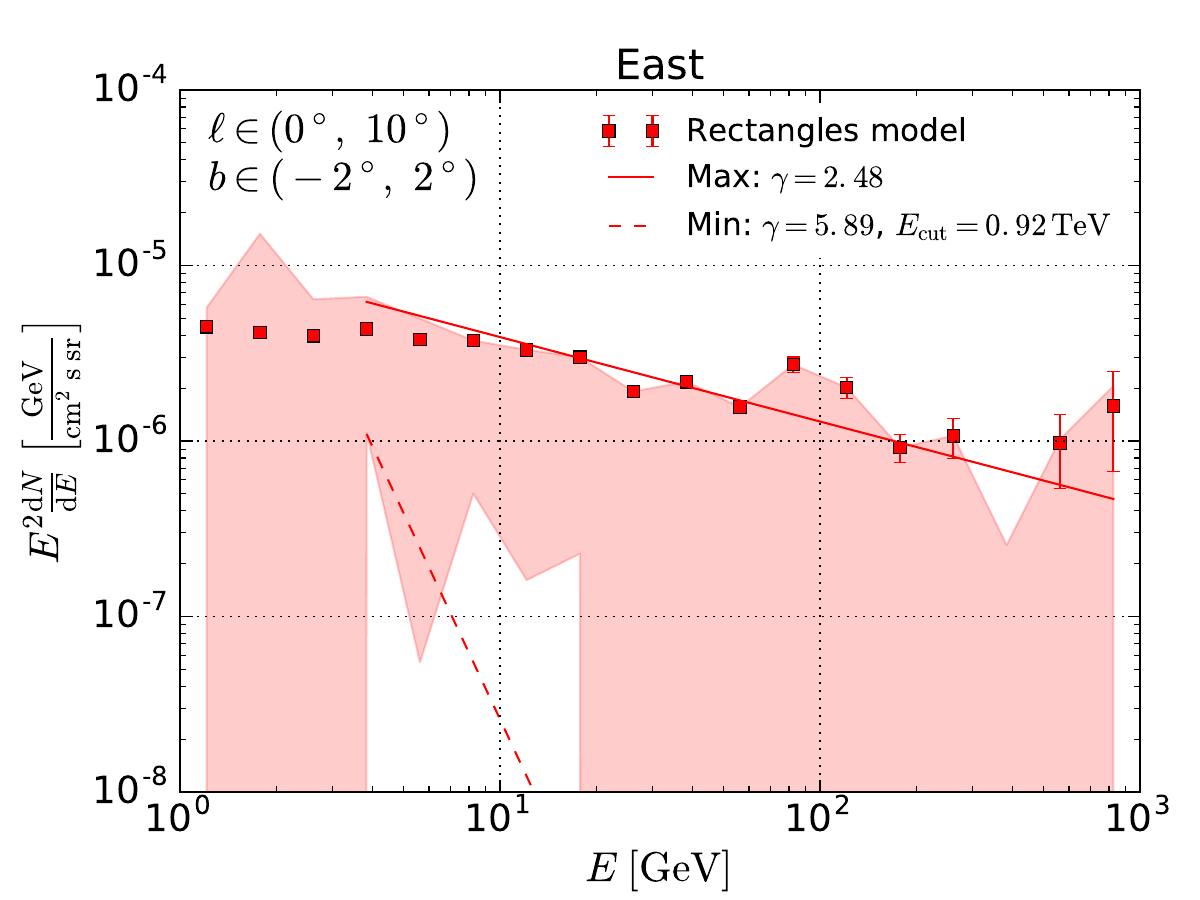}
  \includegraphics[width=\twopic\textwidth]{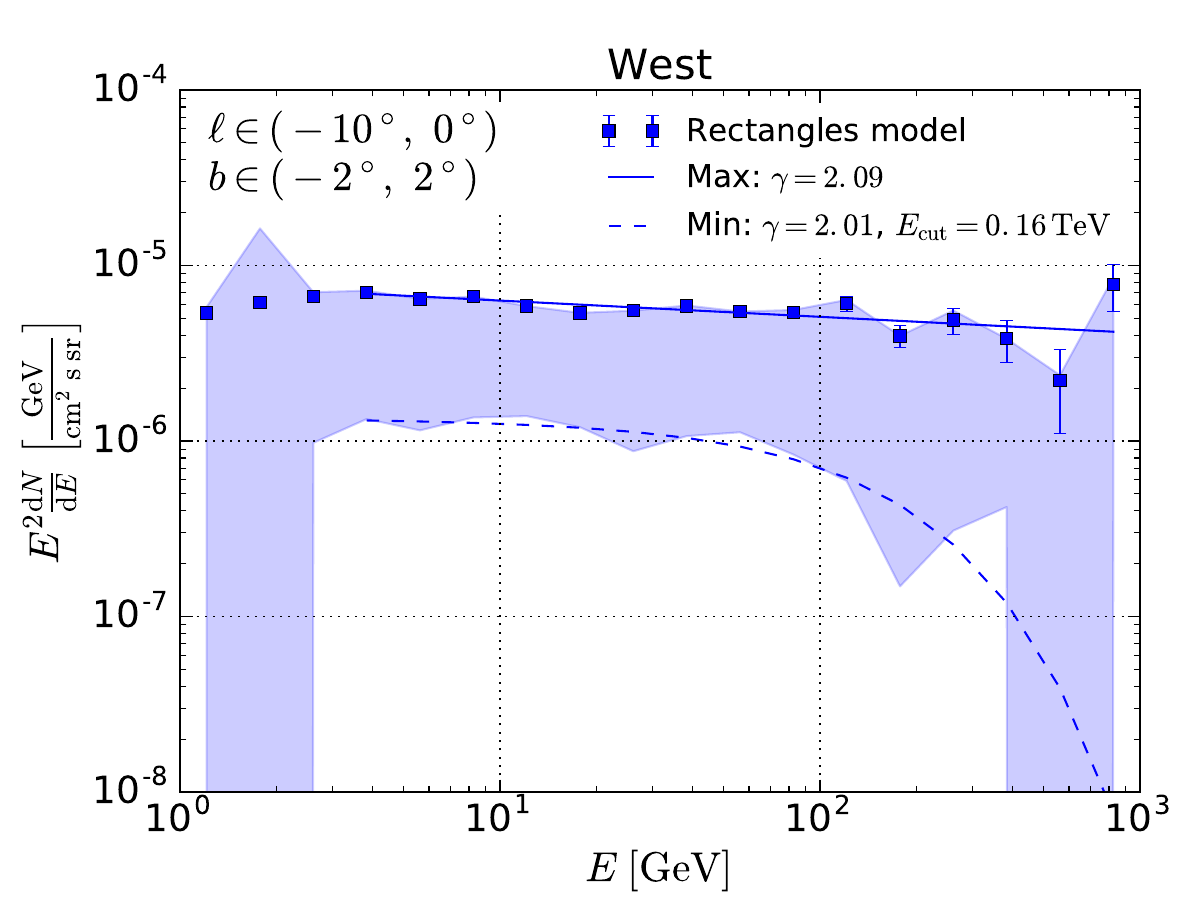}
 \caption{SED of the FBs in the Galactic plane. 
 The shaded areas show the envelope of the FB's spectra in all foreground models considered in the paper
including the changes in the choice of the low-energy range to model the foreground emission
(discussed in Appendix \ref{sec:lowE_syst}).
The lines show the fits to the maximal and minimal points in the envelopes above 3 GeV.
}
 \label{fig:spec_summary}
\end{figure*}

In Figure \ref{fig:spec_summary} we show the envelopes of the gamma-ray spectra at $|b| < 2^\circ$ of the emission at the base of the FBs
for the models of foreground emission that we have considered, including the models where we change the selection of the low energies 
in the definition of the foreground emission model (Appendix \ref{sec:lowE_syst}).
In order to determine the maximal and minimal models of the FBs in the Galactic plane, 
we fit the maximal and minimal points in the envelope above 3 GeV with a power law with a cutoff function
(we fit above 3 GeV because some of the models of the foreground emission in Appendix \ref{sec:lowE_syst} are determined 
for energies between 1 and 2.2 GeV).
The corresponding parameters are reported in the first row of Table \ref{tab:summary}.
We also fit the IC and hadronic models to the maximal and minimal points in the envelope and report the corresponding parameters
in Table \ref{tab:summary}.

\begin{table*}
  \begin{center}
    \caption{Summary of the min and max models for the parametric, 
    IC and hadronic models of the FBs for $|b| < 2^\circ$ and $-10^\circ < \ell < 0^\circ$. 
    For the parametric model, we report the energy spectrum of the gamma rays,
    for the IC model we report the column density of the electrons' spectrum as a function of energy,
    while for the hadronic model -- the column density of the protons' spectrum as a function of momentum.
    The spectra are normalized at $E_0 = 1$ GeV. The last column shows the statistical 95\% confidence lower limit on the cutoff, $E_{\rm cut, 95\%}$.}
    \label{tab:summary}
    \begin{tabular}{| l |c|c|c|c|c|c|c|}
     	\hline
		 {\hspace{2cm}Model} & Type  & norm & index & cutoff & $E_{\rm cut, 95\%}$ \\ 
		       &        &   &  & {\rm\small TeV} & {\rm\small TeV}\\ 
		\hline
  		\multirow{2}{*}{Parametric, $\frac{dN_\g}{dE}$ {\small $\left[{\rm {GeV^{-1}\, cm^{-2}\, s^{-1} }}\right]$}} & max &  $\SI{7.8e-6}{}$ \red{}& 2.09 \red{} &  -- & 0.99 \red{} \\ 
		& min & $\SI{1.4e-6}{}$ \red{}& 2.01 \red{} & 0.16 \red{} & 0.04 \red{} \\ 
 		\hline
  		\multirow{2}{*}{IC, $\frac{d\Sigma_e}{dE}$ {\small $\left[{\rm {GeV^{-1}\, cm^{-2}}}\right]$}} & max & $\SI{3.4e10}{}$ & 2.67 &  -- & 19 \\ 
		& min & $\SI{1.3e10}{}$ & 2.81 &  7.2 & 0.96 \\ 
 		\hline
  		\multirow{2}{*}{Hadronic, $\frac{d\Sigma_p}{dqc}$ {\small $\left[{\rm {GeV^{-1}\, cm^{-2}}}\right]$}} & max & $\SI{9.5e11}{}$ & 2.13 &  -- &65 \\ 
		& min & $\SI{1.1e11}{}$ & 1.98 &  1.8 & 0.23  \\ 
 \hline
    \end{tabular}
  \end{center}
\end{table*}

\section{Intepretation}
\label{sec:Interpretation}

The bright and hard gamma-ray emission at the base of the bubbles can be at any position along the line of sight.
In this section we consider two characteristic scenarios: that the emission is near the GC or that the emission is 
produced by one or a few SNRs closer to us. 

\subsection{Emission near the GC scenario}
\lb{sec:GC_scenario}

In the estimates in this subsection we use the following characteristic sizes: 
the distance to the GC is $\SI{8.5}{kpc}$, 
the size of the region with the enhanced gamma-ray emission interpreted as an additional population of CR:
$-10^\circ < \ell < 0^\circ$ and $|b| < 6^\circ$.
If we assume that the geometry of the region is a simple box, then the size of the box along the GP is $\SI{1.5}{kpc}$,
while the half-size in the vertical direction (distance from the GP to the boundary) is $\Delta h = 0.9$ kpc.
The total gamma-ray luminosity in this case can be estimated 
by summing the data points in the rectangles model in Figure \ref{fig:SED_with_fits}
for $|b| < 6^\circ$, which yields the luminosity of $L = 1.1 \times 10^{37}\:{\rm erg\ s^{-1}}$ above 1 GeV.

\begin{figure*}[h]
\centering
\includegraphics[width=\twopic\textwidth]{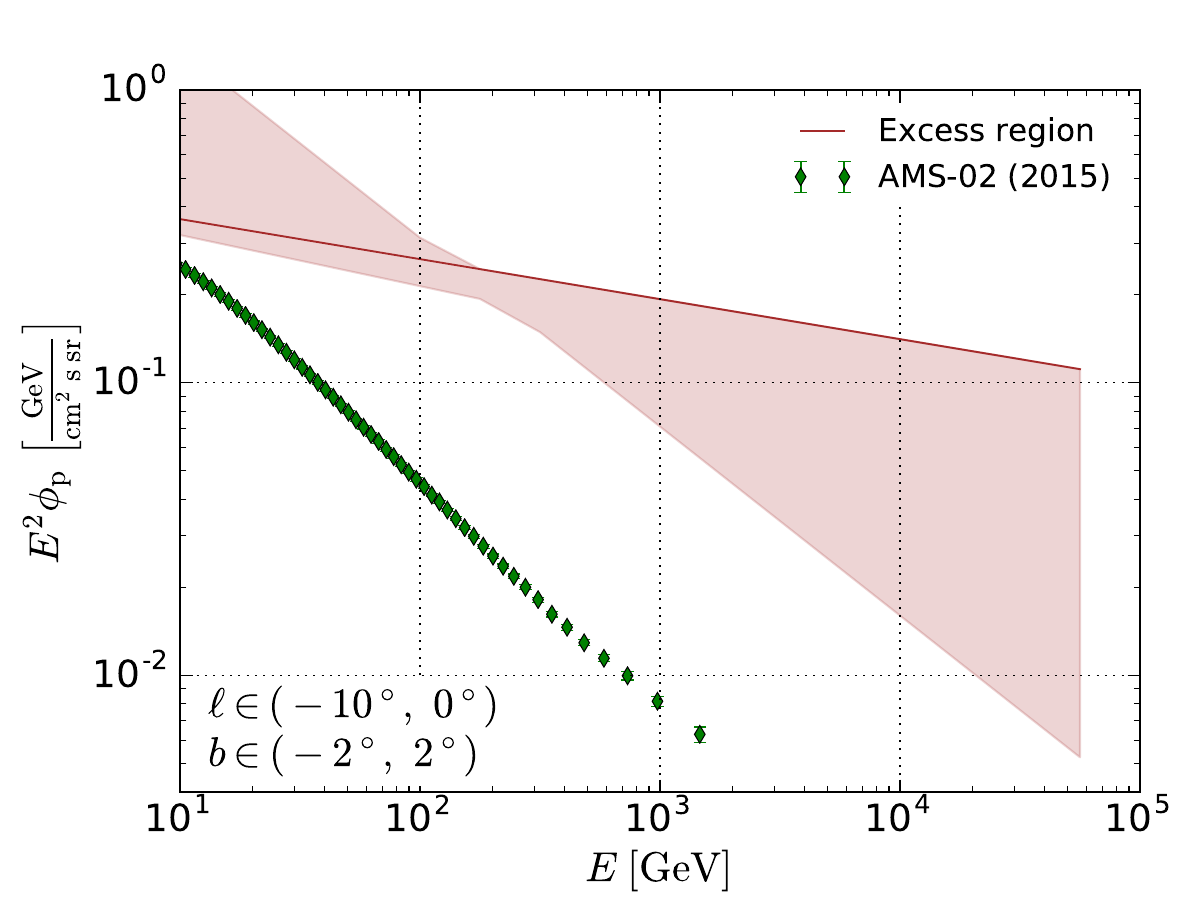}
\includegraphics[width=\twopic\textwidth]{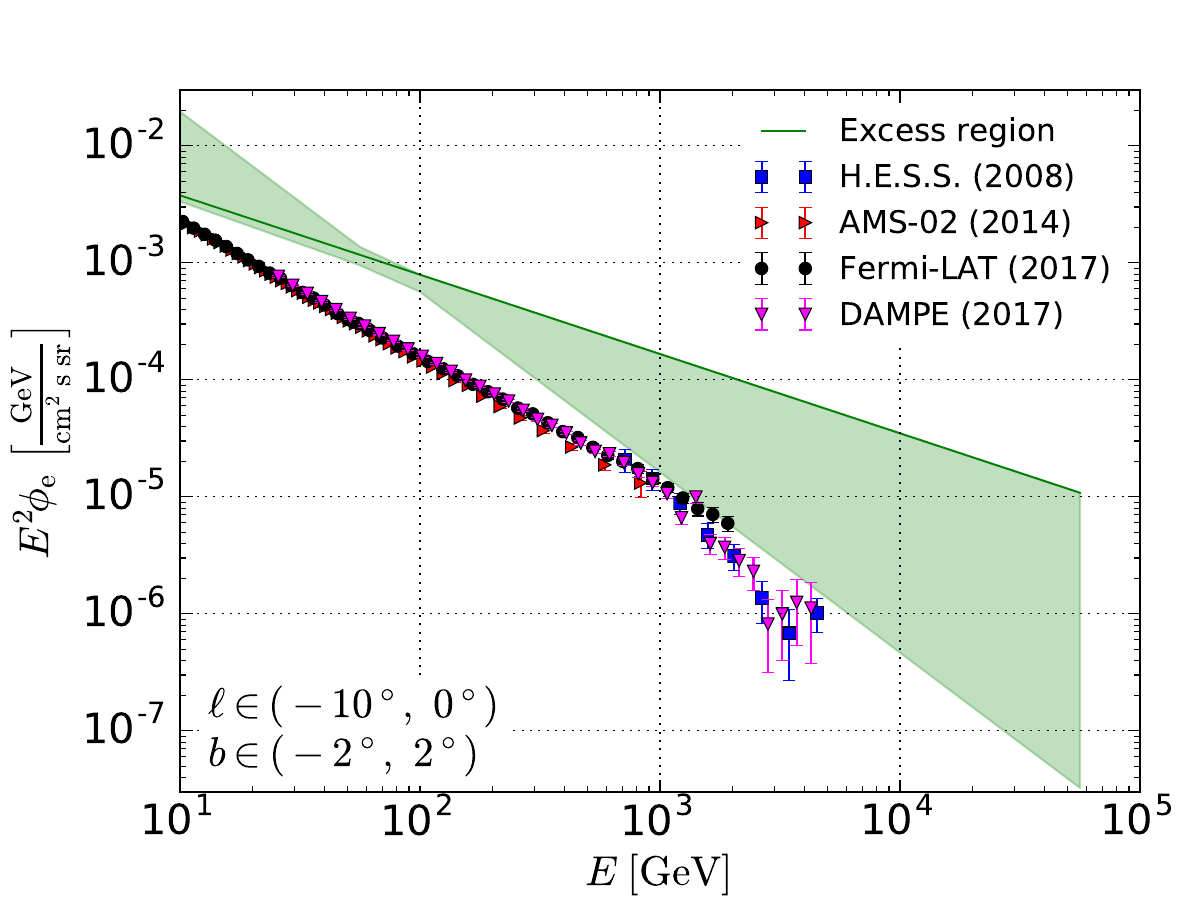}
  	\caption{
	Proton (left) and electron (right) spectra fitted to the gamma-ray emission at the base of the FBs.
	Solid lines show the best-fit spectra in the rectangles model of the FBs.
	The shaded bands represent the systematic uncertainties, 
	estimated from the maximal and minimal best-fit FBs spectra among the foreground models. 
	Uncertainties in the ISRF energy density near the GC result in an additional factor of $\sim$ 2 uncertainties in the CRe energy density 
	(see Appendix \ref{app:ISRF} for details). 
	For comparison with the local CR spectrum, we show the local fluxes of CRp  measured by AMS-02 \citep{2015PhRvL.114q1103A} and 
	CRe measured by H.E.S.S. \citep{2008PhRvL.101z1104A}, AMS-02 \citep{2014PhRvL.113v1102A}, 
	\Fermi-LAT \citep{2017PhRvD..95h2007A}, and \DAMPE \citep{2017Natur.552...63D}.
	The error bars represent the systematic and statistical uncertainties added in quadrature.}
  	\label{fig:Particle_spectra}
\end{figure*}

One of the most intriguing features of the gamma-ray spectrum at the base of the FBs is the absence of a significant cutoff up to 1 TeV and 
an inferred hard spectrum of underlying electrons or protons.
In particular, if the emission originates in hadronic interactions, the spectrum of the CR protons (CRp) is $\sim E^{-2.2}$ which is significantly harder than the propagated spectrum
$\sim E^{-2.5 - 2.7}$ observed both locally 
\citep[e.g.,][]{2015ApJ...806..240C, 2017A&A...601A..78R, 2017A&A...606A..22N, 2018MNRAS.478.2939P}
and throughout the Galactic plane 
\citep[e.g.,][]{2015PhRvD..91h3012G, 2016ApJS..223...26A, 2016PhRvD..93l3007Y, 2018arXiv181112118A}.
Although there is a hardening of the spectrum of CR in the inner rings of the Galaxy and near the GC,
the inferred spectra of the CR at the base of the FBs are significantly harder than the propagated spectra of the CR in the 
inner rings and near the GC \citep{2015PhRvD..91h3012G, 2016ApJS..223...26A, 2016PhRvD..93l3007Y, 2018arXiv181112118A}.
We assume that the CRp spectrum at the base of the bubbles is equal to the injection spectrum unaffected by the 
propagation softening.
This is possible if the CRp were injected relatively recently and had insufficient time to escape from the region of the enhanced emission,
i.e., the age of the CRp is less than the propagation time to cross the vertical distance of 0.9 kpc.
If we assume a spatially constant diffusion coefficient $D(E) = D_0\left(\frac{E}{\SI{1}{GeV}}\right)^\delta$ with 
$D_0 = \SI{3e28}{cm^2/s} = \SI{100}{pc^2/kyr}$ and $\delta = 0.4$ \citep{2007ARNPS..57..285S},
then the escape time for the protons at $E > \SI{6}{TeV}$ is 
\be
T_{\rm esc} < \frac{\Delta h^2}{2 D(E)} \approx \SI{100}{kyr}.
\ee
This gives us an approximate upper bound on the age of the proton CR, 
assuming that the diffusion coefficient near the GC is similar to the diffusion coefficient near Sun.
The escape time can be significantly longer, if the diffusion length near the GC is smaller than the local one
or if the CR are confined by a particular configuration of magnetic fields.
The CRp energy density within $|b| < 6^\circ$ in the rectangles model of the FBs (Section \ref{sec:Pion_model})  
normalized to the reference density of $n_\Hy = \SI{1}{cm^{-3}}$
is $\de E_\tot / \de V = \SI{360}{meV\: cm^{-3}}$ above $\SI{1}{GeV}$.
In Figure \ref{fig:Particle_spectra} (left), we compare the corresponding flux to the local CRp flux.
The total energy in CRp within $|b| < 6^\circ$ is $E_\tot = \SI{7e52}{erg}$.
This population of CRp can be obtained from $\sim$ 700 SNe, 
assuming that on average SNe inject $\sim 10^{50}$ erg in CRp, which corresponds to about 10\% efficiency
of CR acceleration by a SN with kinetic energy $\sim 10^{51}$ erg \citep[e.g.,][]{Spurio2015}.
There are several populations of young stellar objects near the GC \citep{2009ApJ...702..178Y, 2012A&A...537A.121I},
in particular there is a population of stars close to the GC with an age of about 6 Myr \citep{2006ApJ...643.1011P}.
If we assume that the rate of SNe in the Galaxy is about one per century and that the GC region contributes at the level of
a few percent to the total rate, e.g., a few SNRs per 10 kyr, then the 700 SNRs can be produced near the GC
in the past few million years.

The best-fit spectrum of CRe is $\sim E^{-2.7}$, which is softer than the spectrum of the protons.
Nonetheless, this spectrum is harder than the spectrum expected in the presence of cooling
for a stationary population of CRe.
Thus, unless the injection spectrum is harder than $E^{-2}$, the population of the CRe at the base of the 
bubbles is not affected by cooling up to $\SI{3}{TeV}$,
which is the 95\% confidence minimal lower bound on the cutoff in the CRe (Table \ref{tab:IC}). 
The cooling time for the electrons at
$\SI{3}{TeV}$ is $T_{\rm cool} \approx \SI{200}{kyr}$,
which is comparable to the diffusive escape time from the $\pm 900$ pc volume around the GP at $\SI{3}{TeV}$.
The cooling time can have a factor $\sim$ 3 uncertainty due to the uncertainties in the ISRF near the GC (see Appendix~\ref{app:ISRF}).

\begin{figure*}[h]
\centering
 \includegraphics[width=\twopic\textwidth]{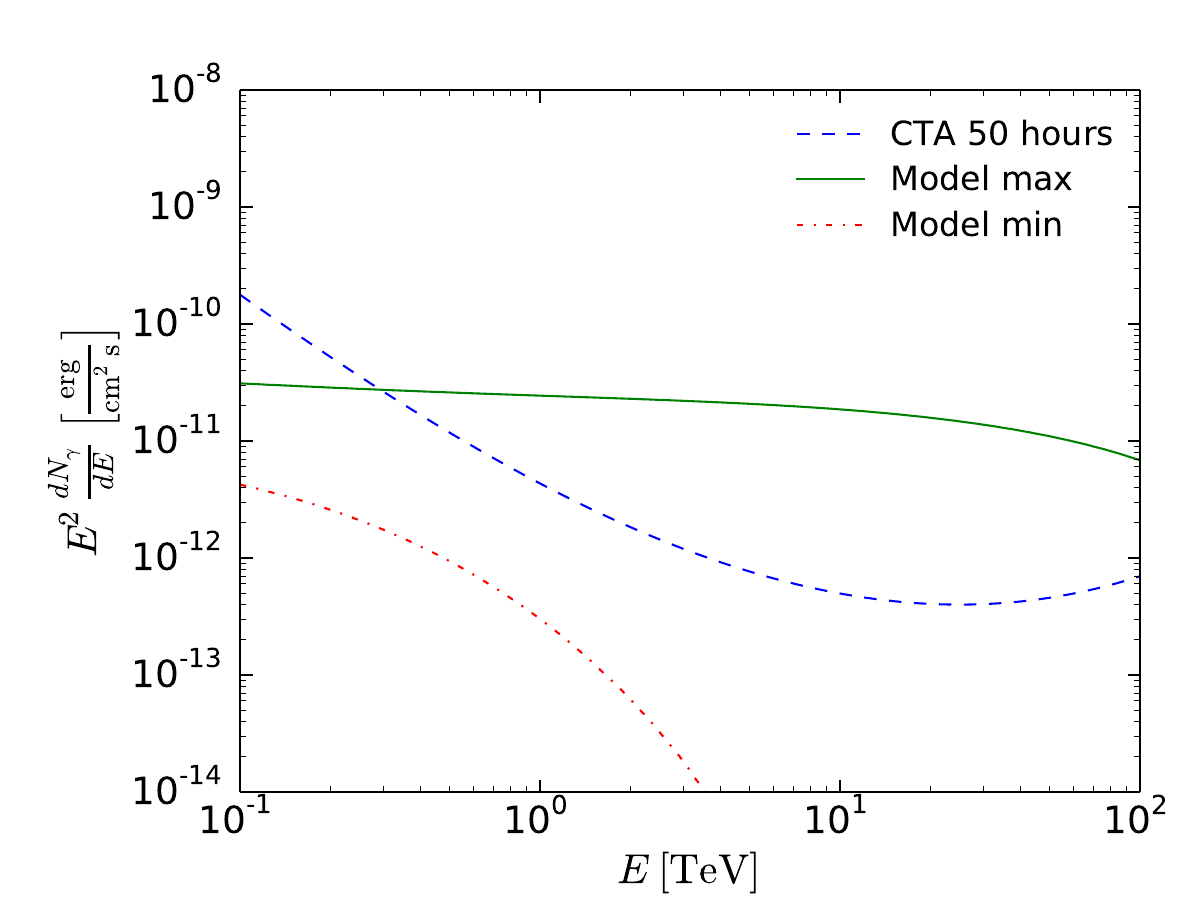}
 \includegraphics[width=\twopic\textwidth]{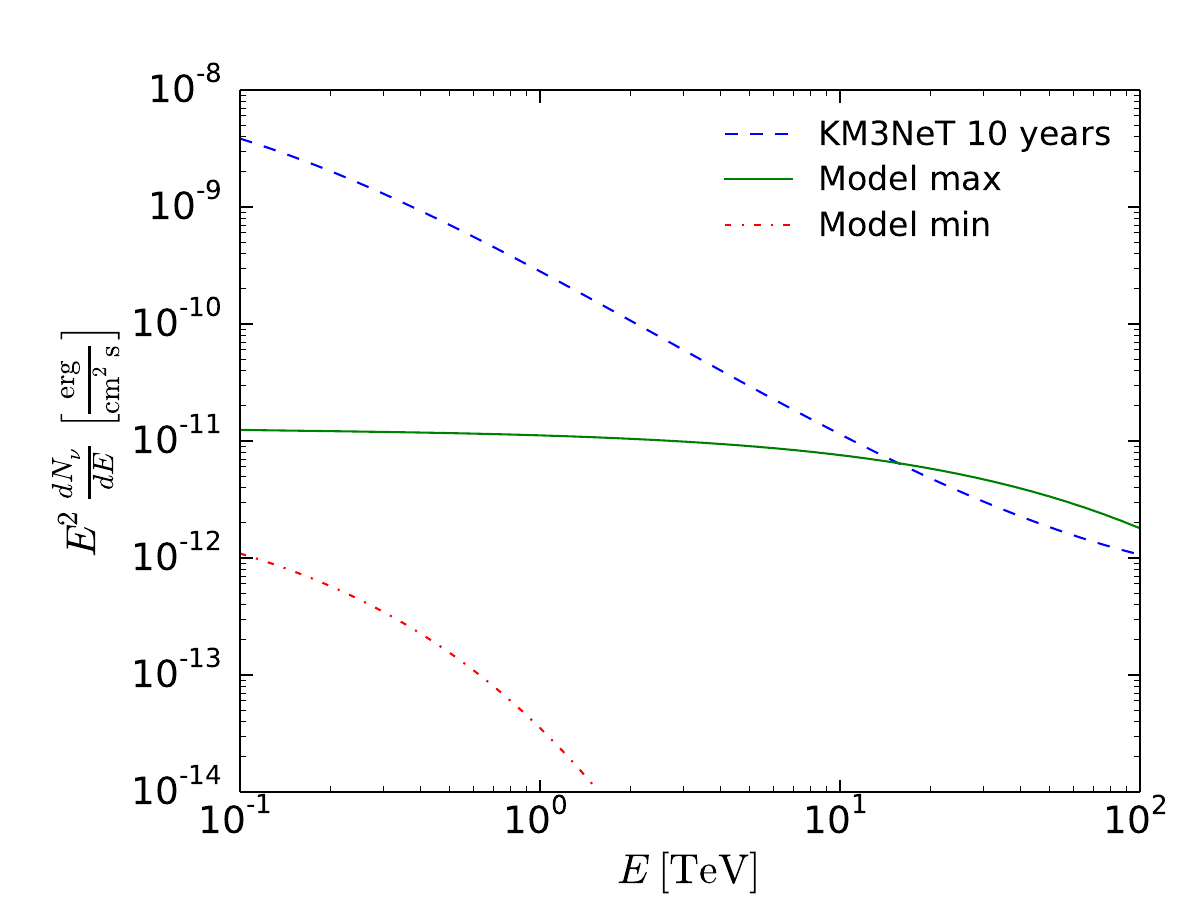}
 \caption{Comparison of the hadronic model of the gamma-ray emission at the base of the FBs
 with the sensitivities of CTA and KM3NeT for a source with $2^\circ$ radius \citep{2018APh...100...69A}.
 The min and max models correspond to the min and max models in
 the hadronic scenario in Table \ref{tab:summary} (for the max model we assume a cutoff in the proton spectrum at 1 PeV)
 integrated over a $2^\circ$ circle to transform intensity to flux.
 The KM3NeT sensitivity is for muon neutrinos only.
 In estimating the sensitivity, we take into account that the GC is below the horizon for about 2/3 of the time
 at the KM3NeT location in the Mediterranean Sea.
 }
 \label{fig:sensitivities}
\end{figure*}

The energy density in electrons with energies above $E_0 = \SI{1}{GeV}$, which is necessary to produce the gamma-ray emission
in the rectangles model of the FBs (Section \ref{sec:IC_model}), is $\SI{4.0}{meV\;cm^{-3}}$.
In Figure \ref{fig:Particle_spectra} (right), we compare the corresponding flux to the local CRe flux.
The total energy content of the ROI in electrons above $\SI{1}{GeV}$ is $E_\tot = \SI{3e51}{erg}$, which corresponds to the CR energy output of 3000 SNe 
assuming a 0.1\% efficiency in converting the SNR kinetic energy to CRe energy.
The ratio of the electron and proton acceleration efficiencies $K_{e/p} = 10^{-2}$ assumed here is relatively optimistic,
given that the ratio of efficiencies can be as low as $\sim 10^{-3}$ \citep[e.g.,][]{2015PhRvL.114h5003P}.
Similar efficiencies of $K_{e/p} \propto 10^{-2}$ can be expected in the acceleration of 
electrons and protons in jets from supermassive black holes \citep[e.g.,][]{2018arXiv180305556B}.
Since the escape time is comparable to the electron cooling time and the required number of SNe in the hadronic
scenario is smaller than the number of SNe in the leptonic scenario (for the ratio of acceleration efficiencies $K_{e/p} \lesssim 10^{-2}$),
we conclude that the majority of the gamma-ray emission near the base of the FBs 
is likely to be produced by the hadronic interactions of CRp with the gas.
The presence of the hadronic production of gamma rays at the base of the FBs, without any sign of a cutoff up to gamma-ray energies of 1 TeV,  
is important
for the detectability of the associated neutrino signal by neutrino telescopes, Figure \ref{fig:sensitivities} \citep[see also][]{2018Galax...6...47R}.

\subsection{A nearby SNR or a superbubble scenario}

In this subsection we discuss models where the hard and bright gamma-ray emission at the base of the FBs is created either by a single SNR
or several SNRs at a distance closer than the distance to the GC.
We start with a single SNR scenario.
For the leptonic model we use the local ISRF and calculate the IC emission as described in Section \ref{sec:IC_model}.
Assuming one SNR, we find a distance of $\SI{50}{pc}$.
The radius of an SNR corresponding to $6^\circ$ circle at a distance of $\SI{50}{pc}$ is about $\SI{5}{pc}$.
Provided that the SNR would be rather close to Earth, it should be detected in radio and X-ray observations,
however no suitable candidates are currently known \citep{2014BASI...42...47G, 2017Green}.
For the hadronic model of the gamma-ray emission, we assume the local gas density of $\SI{1}{cm^{-3}}$.
At the GC, we need about 700 SNRs for this reference gas density; 
thus the distance where a single SNR can explain the excess emission is $d = \frac{8.5\: {\rm kpc}}{\sqrt{700}} \approx 300\:{\rm pc}$.
We can compare the flux from the low-latitude bubbles with the emission of a known SNR, such as Tycho's SNR.
The differential energy flux of Tycho's SNR at 10 GeV is $E^2 {dF}/{dE} \approx 10^{-6}\: {\rm MeV\, cm^{-2} s^{-1}}$ \citep{2017ApJ...836...23A}.
The solid angle of the ROI of $10^\circ \times 12^\circ$ at the base of the FBs is $\Om \approx 0.037$ sr.
If we assume that Tycho's SNR is at a distance of 3 kpc, then the average intensity inside $\Om$ at a distance of 300 pc
is $E^2 {dN}/{dE} \approx 2.7 \times 10^{-6}\: {\rm GeV\, cm^{-2} s^{-1} sr^{-1}}$, which is comparable to the 
FBs emission at latitudes $|b| < 6^\circ$ (e.g., Figure \ref{fig:spec_summary}).
The radius of Tycho's SNR is about 3.5 pc (for the angular diameter of $8'$ at 3 kpc),
while the radius corresponding to $6^\circ$ angular size at 300 pc is about 30 pc, which is almost 10 times larger than the radius of Tycho's SNR.
If we assume that only the core of the emission at the base of the FBs within $|b| < 2^\circ$ corresponds to an SNR,
the corresponding linear size is $\approx 10$ pc (for the same distance of 300 pc).
There are several SNRs in Green's catalog \citep{2014BASI...42...47G, 2017Green} between $-10^\circ < \ell < 0^\circ$,
but all of them have relatively small angular diameters $ \lesssim 30'$.
Thus a single SNR is not a plausible scenario for the emission at the base of the FBs,
unless this SNR is missing in Green's catalog.

\begin{figure}[h]
\hspace{-2mm}
 \includegraphics[width=\onepic\textwidth]{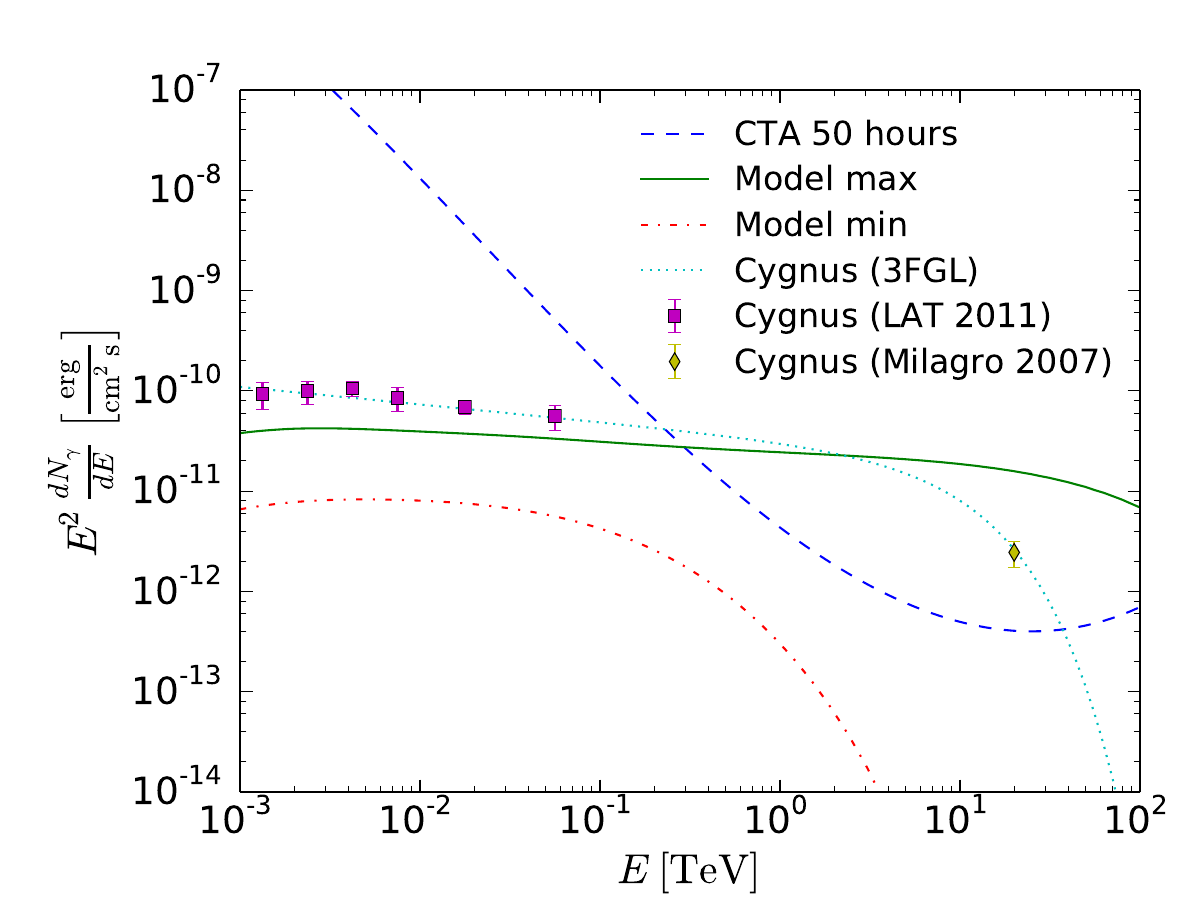}
 \caption{
 Comparison of the min and max models of the residual flux at the base of the FBs with the Cygnus cocoon spectrum. 
 The min and the max models are the same as in Figure \ref{fig:sensitivities}.
 For the 3FGL spectrum we take the power-law parameters from the catalog \citep{2015ApJS..218...23A}
 and add a cutoff in the gamma-ray spectrum at 10 TeV so that the SED is consistent with the Milagro source MGRO J2031+41 
 at 20 TeV \citep{2007ApJ...664L..91A}.
The purple squares represent the measurement of the flux from the Cygnus cocoon by the \Fermi-LAT collaboration
\citep{2011Sci...334.1103A}.
 }
 \label{fig:cygnus}
\end{figure}

Another possibility is that the emission at the base of the FBs is created by several SNRs and / or wind from
massive stars, similar to the Cygnus cocoon \citep{2011Sci...334.1103A} or the Loop I and the local superbubbles 
\citep{2001ApJS..134..283S, 2006A&A...452L...1B, Wolleben:2007pq, 2016Natur.532...73B, 2018Galax...6...26S, 2018Galax...6...56D, 2018Galax...6...62S}.
The Cygnus cocoon \citep{2011Sci...334.1103A} has a radius of about $2^\circ$, 
which at a distance of 1.4 kpc, corresponds to a radius of about 50 pc.
If a Cygnus-like region is responsible for the emission within $|b| < 6^\circ$ at the base of the FBs,
then such region should be at a distance of about 500 pc.
If only the core of the high-intensity region within $|b| < 2^\circ$ is explained by such a cocoon,
then the distance would be similar to the distance to the Cygnus cocoon, i.e., about 1.4 kpc.
In Figure \ref{fig:cygnus} we compare the gamma-ray intensity at the base of the FB integrated over a $2^\circ$ circle
to the flux from the Cygnus cocoon
\cmt{The flux from the Cygnus region around 10 GeV of 
$E^2 {dF}/{dE} \approx 6\times 10^{-5}\: {\rm MeV\, cm^{-2} s^{-1}}$
averaged over a $2^\circ$ circle
corresponds to an intensity of $E^2 {dN}/{dE} \approx 1.6\times 10^{-5}\: {\rm GeV\, cm^{-2} s^{-1}sr^{-1}}$,
which is comparable to the intensity at the base of the FBs within $|b| < 2^\circ$ (Figure \ref{fig:SED_with_fits}),
while the flux from the Cygnus region averaged over a $6^\circ$ circle is comparable to the average intensity
at the base of the FBs within $|b| < 6^\circ$.
Thus the bright gamma-ray emission at the base of the FBs can be explained by a Cygnus-like 
cocoon at a distance of $\sim 500\:{\rm pc} - 1.5 \:{\rm kpc}$ towards the GC.
In Figure \ref{fig:cygnus} we compare the SED in the min and max models at $|b| < 2^\circ$
integrated over a $2^\circ$ circle with the SED measured from the Cygnus cocoon}
\citep{2007ApJ...664L..91A, 2011Sci...334.1103A}.
The flux in the max model is comparable (within a factor of 2) 
to the flux from the Cygnus cocoon between 1 and 100 GeV.

Massive stars, which can inflate cocoons or superbubbles by stellar winds and SN explosions, are usually found in young open stellar clusters.
There are several young open stellar clusters in the direction of the base of the FBs, i.e., within $-10^\circ < \ell  < 0^\circ$.
For example, clusters Trumpler 27 \citep{1977ApJ...215..106M} or NGC 6383 \citep{1978MNRAS.184..661L}
contain more than 10 massive OB stars, have an age of about 10 Myr, and an estimated distance of 1.2 kpc and 1 kpc, respectively,
in the direction of $\ell \approx 356^\circ$, $b \approx 0^\circ$.
There is also a cluster NGC 6405 \citep{1959ZA.....47...15R}, which has a smaller estimated distance of about 500 pc
but an older age $\sim 100$ Myr.
The turbulence created by winds from the massive stars as well as the past SNRs in these clusters can (re)accelerate cosmic rays and
it would also lead to smaller diffusion length, which would prevent the escape of CR on timescales shorter than a few Myr \citep{2011Sci...334.1103A}.
There are relatively few SNRs detected in gamma rays in that area of the sky.
In particular, there is only one SNR, 3FGL J1741.1$-$3053, in the 3FGL catalog \citep{2015ApJS..218...23A}
between $-10^\circ < \ell  < 0^\circ$.
This SNR is associated with the Tornado nebula at a distance of 12 kpc \citep{2013ApJ...774...36C}.
Green's catalog of SNRs
contains several SNRs between $-10^\circ < \ell  < 0^\circ$, but SNRs with measured distances are more than 4 kpc away.
The absence of detected SNRs in the direction of the base of the FBs at distances $\lesssim 1.5$ kpc 
does not necessarily mean that there were no SNRs in the open clusters mentioned above.
A possible reason is that the wind from the massive stars as well as the first SNRs in the cluster create a hot superbubble
with smaller gas density than the average density in the Galactic plane. As a result, the subsequent SNRs quickly expand
in the less dense environment until they reach the common envelope of the superbubble.
For instance, this is observed in simulations of the formation of the local hot superbubble
\citep{2016Natur.532...73B, 2017A&A...604A..81S, 2018Galax...6...26S}.
Whether the stellar clusters can provide the necessary CR injection rate to power the gamma-ray emission at the base
of the FBs deserves a separate detailed study.

An interesting question is whether the FBs themselves can be inflated by a Cygnus-like superbubble,
which accidentally happened along the line of sight towards the GC.
If the FBs are at a distance of about 1 kpc, then their vertical size is also about 1 kpc,
which is 10 times smaller than the size of the FBs at a distance of 8.5 kpc.
In the leptonic scenario of the FBs, the $10^{52}$ erg in CRe at 8.5 kpc \citep{2014ApJ...793...64A}
would correspond to $10^{50}$ erg at 1 kpc.
This power can be provided by several hundred SNe, which is a relatively large number, given the few young stellar 
clusters at distances $\lesssim$ 1 kpc in the direction of the base of the FBs.
This estimate is based on an assumed 0.1\% efficiency in acceleration of CRe by SN shells;
if there is an additional acceleration of the electrons by the turbulence either at the base of the FBs
or inside the high-latitude FB volume, then the required number of SNRs would be smaller.

The electrons can be delivered to the FBs volume by an outflow,
possibly, created by the pressure of the CR themselves \citep[e.g.,][]{2018MNRAS.475..570J}.
The measured red- and blue-shift velocities of the gas outflow in the direction of the FBs are about 200 -- 300 km/s, which imply the gas outflow from the Galactic plane in a biconical model of the FBs with velocities $\gtrsim$ 900 km/s and an age of the outflow at the position above the GC about 6 -- 9 Myr \citep{2015ApJ...799L...7F, 2017ApJ...834..191B}.
If the distance is 10 times smaller, then the age would be $\sim 600 - 900$ kyr,
which is comparable to the cooling time of 1 TeV electrons necessary to explain the 
gamma-ray emission from the FBs at high latitudes \citep{2014ApJ...793...64A}.
Thus the gamma-ray emission at high latitudes could be explained by an IC model with about 100 SNRs
(or fewer if there is re-acceleration of electrons)
created $\sim$ 1 Myr ago, while the bright and hard emission at the base of the bubbles can
be explained by the hadronic emission from about 10 SNRs. 
The actual age of the CR is determined by the confinement within the superbubble and can be also on the order
of 1 Myr or more.
In this scenario,
the star-forming region, which can power the emission at the base of the FBs and, possibly, the FBs themselves
can be situated in the Sagittarius arm at a distance of 1 -- 1.5 kpc in the direction of the GC.

\section{Conclusions}
\lb{sect:concl}

In this paper we use 9 years of \Fermi-LAT Pass 8 Source class data to study the gamma-ray emission
at the base of the FBs.
We use different methods to construct the foreground diffuse gamma-ray components and to determine
the properties of the residual gamma-ray emission at the base of the FBs.
We confirm the earlier findings that the emission at the base of the FB
has a higher intensity than the FBs emission at high latitudes.
The spectrum at the base of the FBs is consistent with a single power law without a cutoff up to about 1 TeV.
The emission is slightly displaced from the GC to negative longitudes,
which favors a starburst scenario of the bubbles formation, unless there is a mechanism to shift the
gamma-ray emission away from Sgr A* in the SMBH scenario.

The gamma-ray emission at the base of the FBs can be explained by either a hadronic or a leptonic model of gamma-ray production
with CR spectra consistent with a power law without a cutoff.
The index of the CRe (CRp) spectrum is 2.6 -- 2.9 (2.0 -- 2.3).
We derive the 95\% confidence lower bound on the cutoff in the CR electron and proton spectra
by selecting the lowest 95\% statistical confidence value among the models of the foreground emission considered 
in Sections \ref{sec:le_data_model} -- \ref{sec:galprop_model}.
Within $|b| < 6^\circ$ and $-10^\circ < \ell < 0^\circ$, the 95\% confidence lower bound on the 
exponential cutoff in the CRe (CRp) spectrum is
about 3 (6) TeV.

If the location of the residual emission is near the GC,
then the total gamma-ray luminosity is $L \approx 10^{37}\ {\rm erg\ s^{-1}}$ and
the required energy of CRe (CRp) above 1 GeV is $E_{\rm e} = \SI{3e51}{erg}$
($E_{\rm p} = \SI{7e52}{erg}$).
Although the total required energy in CRe is smaller than in CRp,
the efficiency of acceleration of hadronic CR is expected to be significantly higher than for leptonic CR.
In particular, with 10\% CRp acceleration efficiency, one needs about 700 SNRs to explain the 
gamma-ray emission in the hadronic scenario near the GC,
while with 0.1\% CRe acceleration efficiency, the required number of SNRs is 3000.
For the GC scenario, the estimated diffusive escape time and the electron cooling time is $\sim 100$ kyr.
In this case one needs a relatively high SN rate of one per $\sim$ 100 yr at the location of the gamma-ray emission.
If the average ISRF energy density is a factor of $\sim$ 2 lower or higher than the ISRF in the reference model (Appendix \ref{app:ISRF}),
then both the cooling time and the required energy density of CRe will be a factor of $\sim$ 2 higher or lower than in the reference model,
which nevertheless requires the same SN rate of one per $\sim$ 100 yr.

If the residual gamma-ray emission is produced at a closer distance, e.g., at about 1 kpc
by a Cygnus-like cocoon or a superbubble in the Sagittarius arm, 
then the required number of SNRs is about two orders of magnitude smaller.
For the superbubble scenario, the characteristic linear size is 10 times smaller, which leads to 100 times shorter
diffusion time of a few kyr. This is shorter than the lifetime of individual SNRs.
Both in the GC and in the superbubble scenarios, the age of the CR population can be greater than the diffusion time if there is an
additional confinement by the SN shells or the shell of a superbubble.

Apart from the question of the location of the emission at the base of the FBs,
one can revisit the question of the location of the FBs themselves:
the FBs could have been inflated by a superbubble about 1 Myr ago
at a distance of 1 kpc from the Sun.
The CR in the past superbubble may create a wind with velocities up to $\sim$ 1000 km/s, 
which is sufficient to inflate a 1 kpc large bubble within 1 Myr
so that the CR electrons in the wind do not cool below 1 TeV and can explain the
gamma-ray emission from the FBs at high latitudes.
The gamma-ray emission at low latitudes is then explained (most naturally) by a hadronic
scenario from a more recent starburst episode at about the same location.

In the future, the bright and hard gamma-ray emission at the base of the FB
should be detectable
with CTA, or a version of the HAWC experiment in the southern hemisphere 
\citep{2017APS..APR.R4005M, 2017arXiv170909624A, 2018Galax...6...47R}.
Observations with Cherenkov telescopes should detect or constrain the cutoff energy in the gamma-ray spectrum in multi-TeV regime.
In the hadronic scenario, for a sufficiently high cutoff value of the CRp spectrum, e.g., around 1 PeV,
the corresponding neutrino emission could be detected with the future KM3NeT telescope.

\subsection*{Acknowledgements}

The authors would like to thank 
Seth Digel, Anna Franckowiak, Stefan Funk, Gudlaugur Johannesson, Tsunefumi Mizuno, Troy Porter, Andrew Strong, and Luigi Tibaldo 
for valuable comments and suggestions.
Also we would like to thank Cristina Popescu for providing us the ISRF data files in fits format.
The Fermi LAT Collaboration acknowledges generous ongoing support from a number of agencies and institutes that have supported both the development and the operation of the LAT as well as scientific data analysis. These include the National Aeronautics and Space Administration and the Department of Energy in the United States, the Commissariat à l'Energie Atomique and the Centre National de la Recherche Scientifique / Institut National de Physique Nucléaire et de Physique des Particules in France, the Agenzia Spaziale Italiana and the Istituto Nazionale di Fisica Nucleare in Italy, the Ministry of Education, Culture, Sports, Science and Technology (MEXT), High Energy Accelerator Research Organization (KEK) and Japan Aerospace Exploration Agency (JAXA) in Japan, and the K. A. Wallenberg Foundation, the Swedish Research Council and the Swedish National Space Board in Sweden.

\newpage
\bibliography{gp_bubbles_papers}  

\begin{appendix}
\section{Modeling and systematic uncertainties}
\lb{sec:lowE_syst}

\begin{figure*}[h]
\centering
\includegraphics[width=\twopic\textwidth]{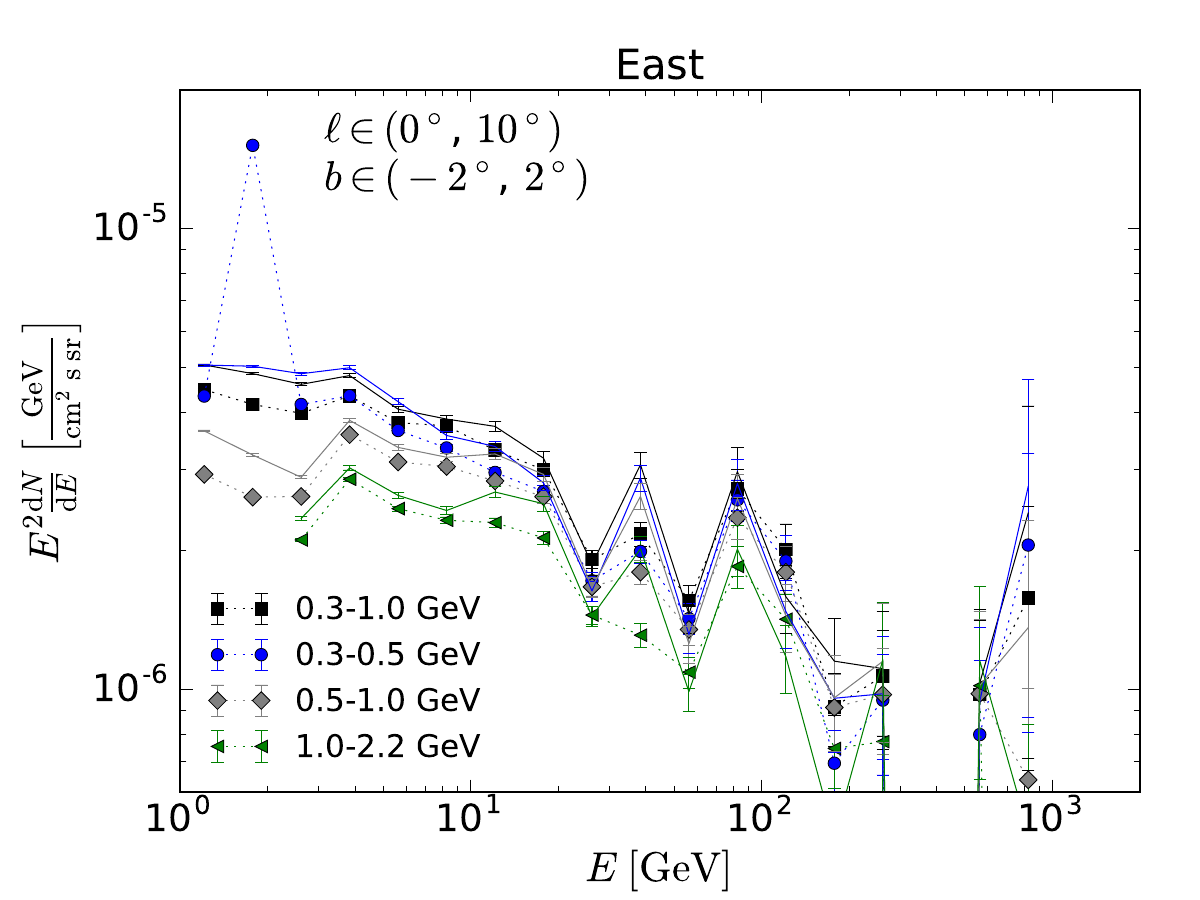}
\includegraphics[width=\twopic\textwidth]{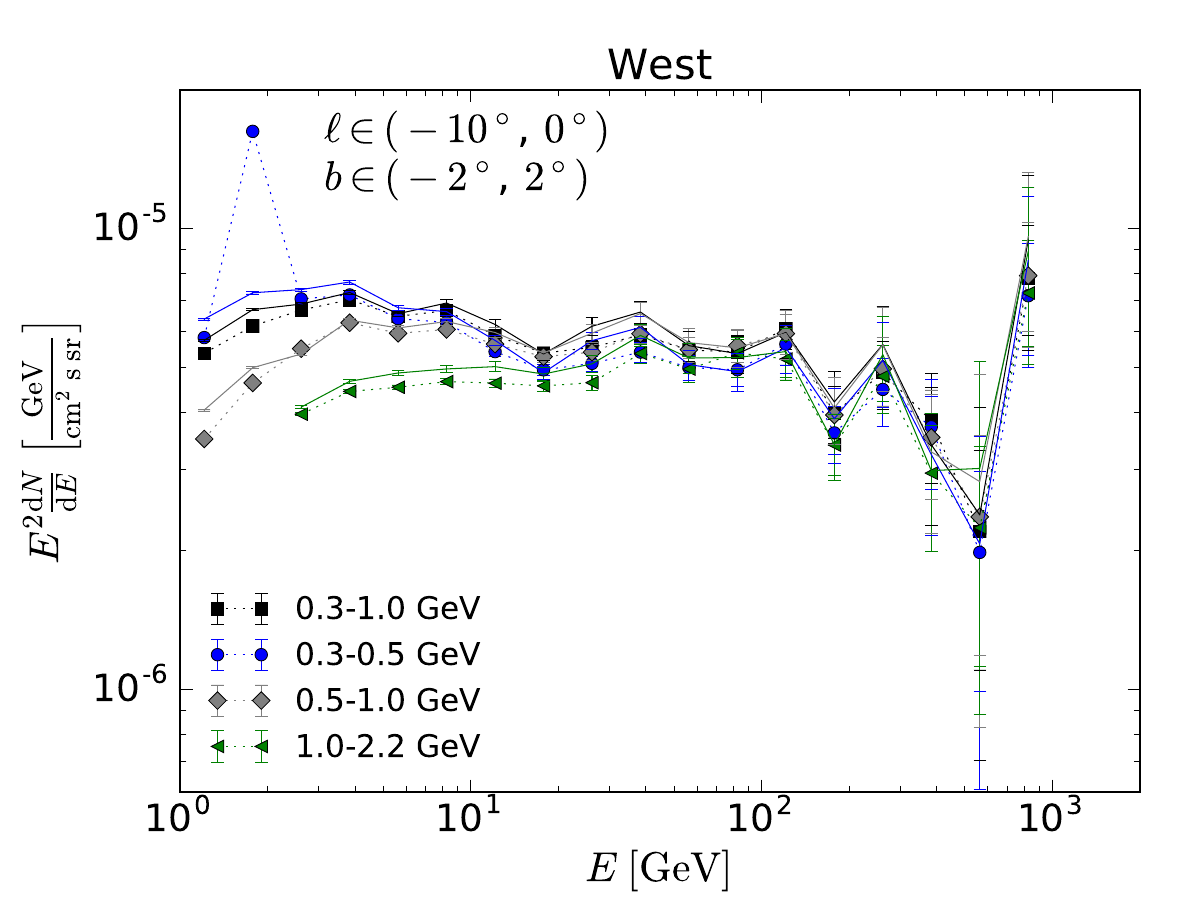}
\caption{SED of the residual in the rectangles model using Source class (dotted lines with markers) and UltraCleanVeto class data (solid lines) 
in four different energy ranges used to determine the foreground diffuse emission template. 
The baseline model, presented in Section \ref{sec:box_model}, has the low-energy range in the definition of the foreground model
$0.3 - \SI{1.0}{GeV}$ (black squares).}
\label{fig:syst_models}
\end{figure*}

In this appendix we study the dependence of the spectrum of gamma-ray emission at the base of the FB
on the selection of the low-energy range in the definition of the diffuse foreground model
and on the choice of the class of the events (UltraCleanVeto vs. Source).
We use the rectangles model of the FB in Section \ref{sec:box_model}
as our baseline.
In this model, the foreground emission template consists of Source data integrated over the energy interval $0.3 - \SI{1.0}{GeV}$. 
To probe the dependence on the choice of the low-energy range in the model, 
we pick three non-overlapping energy ranges $0.3 - \SI{0.5}{GeV}$, $0.5 - \SI{1.0}{GeV}$, and $1.0 - \SI{2.2}{GeV}$ 
and use the same analysis as for the baseline model. 
We also compare with the spectra obtained using UltraCleanVeto data both in the definition of the model
at low energies and in the analysis at high energies.
The residual spectra resulting from the different energy ranges and data classes are shown in Figure \ref{fig:syst_models}. 
The agreement among the models is reasonably good at $E > 10$ GeV, especially for negative longitudes.

As one can see from Figure \ref{fig:syst_models}, 
the highest-energy spectral point for negative longitudes may have an upward statistical fluctuation,
which can influence the conclusions about the lower limit in the cutoff values.
In order to test the dependence on the highest-energy spectral point, we exclude the energy bin $680\;{\rm GeV} - 1\;{\rm TeV}$ and 
repeat the derivation of the best-fit parameters in the rectangles model of the FBs.
The results are presented in Table \ref{tab:param2}.
Without the last data point, the four areas with ``infinite'' cutoff energy have now a finite cutoff,
but with a significance $2\Delta \log L  < 1$, i.e., the models are still consistent with a simple power law.
The 95\% confidence lower limit on the cutoff value is also smaller for the dataset without the last point,
but in this case we are using the data up to 680 GeV rather than 1 TeV.

\begin{table*}
  \begin{center}
    \caption{Best-fit parameters of the rectangles model without the last energy bin: $680\;{\rm GeV} - 1\;{\rm TeV}$.
    The definitions of the columns are the same as in Table \ref{tab:param}.
    \lb{tab:param2}
    }
        \begin{tabular}{|c|c|c|c|c|c|c|c|}
     	\hline
		 Lat & Lon  & $N_0$ & $n$ & $E_{\rm cut}$ &  $2 \Delta \log \La$ & $E_{\rm cut, 95\%}$ & $E_{\rm cut, 95\%}^{\rm min}$ \\ 
		     &        &  {\small $\SI{e-6}{GeV^{-1}cm^{-2}s^{-1} sr^{-1}}$ }&  & {\small $\SI{}{GeV}$ }& &{\small  $\SI{}{GeV}$ }&{\small  $\SI{}{GeV}$ }\\ 
		\hline
  		$(\ang{2}, \ang{6})$ & $(\ang{0}, \ang{10})$ & $1.5 \pm 0.4$ & $1.9 \pm 0.2$ & $45 \pm 22$ & 6.8 &25 & 25\\ 
		& $(\ang{-10}, \ang{0})$ & $3.0 \pm 0.4$ & $2.2 \pm 0.07$ & $720 \pm 1100$ & 0.45 & 210 & 210\\ 
 		\hline
  		$(\ang{-2}, \ang{2})$ & $(\ang{0}, \ang{10})$ & $6.1 \pm 1.0$ & $2.3 \pm 0.08$  & $660 \pm 1000$ & 0.42 & 180 & 2.0 \\ 
		& $(\ang{-10}, \ang{0})$ & $7.5 \pm 0.8$ & $2.1 \pm 0.04$ & $1400 \pm 1500$ & 0.76 & 490 & 460\\ 
 		\hline
  		$(\ang{-6}, \ang{-2})$ & $(\ang{0}, \ang{10})$ & $2.5 \pm 0.3$ & $2.1 \pm 0.07$ & $260 \pm 160$ & 3.4 & 130 & 2.3  \\ 
		& $(\ang{-10}, \ang{0})$ & $3.9 \pm 0.4$ & $2.1 \pm 0.05$ & $790 \pm 820$ & 0.96 & 290 & 290\\ 
 \hline
    \end{tabular}
  \end{center}
\end{table*}

\newpage
\section{Gamma distribution as a likelihood for smoothed data}
\lb{app:gamma}

In this appendix we show that smoothing the data and using gamma distribution instead of the Poisson distribution
is a reasonable procedure (at least in the case when the scale of variations in the diffuse emission is larger than the smoothing radius).
We split the log likelihood into a sum over areas with size approximately equal to the smoothing scale
and neglect the correlation between the different areas after smoothing the data.
Suppose that one of these areas has $n$ pixels with random counts $k_i$ drawn from a Poisson distribution with mean $\ld_0$
(here we assume that the underlying diffuse flux is approximately constant within the smoothing radius).
The likelihood function for parameter $\ld$ is
\be
L(\ld) = \prod_i \frac{\ld^{k_i}}{k_i !} e^{-\ld}.
\ee
The log likelihood is
\be
\log L = \sum_{i = 1}^n (-\ld + k_i \log \ld) + const.
\ee
The maximum likelihood value $\ld_*$ is determined from
\be
0 = \frac{\p \log L}{\p \ld} = -n + \frac{1}{\ld} \sum_{i = 1}^n k_i,
\ee
which gives $\ld_* = \frac{1}{n} \sum_{i = 1}^n k_i$.
The uncertainty is
\be
\frac{1}{\sm^2} = - \left. \frac{\p^2 \log L}{\p \ld^2} \right|_{\ld = \ld_*} = \frac{n}{\ld_*},
\ee
which gives $\sm^2 = \ld_* / n$.

If the smoothing radius is comparable to the size of the region,
then we can approximate the values $k_i$ with the average in the region $\tilde{k}_i = \bar{k} = \frac{1}{n} \sum_{i = 1}^n k_i$
(note that $\tilde{k}_i$ are not integers).
We take the gamma distribution for the likelihood function 
\be
\tilde{L} = \prod_i \frac{\ld^{\td{k}_i}}{\G(\td{k}_i + 1)} e^{-\ld}.
\ee
The log likelihood is
\be
\log \td{L} = \sum_{i = 1}^n (-\ld + \td{k}_i \log \ld) + const.
\ee
The maximum likelihood solution is the same as in the Poisson case $\td{\ld}_* = \bar{k} = {\ld}_*$.
The uncertainty is also the same as in the Poisson case:
\be
\frac{1}{\td{\sm}^2} = - \left. \frac{\p^2 \log \td{L}}{\p \ld^2} \right|_{\ld = \td{\ld}_*} = \frac{n}{\td{\ld}_*} = \frac{1}{\sm^2}.
\ee
Thus, smoothing the data and using the gamma distribution (in this case) gives the same result as using the original Poisson distribution,
which shows that the procedure is well defined from the statistical point of view and it gives reasonable results.

\section{Modeling uncertainty of ISRF near the GC} 
\lb{app:ISRF}

\begin{table*}
  \begin{center}
    \caption{\label{tab:CRe_syst} 
Spectra of CR electrons near the GC relative to the reference model based on the ISRF calculation of \cite{Porter:2008ve} used in GALPROP v54.1.
}
\begin{tabular}{| l |c|c|}
\hline
ISRF model & Relative normalization at 1 GeV  & Index \\
\hline
\cite{Porter:2008ve} & 1 & 2.71 \\ 
\cite{2017ApJ...846...67P} R12 & 0.59 & 2.61 \\ 
\cite{2017ApJ...846...67P} F98 & 1.16 & 2.71 \\ 
\cite{2017MNRAS.470.2539P} & 1.64 & 2.79 \\ 
 \hline
    \end{tabular}
  \end{center}
\end{table*}

\begin{figure*}[h]
\centering
\includegraphics[width=\twopic\textwidth]{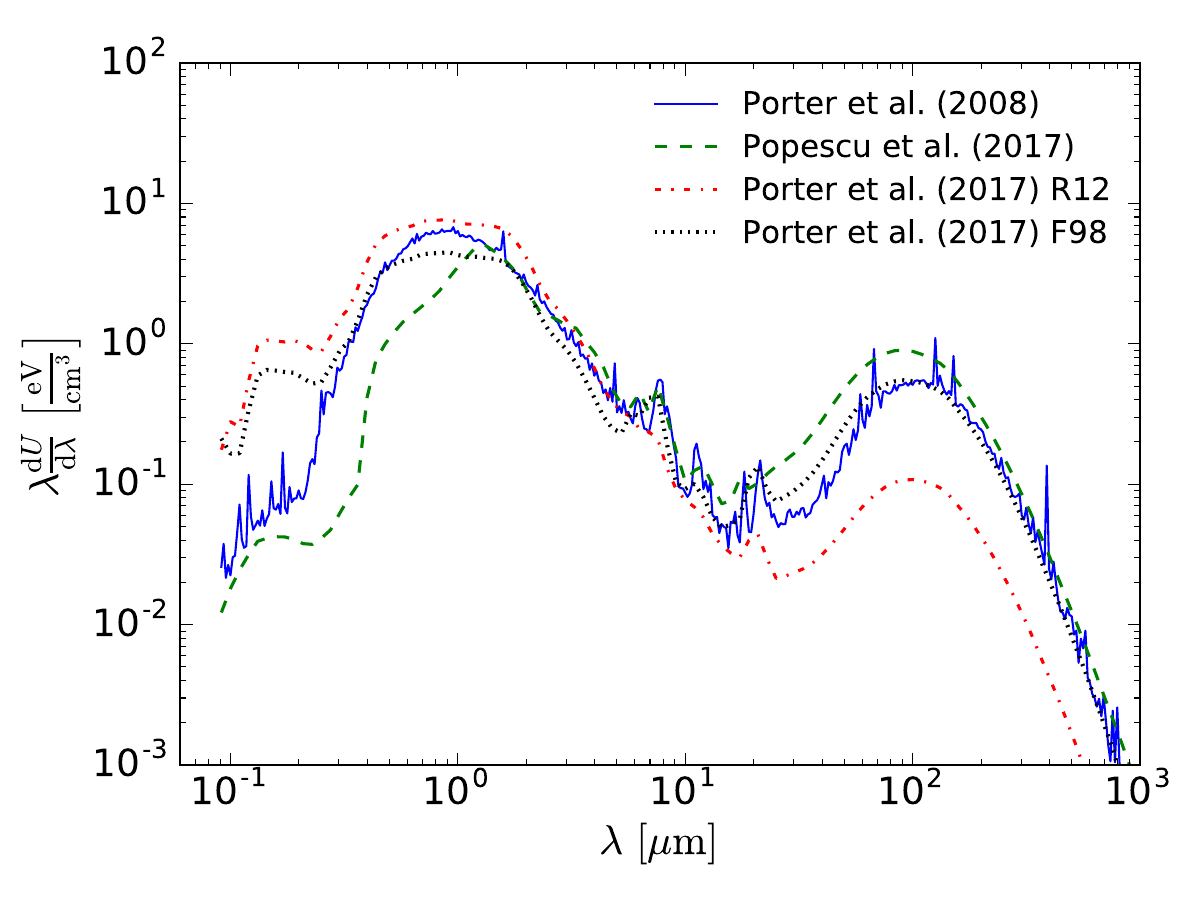}
\caption{Comparison of ISRF SEDs at the GC (see text for more details).}
\label{fig:isrfs}
\end{figure*}

In this appendix we discuss the uncertainty of the IC model of the gamma-ray emission at the base of the FBs 
related to modeling of the ISRF near the GC.
In order to estimate this uncertainty, we compare the ISRF model of 
\cite{Porter:2008ve} (available with the distribution of GALPROP v54.1),
the ISRF model of \cite{2017MNRAS.470.2539P},
and two ISRF models of \cite{2017ApJ...846...67P}: 
R12, based on \cite{2012A&A...545A..39R},
and F98, based on \cite{1998ApJ...492..495F}.
In Figure \ref{fig:isrfs} we show the corresponding densities of ISRFs averaged over the cylinder with the 
radius $R = 1.5$ kpc and the height $z = \pm 0.3$ kpc around the GC, which approximately corresponds 
to the latitude stripe $|b| < 2\degr$ and $|\ell| < 10\degr$.
We notice that there can be up to a factor of 2 differences in the ISRF energy density at the peak 
of the SL emission (around 1 $\mu$m) 
and an order of magnitude differences in the IR wavelengths (around 100 $\mu$m),
which can affect the inferred populations of CR electrons producing the IC gamma rays 
\citep[see also][]{2017ApJ...846...67P, 2019APh...107....1N}.

\begin{figure*}[h]
\centering
\includegraphics[width=\twopic\textwidth]{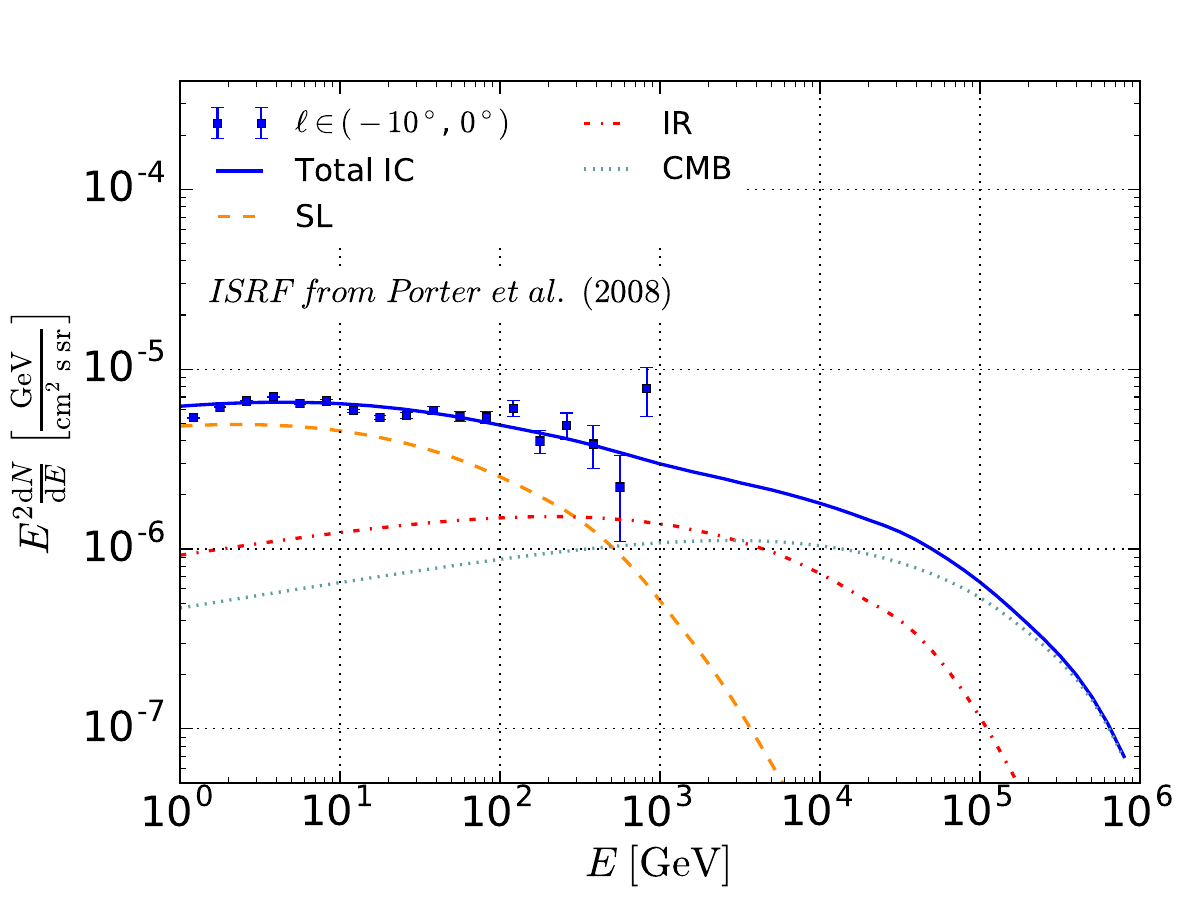}
\includegraphics[width=\twopic\textwidth]{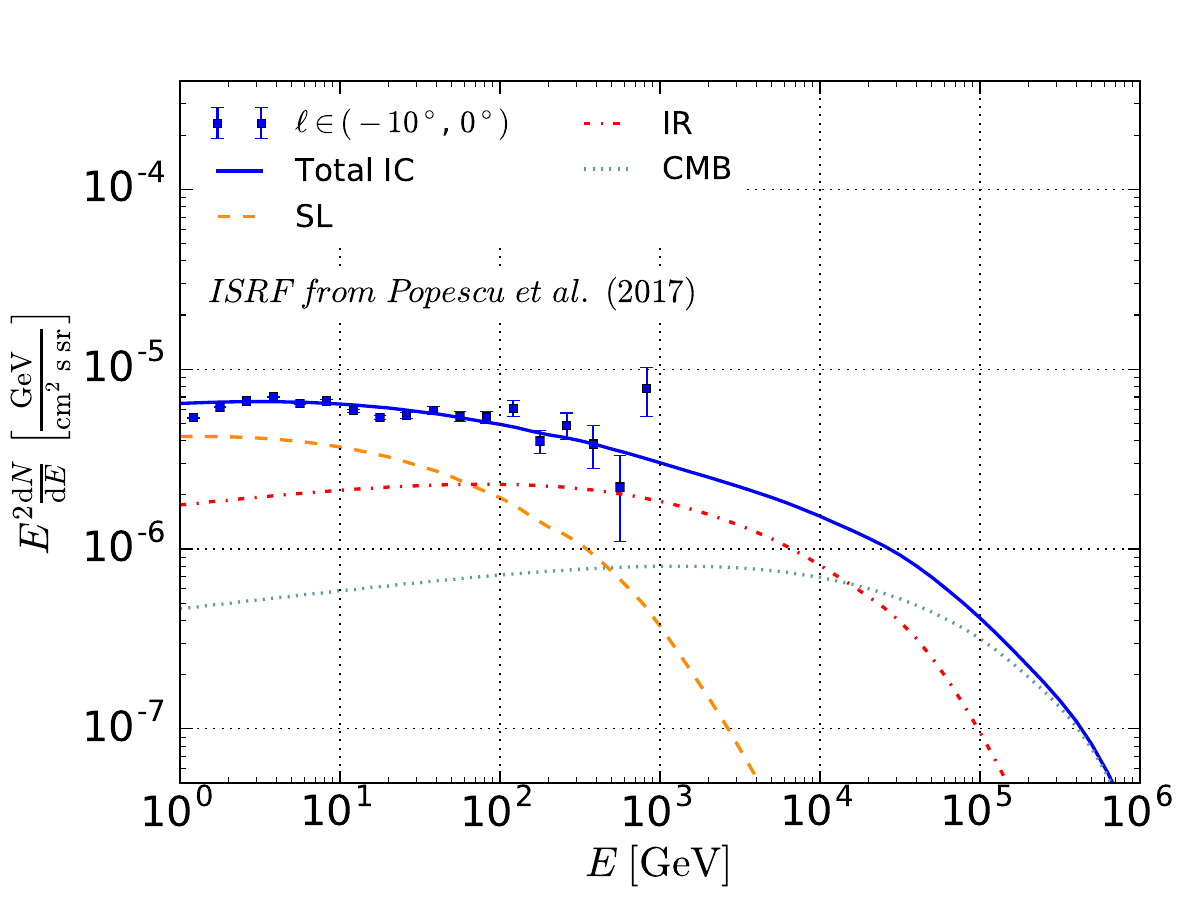}
\includegraphics[width=\twopic\textwidth]{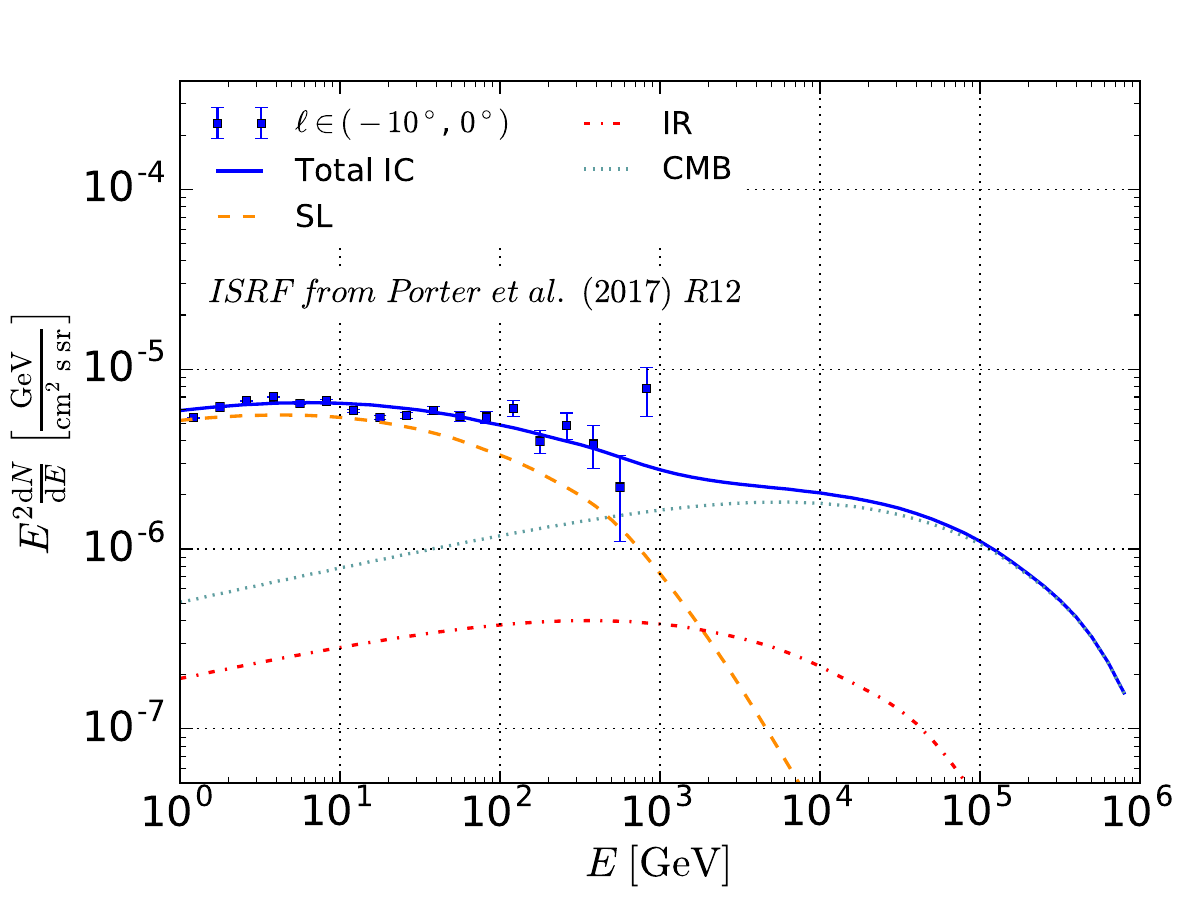}
\includegraphics[width=\twopic\textwidth]{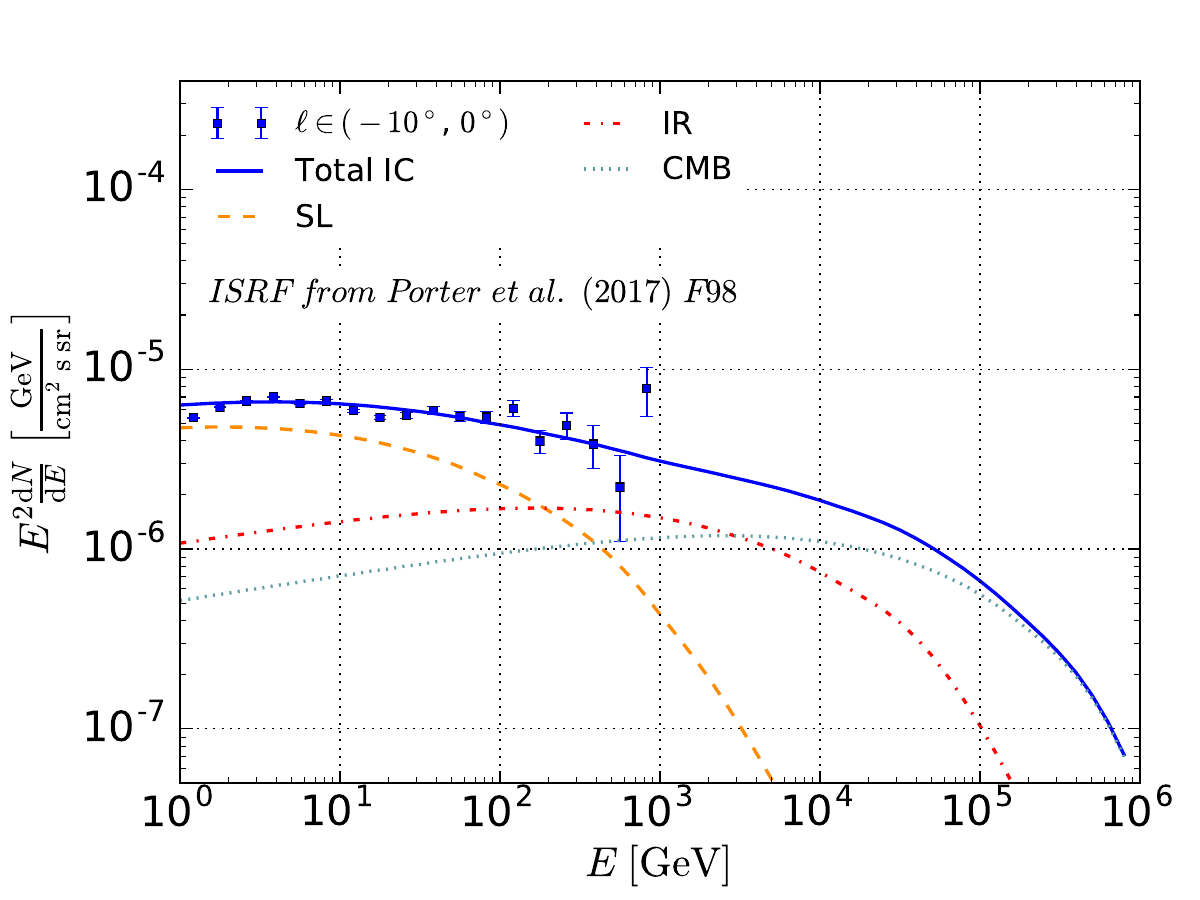}
\caption{Contribution of different components of ISRFs to the IC model of the gamma-ray emission.
The data points correspond to the emission at the base of the FBs in the rectangle $b \in (-2\degr, 2\degr)$, $\ell \in (-10\degr, 0\degr)$
(middle panel in Figure \ref{fig:SED_with_fits}).
The spectrum of CR electrons is modeled as a power-law function with an exponential  cutoff at 1 PeV.
We separate the SL and the IR contributions to the ISRF by formally splitting the ISRF at 0.1 eV.}
\label{fig:GC_CR}
\end{figure*}

In Figure \ref{fig:GC_CR} we show the IC models of the gamma-ray emission 
at the base of the FBs in the rectangle $b \in (-2\degr, 2\degr)$, $\ell \in (-10\degr, 0\degr)$
for the different ISRF models near the GC.
We separate the SL, IR, and CMB contributions.
The SL and IR ISRFs are separated by splitting the radiation field energy densities at 0.1 eV ($\approx 12\ \mu$m).
We model the CR electrons spectra by a power-law function with an exponential cutoff.
We fix the cutoff at 1 PeV and fit the normalization and the spectral index by fitting 
the IC model to the \Fermi-LAT spectral points.
The corresponding parameters are presented in Table \ref{tab:CRe_syst}.
We use the \cite{Porter:2008ve} ISRF as the reference model and show the normalizations of the CRe spectra
relative to the reference model (the normalizations are determined at 1 GeV).
The overall spread in the normalizations is about a factor of 3,
while the overall variation of the index of the CRe spectra is less than about 0.2.

\end{appendix}

\end{document}